%% file: main.tex
\documentclass[journal,twocolumn,10pt,nofonttune]{IEEEtran}

\usepackage{amsmath,amsfonts,amssymb}
\usepackage{algorithmic}
\usepackage{array}
\usepackage{textcomp}
\usepackage{multirow}
\usepackage{stfloats}
\usepackage{url}
\usepackage{verbatim}
\usepackage{graphicx}
\usepackage{xspace}
\usepackage{xcolor}
\usepackage{soul}
\usepackage{booktabs}
\usepackage{listings}
\usepackage{enumitem}
\usepackage{subcaption}
\usepackage{balance}
\usepackage[hidelinks]{hyperref}
\usepackage[acronyms,nonumberlist,nopostdot,nomain,nogroupskip,acronymlists={hidden}]{glossaries}
\usepackage[numbers,sort&compress]{natbib}
\usepackage[font=footnotesize]{caption}
\usepackage[font=footnotesize]{subcaption}

\usepackage{tikz}
\usepackage{pgfplots}
\usetikzlibrary{patterns}
\usetikzlibrary{calc,arrows.meta,positioning}
\usepgfplotslibrary{groupplots}
\pgfplotsset{compat=1.18}

\def\BibTeX{{\rm B\kern-.05em{\sc i\kern-.025em b}\kern-.08em
    T\kern-.1667em\lower.7ex\hbox{E}\kern-.125emX}}

\newglossary[algh]{hidden}{acrh}{acnh}{Hidden Acronyms}
\newcommand{\framework}{MAC-Gyver\xspace}
\newcommand{\macemu}{\texttt{mac-emu}\xspace}

\newcommand{\freqselfixed}{\textsc{FreqSel-Fixed}\xspace}
\newcommand{\freqselrefined}{\textsc{FreqSel-Refined}\xspace}

\newlength\fheight
\newlength\fwidth

\input{acronyms}

\definecolor{codegreen}{rgb}{0,0.6,0}
\definecolor{codegray}{rgb}{0.5,0.5,0.5}
\definecolor{codepurple}{rgb}{0.58,0,0.82}
\definecolor{backcolour}{rgb}{0.95,0.95,0.92}

\lstdefinestyle{mystyle}{
    backgroundcolor=\color{backcolour},
    commentstyle=\color{codegreen},
    keywordstyle=\color{blue},
    numberstyle=\tiny\color{codegray},
    stringstyle=\color{codepurple},
    basicstyle=\ttfamily\footnotesize,
    breakatwhitespace=false,
    breaklines=true,
    captionpos=b,
    keepspaces=true,
    numbers=left,
    numbersep=5pt,
    showspaces=false,
    showstringspaces=false,
    showtabs=false,
    tabsize=2,
}
\setlist[itemize]{leftmargin=*}

\begin{document}

\bstctlcite{IEEEtranBSTcontrol}

\title{MAC-Gyver: Open, Programmable, Scheduling\\for AI-RAN 6G Systems}

% \framework: Open, Programmable, Scheduling for AI-RAN 6G Systems

\author{Maxime~Elkael, Reshma Prasad, Tamerlan Aghayev, Salvatore D'Oro, Michele Polese, Tommaso Melodia%
  \thanks{The authors are with the Institute for Intelligent Networked Systems, Northeastern University, Boston, MA, USA. E-mail: \{m.elkael, r.prasad, aghayev.t, s.doro, m.polese, melodia\}@northeastern.edu.}
  \thanks{This work was supported by the U.S. NSF under award TI-2449452 and by OUSW(R\&E) through Army Research Laboratory
Cooperative Agreement Number W911NF-24-2-0065. The views and conclusions
contained in this document are those of the authors and should not be
interpreted as representing the official policies, either expressed or implied,
of the Army Research Laboratory or the U.S. Government. The U.S. Government is
authorized to reproduce and distribute reprints for Government purposes
notwithstanding any copyright notation herein.}
   }

% \markboth{IEEE Transactions on Mobile Computing}%
% {Elkael: MAC-Gyver}

\maketitle

\begin{abstract}
Cellular networks are integrating \gls{ai} into radio access network control. The MAC scheduler is a promising target because it allocates a limited resource, spectrum, at every slot, under competing latency, throughput, and reliability requirements. However, most learning-based schedulers are evaluated only in simulation. Production schedulers are difficult to modify, and realistic stress tests require more radio hardware than most laboratories can provide. We present \framework, an open-source framework for developing and evaluating scheduling applications that execute directly inside the OpenAirInterface scheduler. It exposes scheduler observations and controls through typed interfaces while preserving the underlying protocol and real-time execution paths. The same applications run over the air and in \macemu, a PHY-less emulator that executes the production OpenAirInterface Layer~2 stack for up to 90 users on one host at real-time slot pace, with a 3GPP-compliant channel model. To showcase the flexibility of \framework, we evaluate two use cases. A proactive \gls{ul} scheduler predicts packet arrivals and roughly halves median round-trip latency. A frequency-selective \gls{ul} scheduler selects contiguous sub-bands from per-PRB sounding observations and is evaluated across mobility and power-limited operating points against an offline scheduling ceiling. Together, they show how the same production stack can be an AI playground that supports implementation, controlled evaluation, and \gls{ota} validation through complementary scheduling use cases.
\end{abstract}

\begin{IEEEkeywords}
5G, 6G, AI-RAN, MAC Scheduler, OpenAirInterface, dApp, Network Emulation.
\end{IEEEkeywords}

\glsresetall
\glsunset{nr} % not an acronym

\section{Introduction}
\label{sec:intro}

Research on \gls{6g} cellular systems increasingly considers \gls{ai} as part of the \gls{ran} design. This is a key principle within the AI-RAN Alliance, which distinguishes three forms of integration: (i)~AI-and-RAN, where \gls{ai} and \gls{ran} workloads share infrastructure; (ii)~AI-on-RAN, where the \gls{ran} serves edge \gls{ai} workloads; and (iii)~AI-for-RAN, where \gls{ai} improves \gls{ran} operation and its \glspl{kpi}~\cite{kundu2025ai}.

We study AI-for-RAN at the \gls{mac} layer. At every slot, the scheduler selects the users to serve, assigns spectrum, and enforces the \gls{qos} requirements of traffic ranging from high-throughput video to \gls{urllc}~\cite{ts38321,anand2020joint}. These decisions trade throughput and fairness against latency and reliability on a sub-millisecond timescale, while considering scarce resources, i.e., bandwidth, and managing user mobility, \gls{mimo}, and other requirements~\cite{lee2009proportional,altam2020leasch,anand2020joint}. Consequently, many studies apply \gls{ai} and \gls{ml} to \gls{mac} scheduling, but most evaluate their algorithms through analysis or simulation~\cite{kela2024simulation}. Only a small subset reaches a production stack~\cite{polese2022colo,doro2022dapps,barker2025real}.

This gap is primarily an engineering problem. In open platforms such as \gls{oai} and OCUDU, the scheduler is intertwined with the protocol stack. Implementing a new policy requires understanding thousands of lines of C/C++ together with their timing and thread interactions.
% and still needs to preserve the existing protocol behavior. 
Observability is also coarse: signals such as per-\gls{prb} \gls{snr} require further modifications to the stack. These costs keep most learning-based schedulers in simulation, where protocol overhead, timing constraints, and radio impairments are abstracted and necessarily approximated.

To address these limitations, we introduce \framework, an open and programmable \gls{mac}-layer framework. First, \framework decomposes the scheduler into a fixed pipeline of typed policy stages. A researcher implements a scheduling policy as a Layer~2 \gls{dapp}~\cite{doro2022dapps,lacava2025dapps} and selects it from the \gls{gnb} configuration at boot. The \gls{dapp} executes synchronously in the scheduler and reads the same state as the default policy. For policies that operate across scheduling opportunities, \framework additionally lets \glspl{dapp} observe transmission outcomes, and retain persistent context. Second, \macemu replaces the radio with a 3GPP channel model while preserving the protocol stack through \gls{l2}. We implement \framework in \gls{oai}, and show how it runs one \gls{gnb} and up to 90 concurrent \glspl{ue} on a single host in real-time. Thanks to this design choice, an \gls{ota} testbed with real radios and \macemu use the same scheduling interfaces, policies, and event logs. Thus, any policy developed in \framework can be ported to the \gls{ota} network without re-implementation. The testbed provides real channels and commercial devices; \macemu provides controlled channels, repeatable experiments, and cell-scale contention. The connection across emulation and real-world deployment is an advantage compared to traditional discrete-event simulators (e.g., ns-3~\cite{riley2010ns}), as they only evaluate schedulers in simulated time. In this way, \framework provides a programmable framework for \gls{l2} prototyping and bridges the gap between research and real-world demonstrations.
% \framework executes the policy deployed at the base station.

The contributions of this paper are as follows:
\begin{itemize}
\item We decompose the per-slot \gls{ul} and \gls{dl} scheduler into typed policy stages that share one candidate structure. The stages expose channel observations, buffer state, \gls{harq} feedback, and the decisions needed for online control and offline analysis. \framework also lets dApps observe transmission outcomes, and retain persistent context across scheduling opportunities. We have implemented this design on the \gls{oai} scheduler, and it has already been contributed upstream~\cite{oaischedrefactor}.
\item We develop \macemu, a \gls{phy}-less emulator that runs one \gls{oai} \gls{gnb} and up to 90 \glspl{ue} on one host in real-time. It uses the same policies and logs as the \gls{ota} deployment.
\item Thanks to \framework's \gls{dapp}-based architecture, we demonstrate stateful, closed-loop scheduling with a proactive \gls{ul} scheduler that learns packet arrivals online and schedules grants before the \gls{ue} requests them. This application was presented in our AUGUSTE conference article~\cite{elkael2026auguste}, of which this paper is an extension. In \gls{ota} experiments with commercial \glspl{ue}, it reduces median \gls{rtt} by about 50\% (20~ms to 10~ms) and one-way \gls{ul} latency from 20~ms to 7~ms, at 7\% overhead.
\item We use \framework and \macemu to design and evaluate a frequency-selective \gls{ul} scheduler from per-\gls{prb} sounding observations
% . Although offline analysis favors a more flexible allocator, execution in the complete scheduling loop shows that the simpler policy performs better. The retained policy 
that improves aggregate throughput by up to 7\%, with gains determined by channel freshness and the users' power-limited operating point.
\end{itemize}

The remainder of this paper is organized as follows. Section~\ref{sec:designspace} presents the design tradeoffs and the scheduler design, while Sec.~\ref{sec:macemu} presents \macemu. Sections~\ref{sec:proactive} and~\ref{sec:freqsel} evaluate the two applications. Section~\ref{sec:related} reviews related work, and Sec.~\ref{sec:conc} discusses the lessons learned and concludes.

\section{\framework Scheduler Architecture}
\label{sec:designspace}

In this section, we first discuss the constraints and tradeoffs involved in scheduler design. We then use these to motivate \framework's scheduler architecture, designed to make scheduling policies easy to modify and extend.

\subsection{Scheduler Design Constraints}
\label{sec:sched_constraints}
% Coupled decisions
% Protocol/timing feasibility
% Application tradeoffs

At each scheduling opportunity (i.e., slot), the scheduler decides which \glspl{ue} to serve and selects their transmission parameters such as the \gls{mcs}, the allocated \glspl{prb}, the \gls{tda} (i.e., which \gls{ofdm} symbols), the beamforming configuration, and the transmit power. Scheduling is therefore a real-time problem, solved every 500~$\mu$s at 30~kHz \gls{scs}, over an action space with several decision variables for each \gls{ue}.

These decision variables are coupled, both with each other and across slots. For example, the \gls{ue}'s power budget ties the number of allocated \glspl{prb} to the achievable \gls{mcs} within a slot, while across slots an \gls{ul} grant must be announced on the \gls{pdcch} during an earlier \gls{dl} slot, so the configured $K_2$ values, the \gls{tdd} pattern, and the available control-channel resources determine which future \gls{ul} slots are reachable. Further constraints, e.g., contiguous \gls{ul} \gls{prb} allocation and the timing of \gls{harq} retransmissions and their \gls{pucch} feedback, restrict the feasible schedules. Consequently, formulations that optimize only data-channel \glspl{prb} can produce allocations that the cellular stack cannot use when considering the constraint of the 3GPP NR protocols.

These characteristics make schedulers hard to design while maintaining flexibility and integrating AI/ML routines. Production implementations, such as the existing \gls{oai} scheduler, entangle every scheduling decision with the management of the aforementioned constraints and of the real-time path. Therefore, changing a single decision, e.g., how the \gls{mcs} is selected, requires understanding and modifying logic across the whole scheduler. At the other extreme, rewriting a scheduler from scratch gives full flexibility, but it means re-implementing the protocol handling, with the risk of re-introducing edge cases and bugs that production code has long since fixed. We therefore propose a modular architecture, which isolates each decision behind a replaceable interface while sharing the feasibility checks and the real-time execution path. We describe it in the next subsection.

\subsection{Architecture}
% Shared candidate structure (light)
% Grant validation and dispatch

\begin{figure}[t]
\centering
\includegraphics[width=0.95\linewidth]{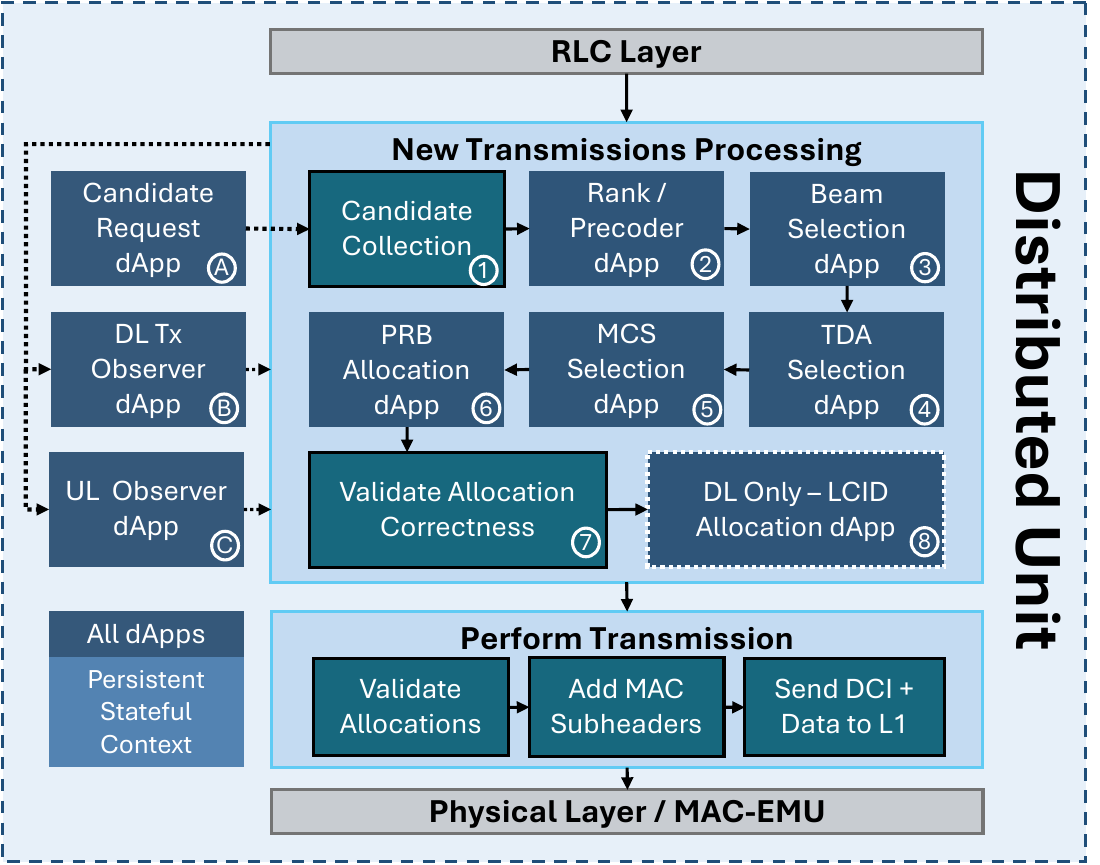}
\caption{\framework dApp architecture. Numbered decision dApps form the scheduling pipeline within one opportunity; lettered dApps observe transmission events or request candidates across opportunities. All dApps may access a persistent stateful context.}
\label{fig:sched_arch}
\end{figure}

\framework designs and implements the \gls{mac} scheduler as a staged pipeline (Fig.~\ref{fig:sched_arch}), where each stage makes one scheduling decision, e.g., \gls{mcs} selection or \gls{prb} allocation. Stages exchange their inputs and decisions through a shared per-\gls{ue} candidate structure. Each stage is a replaceable policy. At \gls{gnb} startup, the configuration selects one implementation per stage and composes a complete scheduler from individual modules, which can be abstracted as dApps. Beyond these stages, \framework lets dApps observe transmission outcomes and request additional \gls{ul} candidates. All dApps may access a persistent stateful context to carry application-defined state across scheduling opportunities. The pipeline order, the candidate structure, and the protocol feasibility checks remain fixed, so dApps change selected scheduling decisions without reimplementing the surrounding protocol handling.

Figure~\ref{fig:sched_arch} shows the pipeline that \framework executes at each scheduling opportunity both in \gls{dl} and \gls{ul}. Candidate collection~(1) populates one candidate structure per schedulable \gls{ue} (Listing~\ref{lst:candidate}), i.e., a \gls{ue} with a pending \gls{harq} retransmission or with new data to send and a free \gls{harq} process.
The \glspl{dapp} then decide the \gls{mimo} rank and which precoder to use~(2), the serving beam~(3), the \gls{tda}~\footnote{in 5G, the \gls{tda} indicates to the \gls{ue} which OFDM symbols it is allowed to use in the associated grant}(4), the \gls{mcs}~(5) and the \gls{prb} allocation~(6). Retransmission candidates traverse the same \glspl{dapp}, but their decisions must preserve the original transport-block size, e.g., a \gls{tda}, \gls{mcs} or \gls{prb} allocation change is accepted only if it reproduces the same block size. 
In deployments without beamforming, the beam \gls{dapp} returns a single default beam and the pipeline operates unchanged. \gls{prb} allocation may also lower the selected \gls{mcs} to satisfy the \gls{ue} power budget. The validation step~(7) then checks each grant against the protocol constraints before dispatch. 
Finally, the \gls{lcid} \gls{dapp}~(8) divides the transport-block bytes among logical channels for the UE on the \gls{dl} only. This multiplexing is a sender-side decision, as the \gls{ue} \gls{mac} performs the same task on the \gls{ul} path.
The \glspl{dapp} receive their input data per \gls{ue} from the candidate data structure of Listing~\ref{lst:candidate} (\texttt{nr\_ul\_candidate\_t} is the \gls{ul} analog). They also output their controls in that same structure. Candidate collection fills the metrics in that structure, and decision fields are written by the \glspl{dapp}, which ensures every \gls{dapp} observes the same state, enriched by the previous \glspl{dapp}' decisions.

Beyond the numbered stages, Fig.~\ref{fig:sched_arch} shows three dApp interfaces for policies that operate outside of the mandatory scheduling chain. The candidate-request dApp~(A) runs during \gls{ue} candidate collection and may request that a \gls{ue} enter the pipeline before its buffer state indicates pending data (this enables e.g., the proactive scheduler of section \ref{sec:proactive}). The \gls{dl} transmission observer~(B) is invoked after a scheduled transport block has been constructed, while the \gls{ul} observer~(C) receives the outcome of each \gls{ul} grant. These two observers enable \framework to expose data such as the amount of padding bytes in a received \gls{ul} \gls{pdu} or the time of arrival of \gls{dl} \glspl{pdu}.
Unlike the numbered stages, these observers do not modify the completed transmission. Instead, their primary role is to expose extra data to enable more intelligent decisions.
Finally, note that all \glspl{dapp} have access to a stateful context, which is (i) shareable across the different \glspl{dapp} and (ii) persisted across slots. This enables \gls{dapp} designers to expose context in new ways and to make \glspl{dapp} stateful if needed (for example, combined with \gls{dapp} (B), it enables tracking the rate of arrival of \gls{dl} \glspl{pdu}).

\begin{lstlisting}[
    language=C,
    float=t,
    caption={\gls{dl} candidate struct.},
    label=lst:candidate]
typedef struct nr_dl_candidate {
  /* Identity and scheduling state */
  uint16_t rnti;
  bool     is_retx;        // HARQ retx pending
  uint32_t pending_bytes;  // RLC buffer occupancy
  float    avg_throughput; // EWMA goodput
  float    bler;           // current BLER estimate
  int      current_mcs, max_mcs;
  uint64_t fiveQI;         // QoS class
  nssai_t  nssai;          // slice ID
  bool     skipped, scheduled; // stage outcomes
  /* Channel observations (read-only) */
  uint16_t cqi;            // wideband CQI
  uint8_t  csi_ri;         // reported rank
  const int16_t *beam_rsrp;    // per-SSB L1-RSRP
  const float *srs_rb_energy;  // per-RB channel
                               // energy from SRS
  /* Decisions, written by the named dApp */
  NR_sched_pdsch_t sched_pdsch; // MCS, PRBs,
                                // layers, TDA
  int alloc_beam_idx;      // selected beam
  int alloc_cce_index;     // PDCCH position
} nr_dl_candidate_t;
\end{lstlisting}

In its OAI implementation, \framework passes each candidate scheduling structure through \gls{oai}'s existing grant-validation path. This path checks, for example, that the selected \glspl{prb} are available, that the \gls{pdcch} has sufficient capacity for the grant, and, on the \gls{dl}, that a \gls{pucch} resource is available for \gls{harq} feedback. When a check fails, the pipeline rejects the grant and reports the cause through the \gls{oai} logs, helping the researcher identify and correct infeasible behavior in the corresponding \gls{dapp}. Only validated grants are committed and sent to Layer~1 through the \gls{fapi} interface.

\framework's scheduler backbone has been upstreamed to \gls{oai}~\cite{oaischedrefactor}, and the complete \framework and \macemu source code is available as open source \footnote{\url{https://github.com/wineslab/MAC-Gyver}}. The same interfaces also support the agentic scheduler control demonstrated in AgentRAN, which builds on \framework's lightweight design and scheduler-state exposure~\cite{elkael2026agentran}.

The two use cases exercise complementary parts of this architecture. Section~\ref{sec:proactive} combines the candidate-request \gls{dapp}~(A), \gls{dl} transmission observer~(B), and \gls{ul} observer~(C) with persistent context to act across scheduling opportunities. Section~\ref{sec:freqsel} uses the rank and precoder produced by \gls{dapp}~(2) to implement frequency-selective \gls{prb} allocation in \gls{dapp}~(6); it also replaces \gls{mcs} selection~(5) independently in the oracle experiment. Before presenting these use cases, the next section introduces \macemu, which preserves these interfaces while providing controlled execution of the complete scheduler.

\section{\macemu: Real-Time L2 Emulation at Scale}
\label{sec:macemu}

Next, we present \macemu, \framework's software-emulation component whose role is to execute the same \gls{oai} scheduler and registered \glspl{dapp} under controlled channel conditions and at cell scale, without requiring radio hardware. \macemu preserves the protocol stack through \gls{l2} and follows the real-time slot clock, so policies exercise the same \gls{ota} scheduling state, protocol constraints, and execution path.

\subsection{Motivation}
\label{sec:macemu_motivation}

Exposing the scheduler through \framework addresses the difficulty of implementation, but evaluation still requires representative channel behavior, cell-scale contention, and repeatable operating conditions. While \gls{ota} and hardware-based testbeds provide high-fidelity evaluation, they lack the scale of real deployments, i.e., tens to a hundred RRC-active \glspl{ue}~\cite{perez2023characterizing}. Simulators such as ns-3~\cite{henderson2008network} provide scale by executing a reimplemented \gls{du} in simulated time and cannot directly host arbitrary applications and traffic. Neither option alone provides controlled, real-time, cell-scale execution of the exact policy deployed in the \gls{du}.

In the current research landscape, scaled real-time experiments happen on hardware based testbeds (e.g., X5G~\cite{villa2024x5g}, Colosseum~\cite{bonati2021colosseum}, OAIC~\cite{upadhyaya2023open}). This gives a great degree of realism, but it comes at the cost of complexity in the infrastructure management and experiment execution. Indeed, managing multiple physical nodes (either for \gls{ran} or \glspl{ue}) requires a high level of automation, multiple servers, and software that is distributed across the network.
% Our extensive experience 
This complexity, combined with use cases that inherently push the boundaries of previous experiments, forces researchers to often redesign their own harness and scripts for new experiments, which requires them to manage edge cases and carefully handle all the potential states of the experiments. 
All this complexity compounds due to the mix of software, hardware and networking elements involved, and with the fact that lab deployments usually are temporary, making reproducibility challenging. 
% For these reasons, while hardware-in-the-loop experimentation should remain as a key validation tool, we suggest the research community also adopts alternative software-based experimentation too, as tools which enable better reproducibility and ease-of-use.
For these reasons, software-based experimentation should complement hardware-in-the-loop validation. Software
platforms enable rapid, reproducible, and scalable exploration, while hardware remains important for validating selected policies under real
radio impairments. Executing the same protocol stack and policy in both environments, as enabled by \framework, makes this transition
meaningful by avoiding differences caused by reimplementation and subtle differences between simulators and real cellular stacks.

One such tool is \gls{oai}'s \gls{rfsim}. It runs the real \gls{oai} stack fully in software, supports application traffic, and can be packaged as a reproducible single-host experiment (via, e.g., Docker Compose). It retains the complete \gls{phy} and exchanges \gls{iq} samples between the \gls{gnb} and software \glspl{ue} via TCP. This makes \gls{rfsim} a good tool for protocol-level evaluation. However, retaining
waveform processing also preserves its computational cost, and the use of TCP makes it non real-time.

Prior work has characterized these limitations. Wei \textit{et al.} show that \gls{phy} processing dominates the
computational cost of an \gls{oai} \gls{gnb}, with the \gls{du} accounting for most of the cost in a disaggregated
deployment~\cite{wei20225gperf}. Rouili \textit{et al.} report that \gls{rfsim} consumes more CPU than their SDR-based
deployment and attribute this overhead to \gls{phy} processing and the exchange of \gls{iq} samples~\cite{rouili2024evaluating}.
More recently, Mamaghani \textit{et al.} benchmark an \gls{rfsim}-based system with up to five \glspl{ue} on an AMD Ryzen~9
7900X host with 32~GB of memory~\cite{mamaghani2026tiny}. After extending \gls{rfsim} with multi-tap channel
convolution, they report a 90th-percentile slot-computation time of 8~ms for five \glspl{ue}; beyond
four \glspl{ue}, round-trip latency frequently reaches hundreds of milliseconds and connections may fail. These results show
that retaining waveform-level processing becomes a bottleneck well before the cell-scale experiments targeted in this work.

\macemu therefore removes waveform-level \gls{phy} processing while retaining the protocol layers and scheduling path that determine \gls{dapp} behavior. This boundary provides the scale of a software abstraction without replacing the \gls{du} scheduler under evaluation. The remainder of this section describes the resulting architecture, channel model, and validation.

\subsection{Architecture}
\label{sec:macemu_arch}

\begin{figure}[t]
\centering
\includegraphics[width=0.95\linewidth]{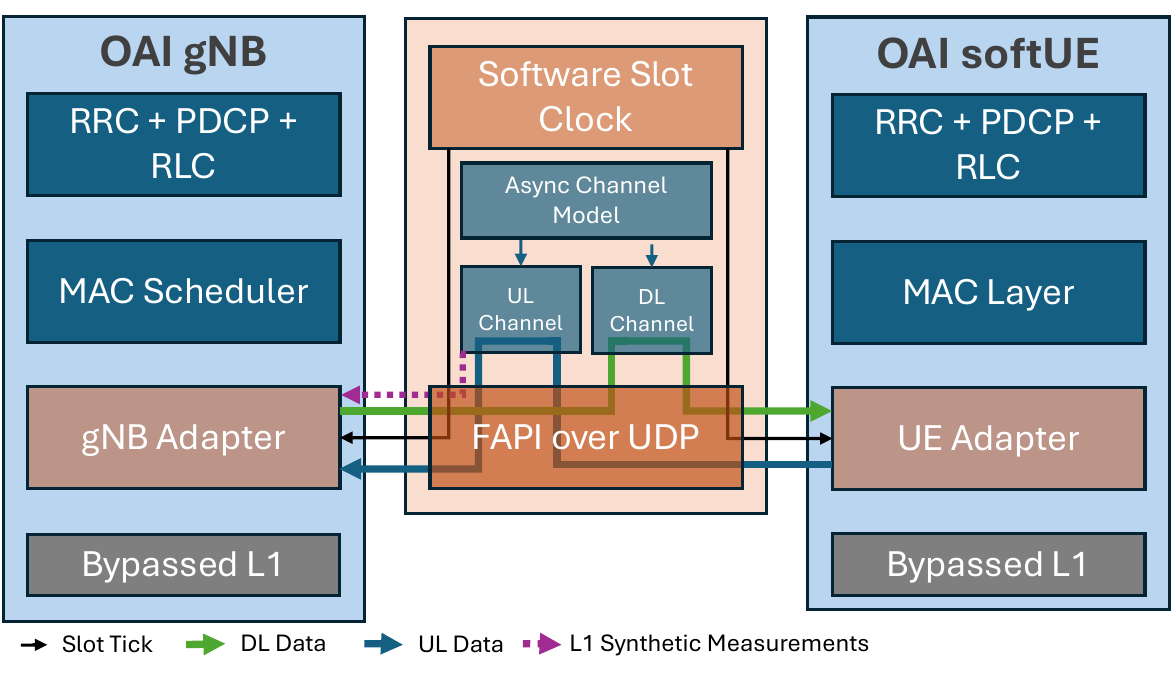}
\caption{\macemu architecture. Unmodified \gls{oai} \gls{l2} stacks run to their \gls{fapi} boundary; adapters and the \macemu runtime bypass the \gls{phy}. The software slot clock paces the \gls{gnb} and the \gls{ue} processes; the purple arrow carries the Layer~1 measurements synthesized by the channel model.}
\label{fig:macemu_arch}
\end{figure}

\macemu runs the production \gls{gnb} and \gls{ue} stacks unmodified down to their \gls{fapi} boundary, and replaces everything below it (i.e., the \gls{phy}). The runtime replaces the \gls{phy} by synthesizing the Layer~1 observations and transmission outcomes presented to the scheduler, abstracting the underlying \gls{phy} implementation, such as waveform generation and \gls{rf} transmission. As shown in Fig.~\ref{fig:macemu_arch}, a thin adapter bypasses the \gls{phy} on each side and exchanges the \gls{fapi} messages with the \macemu runtime over UDP, while \gls{rrc}, \gls{pdcp}, \gls{rlc}, and \gls{mac} run the deployed \gls{oai} code. Because \framework and its \glspl{dapp} execute above this boundary, the same dApp interfaces and policy implementations run in \macemu and over the air; only the source of Layer~1 observations and transmission outcomes changes.

Inside the runtime, a software slot clock paces the system and fires once per slot (500~$\mu$s at the 30~kHz \gls{scs} used in this work) and distributes slot ticks to the \gls{gnb} and every \gls{ue}. At each tick, \gls{dl} \glspl{dci} and data flow from the \gls{gnb} adapter to the \glspl{ue}, while \gls{ul} data and \gls{harq} feedback flow back. Both directions traverse the channel model: for every scheduled transport block, it decides the ACK/NACK outcome that enters the feedback stream, and it synthesizes the Layer 1 measurements that a real \gls{phy} would report to the scheduler (purple arrow in Fig.~\ref{fig:macemu_arch}). Section~\ref{sec:macemu_channel} details this model.

Real-time execution rests on three mechanisms. First, the slot clock runs on a dedicated thread with real-time priority, pinned to an isolated core, and paces itself on absolute deadlines. It does so by sleeping until shortly before the next slot boundary, and by busy-spinning for the remainder, avoiding wake-up jitter. Second, the channel model executes asynchronously, on lower-priority worker threads that cannot preempt the thread dedicated to the slot clock. This means that the critical path stays free for message dispatch and scheduling. Third, each \gls{ue} runs as a separate \gls{oai} softUE process whose Layer 1 is bypassed by the \gls{ue} adapter, so a single host executes the \gls{gnb} and tens of \glspl{ue} concurrently. Section~\ref{sec:macemu_validation} quantifies the resulting deadline behavior at up to 90 \glspl{ue}. We observe no computational bottleneck at this count; rather, 90 is the maximum supported by \gls{oai} in our 40~MHz configuration because of its current \gls{pucch}-resource allocation.

\subsection{Channel Model}
\label{sec:macemu_channel}

\macemu abstracts waveform processing but must reproduce its consequences on the scheduler. For each scheduled transport block, its configurable channel model determines the ACK/NACK outcome returned through the protocol stack, and synthesizes the Layer~1 measurements exposed to the scheduler and its \glspl{dapp}. The model is defined per \gls{ue} in a YAML file that specifies the path-loss model, shadowing, small-scale fading, antenna gains, and mobility. We implement free-space, \gls{umi}, and \gls{uma} path loss as well as TDL and CDL fading from the 3GPP channel model technical report 38.901~\cite{tr38901}. Mobility combines a Doppler frequency with stateful patterns, e.g., a \gls{ue} that alternates between moving and standing still. As the system is focused on \gls{tdd}, we assume that \gls{dl} and \gls{ul} channels are reciprocal.
% : \gls{dl} and \gls{ul} share one realization per \gls{ue}.

The ACK/NACK decision proceeds in three steps. First, a link budget and the fading realization give per-subcarrier \glspl{snr},
\begin{equation}
\gamma_k(t) = \bar{\gamma}\,\lvert H_k(t)\rvert^2,
\label{eq:persc_snr}
\end{equation}
where $\bar{\gamma}$ is the average \gls{snr} from transmit power, antenna gains, path loss, shadowing, and thermal noise, and $H_k(t)$ is the channel frequency response at subcarrier $k$. Second, \gls{eesm} compresses the allocated subcarriers into one effective \gls{snr}~\cite{lagen2020},
\begin{equation}
\gamma_{\mathrm{eff}} = -\beta_m \ln\!\Big(\tfrac{1}{N}\sum_{k=1}^{N} e^{-\gamma_k/\beta_m}\Big),
\label{eq:eesm}
\end{equation}
where $N$ is the number of allocated subcarriers and $\beta_m$ is the calibration factor of \gls{mcs} $m$ from~\cite{lagen2020}. \gls{harq} retransmissions add their effective \glspl{snr} in the linear domain (Chase combining). Third, a three-parameter Richards curve maps the combined \gls{snr} to an error probability,
\begin{equation}
\mathrm{BLER}_{m,w}(\gamma) = \big(1 + e^{a(\gamma - b)}\big)^{-c},
\label{eq:richards}
\end{equation}
where $(a, b, c)$ control the steepness, position, and asymmetry of the waterfall for \gls{mcs} $m$ and number $w$ of allocated \glspl{prb}; the outcome is one Bernoulli draw per transport block.

Parameters $(a, b, c)$ are fitted offline for each link direction (\gls{dl} or \gls{ul}), \gls{mcs}, and number of \glspl{prb} allocated to the transmission. We do this using link-level simulations of the \gls{oai} \gls{ue} and \gls{gnb} receivers. \gls{oai} exposes each receive chain in isolation as a link-level simulator, \texttt{nr\_dlsim} for the \gls{ue}'s \gls{pdsch} receiver and \texttt{nr\_ulsim} for the \gls{gnb}'s \gls{pusch} receiver: each run encodes transport blocks, applies \gls{awgn}, executes the full demodulation and \gls{ldpc} decoding chain, and reports the block error rate.
We sweep \gls{snr} for the \gls{awgn} channel and fit Eq.~\eqref{eq:richards} to the resulting waterfall. Each possible number of \glspl{prb} requires its own curve because it sets the transport-block size and therefore the \gls{ldpc} code-block segmentation, which changes both the position and the steepness of the waterfall. We run the parameter sweep offline for a given carrier configuration, producing one curve per direction, \gls{mcs}, and number of allocated \glspl{prb}. For the 40~MHz carrier used in this work (106 \glspl{prb} and 29 \gls{mcs} indices per direction), it yields 6,148 curves.

By construction, these curves reproduce the \gls{ldpc} receivers exactly. A real link, however, deviates from them due to radio-frequency impairments, such as transmitter error-vector magnitude and phase noise, as well as implementation differences between commercial receivers and \gls{oai} (e.g., quantization in the ASIC of a \gls{ue}).

\begin{figure}[t]
  \centering
  \includegraphics[width=0.95\linewidth]{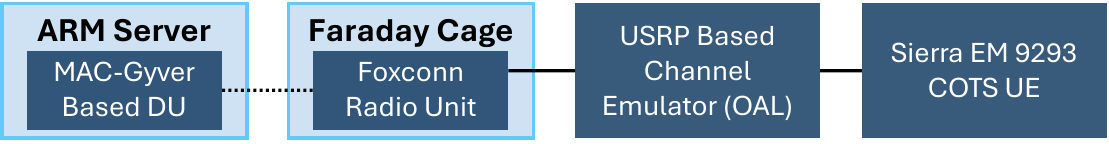}
  \caption{\gls{oal} hardware reference setup.}
  \label{fig:oal_setup}
\end{figure}

We use the hardware reference setup in Fig.~\ref{fig:oal_setup} to calibrate and validate \macemu. An \gls{oai} \gls{gnb} drives a Foxconn RPQN \gls{ru}, which connects through the \gls{oal} programmable channel emulator~\cite{deshpande2024openairlink} to a Sierra Wireless EM9293 \gls{cots} \gls{ue}. \gls{oal} applies controlled attenuation and 3GPP fading profiles to the bidirectional radio link. Measurements from this setup parameterize hardware-specific effects in the channel model, while Sec.~\ref{sec:macemu_validation} uses the same physical chain for closed-loop validation.
Note that at the time of writing, our \gls{oal} setup only supports a 100~MHz (273-\gls{prb}) cell configuration, while \macemu experiments are at 40~MHz (106-\gls{prb}) throughout this work. We consider 106 \glspl{prb} allocations throughout this section, which equalizes transport-block sizes.

\begin{figure}[t]
\centering
\begin{minipage}{0.98\linewidth}
  \centering
  \scriptsize
  \input{figures/macemu_floor_characterization}
\end{minipage}
\caption{High-\gls{snr} initial-transmission error floor across (a)~\gls{mcs} at 106~\glspl{prb} and (b)~the number of allocated \glspl{prb} at \gls{mcs}~8. Bars aggregate the retained stationary measurements; error bars show 95\% Wilson intervals, and dashed lines show the average error floor in each panel.}
\label{fig:macemu_floor}
\end{figure}
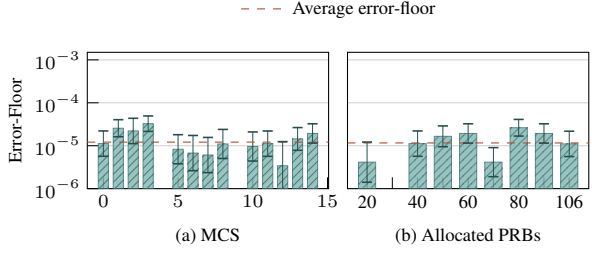

We take long fixed-\gls{mcs} measurements at \glspl{snr} well above each waterfall, with \gls{harq} retransmissions disabled so that every logged block is an initial transmission. Where the fitted curves predict a vanishing \gls{bler}, the hardware instead settles on an error floor. Figure~\ref{fig:macemu_floor} tests whether this floor depends on either \gls{mcs} or the number of allocated \glspl{prb}. Panel~(a) varies the \gls{mcs} over indices 0--14 while capping each allocation at 106~\glspl{prb}. After retaining the stationary half of each run, the resulting 9.6 million blocks yield an average error floor of $1.21\times10^{-5}$ with no systematic dependence on \gls{mcs}. We restrict this experiment to the contiguous range 0--14 because, at higher indices, the setup did not consistently provide stationary measurements sufficiently above the decoding waterfall. Panel~(b) fixes \gls{mcs}~8 and varies the allocation from 20 to 106~\glspl{prb}. Across 7.2 million blocks, the average is $1.16\times10^{-5}$, with no dependence on the number of allocated \glspl{prb}.
We therefore set $\varphi=1.2\times10^{-5}$ and extend the ACK/NACK decision as
\begin{equation}
P(\mathrm{NACK}) = \varphi + (1-\varphi)\,
\mathrm{BLER}_{m,w}\big(\gamma_{\mathrm{eff}}\big),
\label{eq:mismatch}
\end{equation}

Beyond this data-plane focused model, we also model the effect of channel impairments on the control channel that carries the \gls{dl} \gls{harq} feedback. Our experiments reveal that, on real hardware, the \gls{gnb} occasionally fails to detect feedback that the \gls{ue} did transmit, which the \gls{gnb} conservatively treats as a NACK, scheduling a retransmission. We measure this effect over three 3GPP TDL fading profiles of increasing delay spread (TDL-A30, TDL-B100, and TDL-C300, i.e., 30 to 300~ns of delay spread~\cite{tr38901}), reproduced by the channel emulator: the rate of missed detection is strongly ordered by the delay spread, at $0.06\%$, $0.77\%$, and $5.5\%$ of feedback occasions, respectively.

The delay-spread dependence follows from the encoding of the \gls{harq} feedback, which goes through \gls{pucch} format~0. \gls{pucch} format~0 encodes ACK/NACK as cyclic shifts of a base sequence. The detector therefore discriminates hypotheses in the delay domain and tolerates a bounded delay error around the expected arrival time. For feedback occasion $i$, \macemu computes the instantaneous power-weighted delay centroid
\begin{equation}
\bar{\tau}_i =
\frac{\sum_{\ell=1}^{L} |h_\ell(t_i)|^2 \tau_\ell}
     {\sum_{\ell=1}^{L} |h_\ell(t_i)|^2},
\label{eq:pucch_centroid}
\end{equation}
where $L$ is the number of channel taps, $h_\ell(t_i)$ is the complex coefficient of tap $\ell$ at time $t_i$, and $\tau_\ell$ is its delay.

We use a receiver-level abstraction of the \gls{ul} timing-tracking loop. The \gls{gnb} estimates each \gls{ue}'s \gls{ul} arrival time and sends timing-advance commands, which the \gls{ue} applies to subsequent transmissions. Since the loop retains state, a residual timing error remains at the \gls{gnb}'s \gls{pucch} detector and evolves across feedback occasions. \macemu represents this correlated error with the Gauss--Markov state $n_i$. The state and the resulting \gls{dtx} decision are
\begin{align}
n_i &= \rho_i n_{i-1} + \sqrt{1-\rho_i^2}\,\epsilon_i,
& \rho_i &= e^{-\Delta t_i/t_{\mathrm{corr}}}, \label{eq:pucch_tracking}\\
d_i &= \mathbf{1}\!\left\{\bar{\tau}_i+\sigma n_i>W\right\},
\label{eq:pucch_dtx}
\end{align}
where $n_i$ is the unit-variance Gauss--Markov timing-error state, $\epsilon_i\sim\mathcal{N}(0,1)$ is a new Gaussian random sample drawn at feedback occasion $i$, independently of the previous samples, $\Delta t_i$ is the time since the previous feedback occasion, and $t_{\mathrm{corr}}$ is the tracking-loop correlation time. The parameter $\sigma$ scales the timing error, $W$ is the detector window, and $d_i=1$ indicates a \gls{dtx}. The stationary distribution of $n_i$ is independent of $t_{\mathrm{corr}}$. Consequently, $W$ and $\sigma$ determine the \gls{dtx} rate, while $t_{\mathrm{corr}}$ determines how events cluster in time.

We calibrate the model against the \gls{dtx} statistics measured on the \gls{oal}-based testbed: $W$ and $\sigma$ by maximum likelihood on the per-channel \gls{dtx} rates, and $t_{\mathrm{corr}}$ on the burstiness of the resulting retransmissions. We obtain $W = 475$~ns, $\sigma = 140$~ns, and $t_{\mathrm{corr}} = 100$~ms, which yields \gls{dtx} rates of $0.07\%/0.28\%/5.4\%$ on TDL-A30/-B100/-C300, against the $0.06\%/0.77\%/5.5\%$ measured on the hardware testbed.

\begin{figure*}
\centering
\definecolor{crimson2143940}{RGB}{214,39,40}
\definecolor{darkgrey176}{RGB}{176,176,176}
\definecolor{steelblue31119180}{RGB}{31,119,180}
\pgfplotslegendfromname{ollaLegend}\par\vspace{0.2em}
\begin{subfigure}[t]{0.49\linewidth}
    \centering
    \scriptsize
    \input{figures/olla_mcs}
    \caption{\gls{mcs} of initial transmissions.}
    \label{fig:macemu_olla_mcs}
\end{subfigure}%
\hfill
\begin{subfigure}[t]{0.49\linewidth}
    \centering
    \scriptsize
    \input{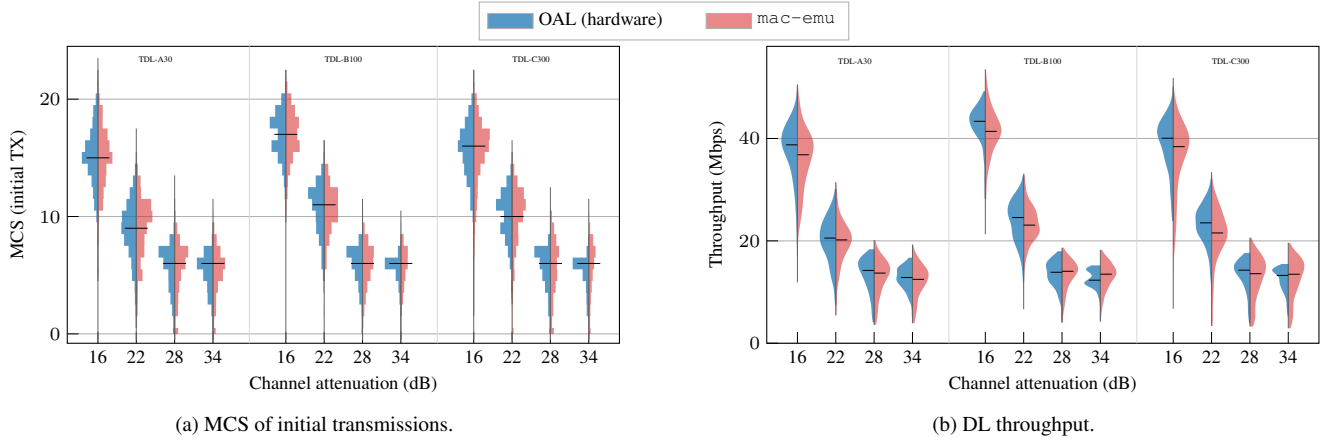}
    \caption{\gls{dl} throughput.}
    \label{fig:macemu_olla_tput}
\end{subfigure}
\caption{Closed-loop link adaptation in \macemu and on \gls{oal}, one \gls{ue}, saturated \gls{dl} traffic.}
\label{fig:macemu_olla}
\end{figure*}

\begin{figure*}[t]
\centering
\input{figures/bler_grid}
\caption{Initial-transmission \gls{bler} versus \gls{mcs} for the same runs as Fig.~\ref{fig:macemu_olla}.}
\label{fig:macemu_olla_bler}
\end{figure*}
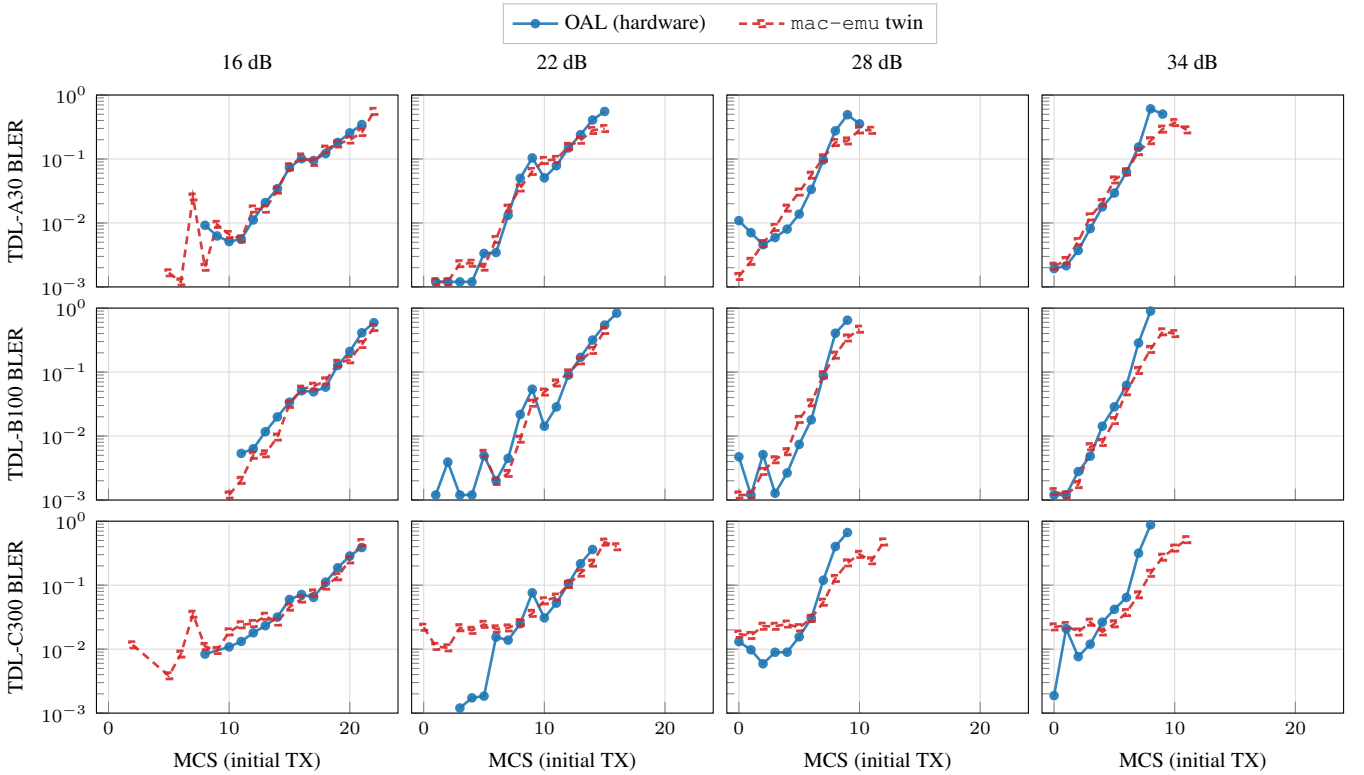

Besides the ACK/NACK results, \macemu also synthesizes the measurements consumed by the scheduler. The wideband \gls{cqi} follows the standard reporting rule: the report carries the highest \gls{cqi} index whose predicted \gls{bler}, evaluated through Eq.~\eqref{eq:richards}, does not exceed the 10\% target (TS~38.214, Clause~5.2.2.1)~\cite{ts38214}.
The scheduler also receives the per-\gls{prb} \gls{srs} channel matrix. On a real \gls{gnb} this matrix comes from a channel estimator (e.g., least squares in \gls{oai}) and therefore carries estimation error.
To model this error, we measure it offline with the \gls{srs} receiver in isolation (\gls{oai}'s \texttt{nr\_srssim}) and inject noise of matching variance into the synthesized matrix, which is derived from the current channel state; the calibrated synthesis matches the simulator with an \gls{rmse} of $0.36$~dB.
The synthesized \gls{cqi} and \gls{srs} measurements follow the purple path in Fig.~\ref{fig:macemu_arch}, where \framework exposes them to \glspl{dapp} through the candidate structure. Section~\ref{sec:freqsel} uses the per-\gls{prb} \gls{srs} measurements for frequency-selective allocation.

\subsection{Validation}
\label{sec:macemu_validation}

We validate the two properties required for using \macemu as \framework's controlled execution environment: whether the scheduler behaves the same with \macemu as with a hardware link, and whether it sustains real-time execution under cell-scale contention. The preceding subsection used the \gls{oal} setup to estimate the error-floor and \gls{pucch} parameters. We now use the same physical testbed to compare the closed-loop \gls{mcs}, throughput, and \gls{bler} observed in \macemu and on hardware for the default \gls{oai} scheduler. We then verify whether the emulator sustains its real-time budget at scale. All hardware comparisons use a single-antenna link.

We compare the link-level \glspl{kpi} of a single-\gls{ue} deployment on \macemu and on the hardware testbed under saturated \gls{dl} \gls{udp} traffic, across the three TDL profiles of Sec.~\ref{sec:macemu_channel} at four channel attenuations each (16, 22, 28, and 34~dB). The emulated \gls{gnb} mirrors the hardware configuration, which is checked automatically at startup. Notably, the testbed reports no \gls{cqi} (\gls{csirs} is disabled), so \gls{dl} link adaptation is purely \gls{bler}-driven on both platforms.
Since the hardware testbed's absolute attenuation-to-\gls{snr} mapping is not recoverable exactly (the channel emulator applies a programmable attenuation between transmitter and receiver rather than an explicit path-loss model, so the absolute link budget depends on unmeasured gains along the chain), we calibrate \macemu's path loss per operating point, rerunning the experiment to find the closest \gls{mcs}-distribution match with the hardware setup (via the Wasserstein distance). This enables us to evaluate how closely \macemu reproduces the hardware setup despite this limitation.
For each operating point, we run the experiment five times at the best-matching path loss, on both platforms.

Figure~\ref{fig:macemu_olla} reports the \gls{mcs} and throughput distributions, and Fig.~\ref{fig:macemu_olla_bler} the per-\gls{mcs} initial-transmission \gls{bler}. Across the 12 points, the median Wasserstein distance between the pooled \gls{mcs} distributions is $0.35$ \gls{mcs} steps (worst $0.94$), the mean-\gls{mcs} error stays below $0.35$ everywhere, and median throughput agrees to $-4\%$ (all points within $13\%$, which is similar to the variability we observe between runs of the same experiment on hardware)
We attribute the consistently lower throughput of \macemu to the \glspl{prb} number mismatch (106 (40~MHz) on \macemu, 273 (100~MHz) on \gls{oal}): while we limit \gls{oal} to 106 \glspl{prb} in the scheduler, the \gls{ssb} is scheduled after those \glspl{prb}. This is not the case for \macemu, which loses $\approx 4\%$ of schedulable capacity (1 SSB every $20~ms$, which caps the number of contiguous \glspl{prb} to 40 for that slot in our setup).

The \gls{bler}-versus-\gls{mcs} curves closely follow each other through the band the outer loop targets. The remaining \gls{bler} mismatch concentrates on TDL-C300 at low \gls{mcs}, where missed PUCCH-driven errors dominate: the model injects detector misses independently of the instantaneous channel state, yielding a flat \gls{bler} contribution across the low \glspl{mcs} (absent on the other profiles, whose rates of missed PUCCH are negligible), whereas the hardware's low-\gls{mcs} \gls{bler} retains more structure. Because the link-adaptation loop rarely selects these low \gls{mcs} values in steady state, this discrepancy has no visible effect on the resulting \gls{mcs} and throughput distributions.

\begin{figure}[t]
\centering
\setlength\fwidth{0.95\linewidth}
\setlength\fheight{5.2cm}
\input{figures/macemu_throughput}
\caption{Aggregate cell throughput versus concurrent \gls{ue} count.}
\label{fig:macemu_throughput}
\end{figure}
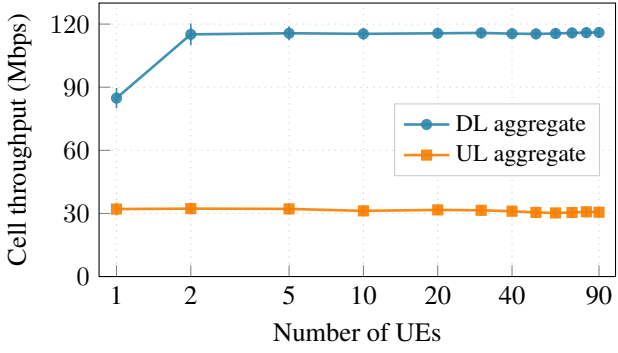

For the scalability evaluation, we focus on verifying that \macemu maintains the real-time guarantees at scale. To do this, we gradually increase the number of \glspl{ue} attached to a 40~MHz SISO cell with very good channel conditions (all \glspl{ue} reach \gls{mcs} 28), and saturate both the \gls{ul} and \gls{dl} using mgen~\cite{mgen}. We evaluate up to 90 \glspl{ue}, the maximum supported by \gls{oai} in this configuration because of its current \gls{pucch}-resource allocation.
For this experiment, the \gls{gnb} thread runs on one isolated core, while the \gls{ue} processes are distributed across ten additional cores (the whole workload runs on an NVIDIA DGX Spark with 20 CPUs, out of which 10 are performance cores, which we use for \macemu).
We report the aggregate throughput in Fig.~\ref{fig:macemu_throughput}, where we observe that \macemu sustains the maximum capacity of the cell in both directions ($115$ and $32$ $\pm 0.3$~Mbps). In particular, \macemu shows no computational saturation at the 90-\gls{ue} limit. The only case where we reach a lower throughput is with a single \gls{ue}. This is caused by a separate \gls{oai} limitation: its rigid \gls{pucch}-allocation policy limits the maximum number of simultaneous \gls{harq} processes in our configuration.

\begin{figure}[t]
\centering
\begin{subfigure}[t]{\linewidth}
    \centering
    \begin{minipage}{0.82\linewidth}
        \centering
        \scriptsize
        \input{figures/scaling_stage}
    \end{minipage}
    \caption{Slot-thread execution time.}
    \label{fig:macemu_scaling_stage}
\end{subfigure}
\\[0.4ex]
\begin{subfigure}[t]{\linewidth}
    \centering
    \begin{minipage}{0.82\linewidth}
        \centering
        \scriptsize
        \input{figures/scaling_violation}
    \end{minipage}
    \caption{Deadline-violation probability.}
    \label{fig:macemu_scaling_violation}
\end{subfigure}
\caption{\macemu slot-thread scaling under saturated bidirectional traffic, versus concurrent \gls{ue} count.}
\label{fig:macemu_scaling}
\end{figure}
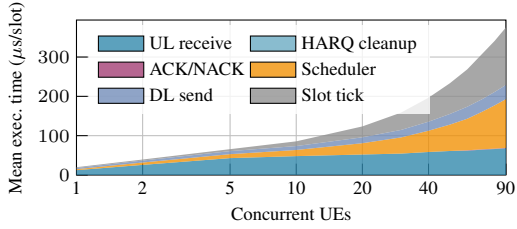
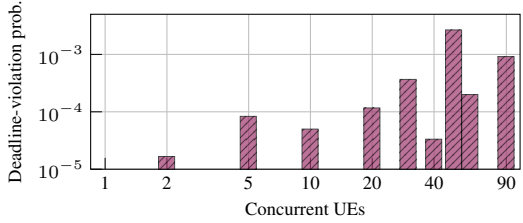

We also collect a breakdown of the runtime per slot in this experiment, which is reported in Fig.~\ref{fig:macemu_scaling}. We observe that as the number of \glspl{ue} increases, the two main contributors to the runtime are the execution of the scheduler and the distribution of the slot ticks to each \gls{ue}, which exhibit clear non-linear scaling. Despite this, the average runtime always stays below the critical value of 500~$\mu s$, i.e., the slot duration. The deadline violation probability is very low (around $10^{-3}$ at most), meaning that we near-perfectly respect the real-time deadline.

These results establish \macemu as the controlled execution component of \framework. Its channel model reproduces the observations and transmission outcomes that enter the scheduler loop, while its real-time architecture sustains the complete \gls{gnb} \gls{l2} stack at cell scale. The following sections use the resulting framework in two complementary ways: proactive scheduling combines persistent state and event-driven \glspl{dapp} in an \gls{ota} deployment, whereas frequency-selective scheduling composes per-slot decision stages and uses \macemu to evaluate their interactions under controlled channel conditions.

\section{Use Case: Proactive URLLC Scheduling}
\label{sec:proactive}

Our first use case uses \framework to implement a closed-loop proactive \gls{ul} scheduler. The \gls{dapp} predicts when a \gls{ue} will have data and schedules its \gls{ul} grant before the \gls{ue} sends an \gls{sr}. It adds the \gls{ue} to the scheduler's candidate set while leaving resource allocation and grant validation unchanged. This use case exposes the trade-off between access latency and unused proactive grants.

\subsection{Motivation}
\label{sec:proactive-motivation}

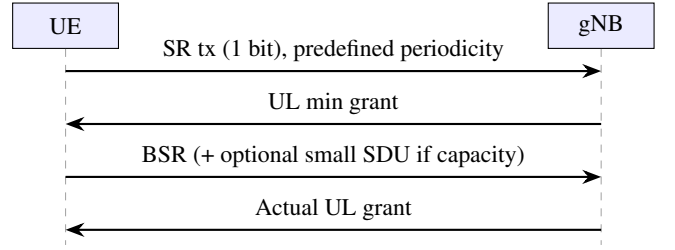
\begin{figure}[t]
    \centering
    \input{figures/schedReq.tex}
    \caption{Scheduling request procedure.}
    \label{fig:schedReq}
\end{figure}

Our main motivation for proactive scheduling is the \gls{sr} procedure, which is a major source of \gls{ul} latency in 5G systems. As shown in Fig.~\ref{fig:schedReq}, when it needs to transmit in the \gls{ul}, a \gls{ue} first waits for its periodic \gls{sr} opportunity on the \gls{pucch}, at which point it transmits a \gls{sr} bit, indicating to the \gls{gnb} it needs to transmit data. It then receives a small grant through \gls{dci}, returns a \gls{bsr}, containing the size of the payload. This enables the \gls{gnb} to then schedule \gls{ul} in further slots. The initial wait depends on the configured \gls{sr} periodicity, typically 10--40~ms and up to 640~ms. Limited \gls{pucch} resources require longer periodicity as the number of \glspl{ue} grows. Consequently, the full procedure adds 15--50~ms and increases jitter, compared with \gls{urllc} targets of 1--10~ms~\cite{ts22261,popovski2018wireless}.

Note that the 5G standard provides \glspl{cg} to avoid the \gls{sr} delay, in which the \gls{gnb} pre-configures periodic grants at fixed periodicity and offsets via \gls{rrc}~\cite{ts38321,le2021urllc}. However, \glspl{cg} assume the traffic is periodic and its characteristics are known in advance, which does not hold for event-driven flows, multi-periodic traffic (e.g., a robot with multiple sensors) and more generally, unknown sparse traffic. Our approach targets these broader arrival patterns by learning from events visible by the scheduler, without requiring \gls{rrc} reconfiguration.

\begin{figure}[t]
\centering
    \setlength\fwidth{0.9\linewidth}
    \setlength\fheight{4cm}
    \input{figures/baseline}
    \caption{\gls{rtt} with and without always-on \gls{ul} scheduling.}
    \label{fig:sr-vs-always}
\end{figure}
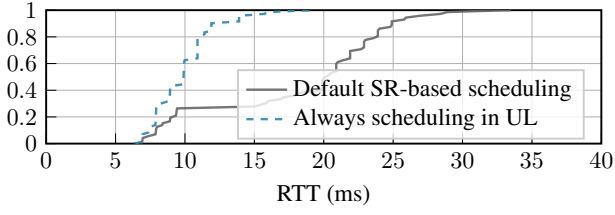

To quantify how much latency the \gls{sr} procedure adds, we compare the default scheduler of \gls{oai} with a simple always-on policy that grants the \gls{ue} in every slot without waiting for an \gls{sr}.
Figure~\ref{fig:sr-vs-always} evaluates both policies over the air with a single \gls{ue} performing a `ping` every 100~ms. The default scheduler produces an \gls{rtt} of 15--25~ms with a median near 20~ms. Always-on scheduling reduces this range to 7--12~ms and removes most of the variation, confirming that the \gls{sr} accounts for roughly half of the observed latency. This however comes at a spectrum efficiency cost, as most grants carry no data under such sparse traffic, and the cost grows linearly with the number of \gls{urllc} \glspl{ue}. We therefore use always-on scheduling as the minimum-latency reference and seek a policy that approaches its latency with fewer unused grants.

\subsection{Problem Formulation}
\label{sec:proactive-problem}

We formalize proactive scheduling as a multi-objective optimization over \gls{ul} scheduling strategies. Within a \gls{urllc} slice of \glspl{ue} $\mathcal{N}$ with homogeneous \gls{qos} requirements, we seek scheduling strategies that balance three objectives: worst-case expected latency, worst-case jitter, and total slice resource consumption. The resource-consumption ratio for
\gls{ue} $i$ is
\begin{equation}
U_i(S) \triangleq
\frac{\mathbb{E}[R_i(S)]}{R_{slice}/|\mathcal{N}|}.
\end{equation}
The resulting problem is
\begin{equation}\label{eq:multiobj}\tag{P-URLLC}
\underset{S \in \mathcal{S}}{\text{minimize}} \quad
\begin{bmatrix}
\max_{i \in \mathcal{N}} \mathbb{E}[L_{sched}^i(S)] \\
\max_{i \in \mathcal{N}} V(L_{sched}^i(S)) \\
\sum_{i \in \mathcal{N}} U_i(S)
\end{bmatrix}.
\end{equation}
\noindent
where $\mathcal{N}$ is the set of the $|\mathcal{N}|$ \glspl{ue} in a \gls{urllc} slice, assumed to
share homogeneous \gls{qos} requirements. $\mathcal{S}$ is the set of feasible scheduling strategies, with $S\in\mathcal{S}$ being
one such strategy. For each \gls{ue} $i\in\mathcal{N}$, $L_{sched}^i(S)$ is the latency induced by $S$, $V(L_{sched}^i(S))$ is the latency variance (jitter), $U_i(S)$ is the resource-consumption ratio, $R_i(S)$ is the random variable representing the instantaneous resources allocated to $i$, and $R_{slice}$ is the total \gls{prb} budget reserved for the \gls{urllc} slice.

% where:
% \begin{itemize}[leftmargin=*]
% \item $\mathcal{N}$ is the set of \glspl{ue} in a \gls{urllc} slice, assumed to share homogeneous \gls{qos} requirements;
% \item $\mathcal{S}$ is the set of feasible scheduling strategies, and $S$ is one such strategy;
% \item $i$ indexes a \gls{ue} in $\mathcal{N}$;
% \item $L_{sched}^i(S)$ is the scheduling-induced latency for \gls{ue} $i$;
% \item $\mathbb{E}[L_{sched}^i(S)]$ is the expected scheduling latency for
% \gls{ue} $i$;
% \item $V(L_{sched}^i(S))$ is the latency variance (jitter) for \gls{ue} $i$;
% \item $U_i(S)$ is the resource-consumption ratio for \gls{ue} $i$;
% \item $R_i(S)$ is the random variable of instantaneous resources allocated to
% \gls{ue} $i$;
% \item $\mathbb{E}[R_i(S)]$ is the time-averaged resource allocated to
% \gls{ue} $i$;
% \item $R_{slice}$ is the total \gls{prb} budget reserved for the
% \gls{urllc} slice;
% \item $|\mathcal{N}|$ is the number of \glspl{ue} in the slice.
% \end{itemize}
The vector objective in Eq.~\ref{eq:multiobj} is minimized in the Pareto sense where a feasible strategy is
Pareto-optimal if no other strategy achieves values no larger in all three
objectives and a strictly smaller value in at least one.
Solving Eq.~\eqref{eq:multiobj} requires the distribution of scheduling latency, which depends on the unknown \gls{ul} arrival process. We therefore use an online policy that estimates this process from events available inside the \gls{mac} scheduler. Two parameters, the tolerance window $\mathrm{TW}$ and slot restriction $N$, control how closely proactive grants follow each prediction and set the policy's position on the latency-overhead trade-off. A state machine alternates between collecting unbiased arrival samples and scheduling from the resulting predictions.

\subsection{Architecture and State Machine}
\label{sec:proactive-arch}

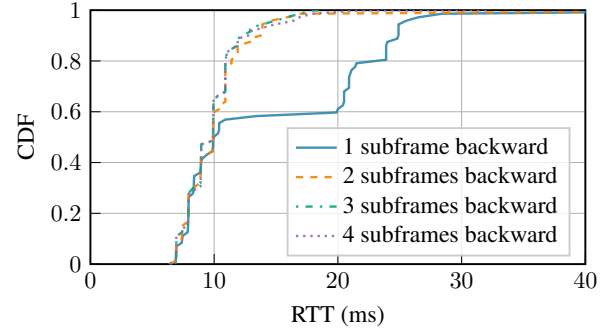
\begin{figure}[t]
\centering
    \setlength\fwidth{0.9\linewidth}
    \setlength\fheight{2cm}
    \input{figures/backward_lookup}
    \caption{\gls{rtt} with a \gls{dl}-aware proactive scheduler at varying lookback windows.}
    \label{fig:backward_lookup}
\end{figure}

We first evaluate a simple prediction rule before presenting the full architecture. For Fig.~\ref{fig:backward_lookup}, we implement a scheduling \gls{dapp} that issues a proactive \gls{ul} grant whenever the scheduler has transmitted \gls{dl} data to the \gls{ue} within the preceding one to four \gls{tdd} periods. The figure reports \gls{rtt} for each lookback window. Extending the window increases the number of proactive grants linearly. A two-period lookback reduces the median \gls{rtt} to approximately 9~ms, close to always-on scheduling. Unlike always-on scheduling, this rule issues grants only during the two periods following each \gls{dl} transmission.

\begin{figure}[t]
    \centering
    \includegraphics[width=\linewidth]{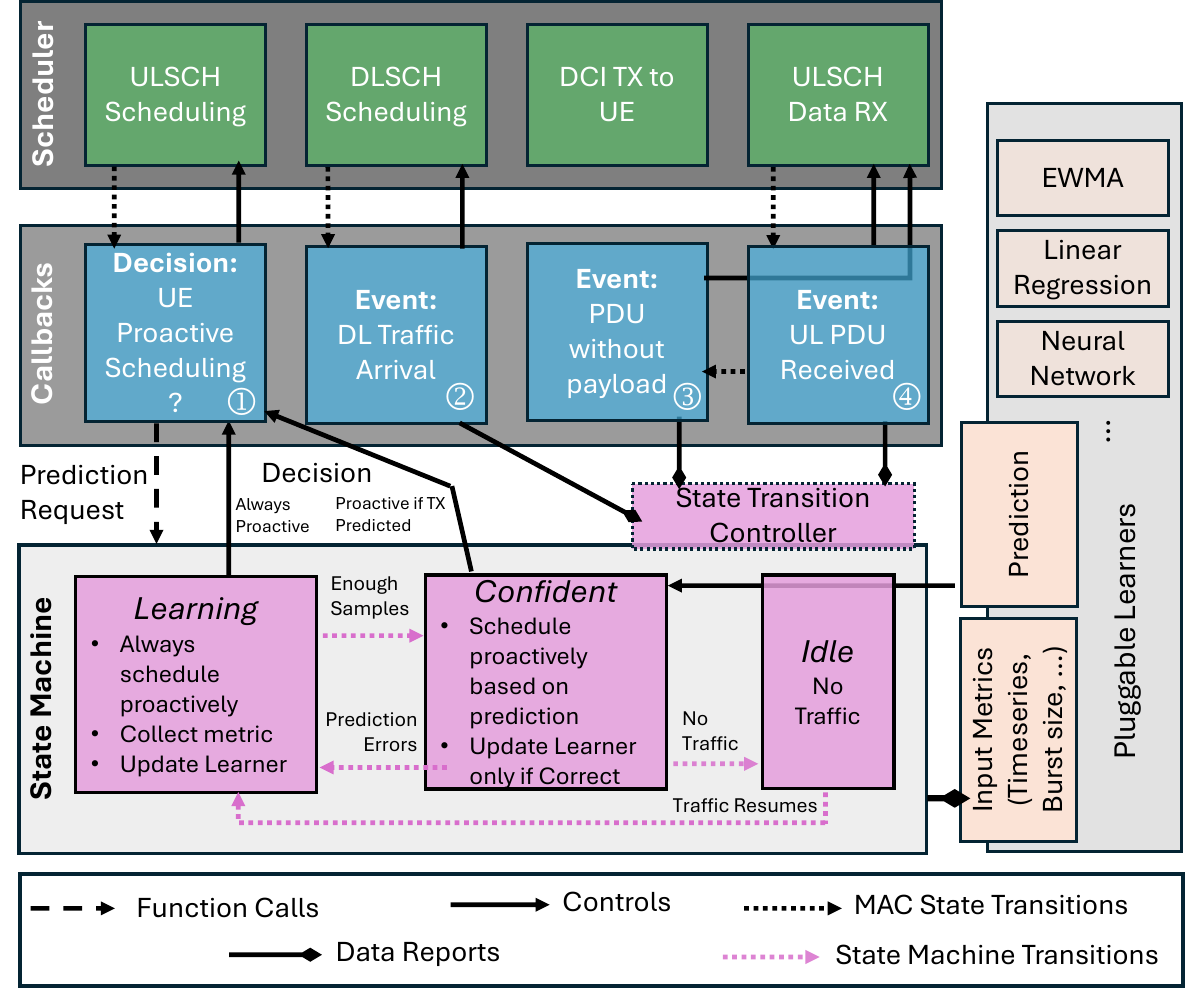}
    \caption{Proactive scheduler architecture.}
    \label{fig:proactive_arch}
    \vspace{-0.5cm}
\end{figure}

We now generalize this basic principle beyond a fixed lookback window. As shown in Fig.~\ref{fig:proactive_arch}, our solution trains a per-\gls{ue} learner which predicts the next \gls{ul} arrival from \gls{mac} events, and a state machine determines when to use the prediction for proactive scheduling.

We implement the proactive scheduler as a combination of three \glspl{dapp} using the interfaces shown in Fig.~\ref{fig:sched_arch}: candidate request~(A), \gls{dl} transmission observation~(B), and \gls{ul} observation~(C). The \gls{dl} observer \gls{dapp} reports each constructed transport block, while the \gls{ul} observer \gls{dapp} reports the outcome of each \gls{ul} grant, including its decoding outcome, received payload, padding, grant parameters, and whether the candidate was requested by a \gls{dapp}. A received payload and an unused proactive grant are therefore two outcomes of the same typed event reported by the \gls{ul} observer \gls{dapp}~(C). Together, these observations update the per-\gls{ue} predictor maintained in the \glspl{dapp}' shared persistent context.
During candidate collection, the candidate-request \gls{dapp} is invoked once for each structurally eligible \gls{ue}. When the online learner predicts that a proactive grant is due, this \gls{dapp} requests that the \gls{ue} enter the candidate set. Because no \gls{bsr} is available before the \gls{sr}, the buffer state remains unknown. In our configuration, the independently selected \gls{prb}-allocation \gls{dapp} therefore assigns the minimum grant, after which the candidate follows the standard allocation and validation path.
The architecture accepts any arrival predictor that outputs a predicted slot. We evaluate an \gls{ewma} and linear regression in this work. More expressive models, including \glspl{nn} and gradient-boosted trees, can use the same interface; we leave their evaluation to future work.

\textbf{State machine.} A state-machine instance is associated with each group of \glspl{ue} that shares an arrival distribution. When no such prior is available, each \gls{ue} receives its own instance. A group may share one instance when its members are known to have the same statistics, such as identical sensors on a fixed reporting schedule. We describe one instance below. The scheduler runs all instances independently.

At initialization, the instance is in the \emph{Learning} state, where it schedules proactively in every slot. This both masks the \gls{sr} delay and collects an unbiased dataset of true \gls{ul} arrival times. In parallel, the learner is updated from any subset of scheduler-observable features, including past \gls{ul} inter-arrival times, \gls{dl} event sizes and timestamps, \gls{dl}-to-\gls{ul} response times, and traffic class. The learner emits a predicted slot offset for the next \gls{ul} arrival. The instance enters \emph{Confident} once a learner-specific readiness condition is met, such as a minimum sample count or a bound on prediction variance.

In the \emph{Confident} state, scheduling follows the learner's prediction through two operator-tunable parameters. The tolerance window $\mathrm{TW}$ admits scheduling within $\pm\mathrm{TW}$ \gls{tdd} periods of the prediction. The slot restriction $N$ limits proactive grants to the last $N$ slots of each \gls{tdd} period. Smaller $(\mathrm{TW},N)$ reduce overhead by committing more closely to the predicted slot. Larger values widen the safety window at the cost of additional grants. In the language of Eq.~\eqref{eq:multiobj}, \emph{Learning} realizes the minimum-latency extreme at maximum resource use, while $(\mathrm{TW},N)=(0,1)$ approaches the minimum-overhead prediction-driven configuration. Intermediate values trace the empirical latency-overhead frontier. The achievable trade-off depends on prediction accuracy.

Two online counters determine whether the instance remains \emph{Confident}. Consecutive reactive transmissions count missed predictions, while unused proactive grants determine the false-positive rate. Crossing a configured threshold, such as three consecutive misses or a 10\% false-positive rate, returns the instance to \emph{Learning}. A learner may also emit $(\mathrm{TW},N)$ dynamically. This lets it absorb drift by temporarily widening the safety window without returning fully to \emph{Learning}. A third \emph{Idle} state pauses prediction after extended inactivity, such as 100 slots without \gls{ul} traffic. When traffic resumes, the instance returns to \emph{Learning}.

Together, the three states balance latency, jitter, and overhead under unknown or drifting traffic without \gls{rrc} reconfiguration or prior knowledge of the arrival process.

\subsection{Evaluation}
\label{sec:proactive-eval}

\begin{figure}[t]
    \centering
    \includegraphics[width=0.95\linewidth]{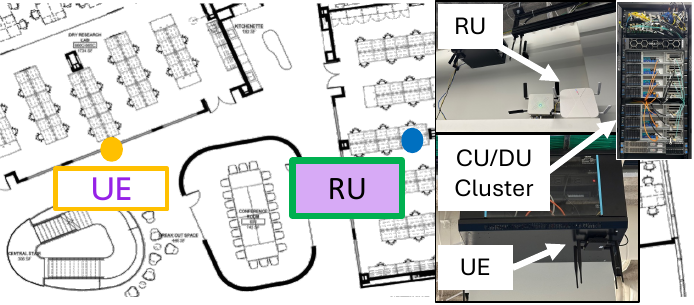}
    \caption{Over-the-air testbed for proactive-scheduler evaluation.}
    \label{fig:proactive_testbed}
    \vspace{-0.5cm}
\end{figure}

To exercise \framework in an \gls{ota} deployment, we evaluate the proactive scheduler on the testbed shown in Fig.~\ref{fig:proactive_testbed}. The deployment is built on the X5G testbed~\cite{villa2024x5g} and the AutoRAN framework~\cite{maxenti2025autoran}. The \gls{oai} \gls{gnb} runs on a Dell PowerEdge R760 server connected to a Foxconn RPQN \gls{ru} operating in band n78 with 100~MHz bandwidth and numerology $\mu=1$. A Sierra Wireless EM9293 \gls{ue} communicates over the air under fair channel conditions, with a median \gls{cqi} of 13. The network uses the DDDDDDDSUU \gls{tdd} pattern with a 10-slot period.

We design three complementary scenarios in which proactive scheduling can leverage different forms of temporal structure.

\textbf{Request-response.} We generate ICMP echo requests with a 300-byte payload from the \gls{upf} to the \gls{ue} at controlled intervals. This scenario represents query-response applications such as industrial sensors responding to polling or roadside infrastructure querying vehicle status. The predictable \gls{dl}-to-\gls{ul} pattern allows us to evaluate \gls{rtt}-based learning, where the scheduler predicts \gls{ul} responses based on prior \gls{dl} transmissions. We configure the scheduling \gls{dapp} with an \gls{ewma} learner.

\textbf{Size-dependent edge inference.} We emulate workloads in which a central node queries an edge device, which runs local \gls{ml} inference and returns the result. Both the inference latency and the resulting \gls{ul} payload may depend on the incoming \gls{dl} traffic. We instantiate this scenario by running an Arctic Embed XS text-embedding model on the \gls{ue}. It receives a variable-length prompt over \gls{dl}, performs inference whose duration depends on the input length, and returns the embedding over \gls{ul}. The server at the \gls{ue} and the client at the \gls{upf} communicate over \gls{udp}. We vary the input using random word sequences of different lengths, which produce different burst sizes at the \gls{mac} layer. We configure the scheduling \gls{dapp} to use linear regression on the \gls{dl} burst size.

\textbf{Periodic autonomous reporting.} We generate periodic \gls{ul}-only traffic at fixed intervals between 10 and 100~ms, with no preceding \gls{dl} trigger. This represents workloads such as heartbeat messages and autonomous sensor reporting. Since there is no \gls{dl} correlation to leverage, the scheduler must learn the inter-arrival pattern. We use MGEN to send 97 packets per second. We select this prime-number rate so that the position of packets within the \gls{tdd} period does not repeat. It produces a 10.31~ms inter-arrival time, between the standardized \gls{cg} periodicities of 10 and 20~ms. We configure the scheduling \gls{dapp} with an \gls{ewma} learner.

For evaluation, we disable the Confident-to-Learning fallback transition that would normally be triggered by accumulated prediction errors. Otherwise, the measured latency and overhead would mix the predictor's accuracy in the Confident state with the always-schedule safety net of the Learning state, which by construction occupies the 100\%-overhead, minimum-delay corner. Disabling the fallback isolates the first effect, so the reported results reflect Confident-state prediction quality alone. Our baseline is the default \gls{oai} scheduler, which does not perform proactive scheduling.

For each scenario, we evaluate the latency CDFs across the $(\mathrm{TW},N)$ sweep. For the request-response scenario, we also evaluate the overhead, defined as the fraction of \gls{ul} slots in which the \gls{ue} is scheduled. The sweep traces an empirical latency-overhead trade-off. A sharp knee occurs when one $(\mathrm{TW},N)$ configuration achieves near-minimum overhead and near-minimum latency, while more aggressive configurations provide no further latency reduction. The location of this knee also indicates prediction quality. For example, a knee at $(\mathrm{TW},N)=(0,1)$ means that the learner localizes \gls{ul} arrivals to one \gls{tdd} period and one eligible \gls{ul} slot.

\textbf{Request-response results.} We vary $\mathrm{TW}$ between 0, which schedules proactively only when the \gls{tdd} period matches the prediction exactly, and 3, which covers $\pm3$ periods around the prediction. We vary $N$ between 1 and 3 because the configured pattern has at most three \gls{ul} transmission opportunities per \gls{tdd} period, comprising two U slots and one mixed S slot.

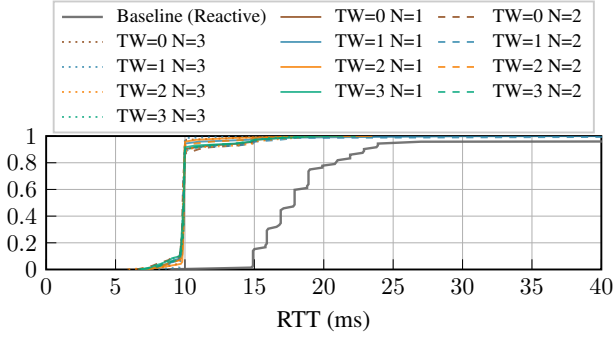
\begin{figure}[t]
\centering
    \setlength\fwidth{0.9\linewidth}
    \setlength\fheight{4cm}
    \input{figures/ewma_comparison_all}
    \caption{\gls{rtt} for the request-response scenario.}
    \label{fig:latency-rtt}
    \vspace{-0.5cm}
\end{figure}

\begin{figure}[t]
\centering
    \setlength\fwidth{0.9\linewidth}
    \setlength\fheight{4cm}
    \input{figures/rtt_overhead}
    \caption{Request-response overhead; bottom labels show $N$.}
    \label{fig:overhead}
    \vspace{-0.5cm}
\end{figure}
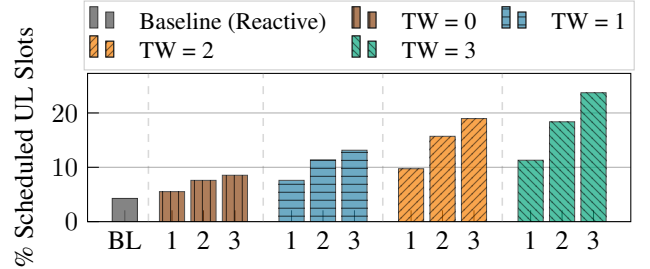

Figure~\ref{fig:latency-rtt} presents the latency results. For every $(\mathrm{TW},N)$ configuration, the distribution improves substantially over the baseline. The median drops from approximately 20~ms with a long tail to 10~ms. The 90th percentile is also 10~ms, and the long tail disappears.
As shown in Fig.~\ref{fig:overhead}, the overhead, measured as the fraction of \gls{ul} slots with some resources scheduled either proactively or reactively, scales linearly with $\mathrm{TW}$ and $N$. The optimal configuration for this scenario is $\mathrm{TW}=0$ and $N=1$, where the overhead is 7\%. The baseline has the lowest overhead, with 4\% of slots scheduled. We present detailed overhead results only for request-response. Overhead scales similarly with $\mathrm{TW}$ and $N$ in the other scenarios, with absolute values proportional to their traffic rates.

\begin{figure}[t]
\centering
    \setlength\fwidth{0.9\linewidth}
    \setlength\fheight{4cm}
    \input{figures/ml_inference}
    \caption{End-to-end latency for the size-dependent edge inference scenario.}
    \label{fig:ml_inference}
    \vspace{-0.5cm}
\end{figure}
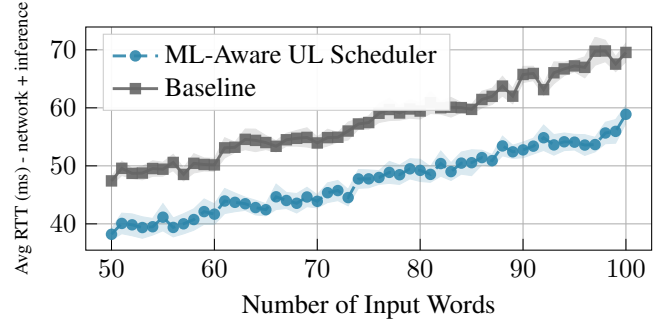

\textbf{Size-dependent edge inference results.} We use $\mathrm{TW}=2$ and $N=1$. As shown in Fig.~\ref{fig:ml_inference}, the scheduler predicts the \gls{ul} arrival time from the \gls{dl} burst size, reducing end-to-end delay by approximately 10~ms over the baseline across all input sizes. This is consistent with the network latency saved in the request-response scenario.

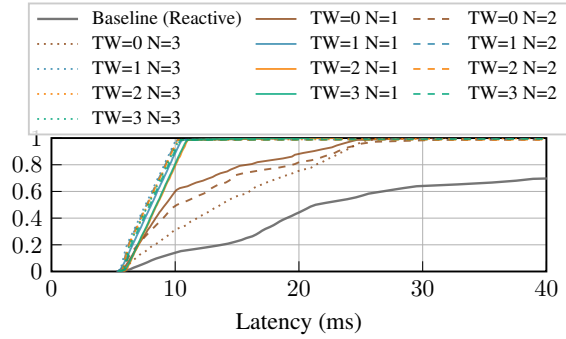
\begin{figure}[t]
\centering
    \setlength\fwidth{0.9\linewidth}
    \setlength\fheight{4cm}
    \input{figures/mgen_comparison_all}
    \caption{One-way \gls{ul} latency for the periodic autonomous reporting scenario.}
    \label{fig:oneway}
\end{figure}

\textbf{Periodic autonomous reporting results.} Figure~\ref{fig:oneway} presents the results for the third scenario. We vary $\mathrm{TW}$ between 0 and 3 and $N$ between 1 and 3. A tolerance window of $\mathrm{TW}=0$ already improves the CDF over the baseline, but the best results appear from $\mathrm{TW}=1$, with a median one-way latency of 7~ms and a 99th percentile of 10~ms. The shift in the optimal $\mathrm{TW}$ reflects the traffic pattern. The 97-packet/s rate produces a 10.31~ms inter-arrival time that is not a multiple of the \gls{tdd} period, so successive packets move across slot boundaries within the pattern. A predictor with $\mathrm{TW}=0$ commits to one \gls{tdd} period and misses the arrival as drift accumulates. A tolerance of $\mathrm{TW}=1$ absorbs the drift at low overhead. A \gls{cg} cannot apply the same optimization because its fixed periodicities of 10 and 20~ms lock the grant to one phase. The same drift would steadily desynchronize the \gls{cg} schedule from the actual arrivals. The difference between $N=1,2,3$ is at most 0.1~ms. For this regular traffic, scheduling a few slots earlier has only a small effect on \gls{ul} latency. The optimal operating point is therefore traffic-dependent, and the architecture allows the learner to emit $(\mathrm{TW},N)$ from the observed statistics.

Overall, this use case demonstrates how \framework composes observations, persistent state, and candidate requests across scheduling opportunities. The \gls{dl} and \gls{ul} observer \glspl{dapp} update the predictor, while the candidate-request \gls{dapp} acts on its output; resource allocation and grant validation remain unchanged. The resulting policy executes inside the \gls{oai} scheduler over the air. We next exercise the complementary set of interfaces: per-slot decision stages evaluated under controlled channel conditions with \macemu.

\section{Use Case: Frequency-Selective UL Scheduling}
\label{sec:freqsel}

Our second use case uses \framework to implement a frequency-selective \gls{ul} scheduler. The scheduler leverages frequency-domain channel variations to assign each \gls{ue} a favorable contiguous sub-band. We show how \framework enables us to leverage \gls{srs} pilots for resource allocation, and how \macemu enables evaluation across large-scale mobility and power-limited scenarios.

\subsection{Motivation and Problem}
\label{sec:freqsel_problem}

The potential gain from frequency-selective scheduling comes from multi-user frequency diversity, i.e., multipath propagation produces different strong and weak frequency regions for different \glspl{ue}. This creates the opportunity of matching each grant to a favorable region, thereby increasing aggregate throughput within the same spectrum budget.
To realize this gain, channel estimates need to be fresh. Indeed, the frequency response estimate (measured through \gls{srs} in the \gls{ul} or through \gls{cqi} in the \gls{dl}) that the \gls{gnb} observes must be close enough to the current channel state. As Doppler increases, the channel deviates from the pilot estimates faster, and the real ordering of candidate sub-bands may change compared to the stale information available at the \gls{gnb}~\cite{okvist2011}. In the \gls{ul}, the available gain may also depend on the \gls{ue}'s operating point (i.e., its power budget and the state of the power-control loop)~\cite{ts38213}.

These considerations have motivated a broad literature on frequency-selective resource allocation, primarily on the \gls{dl}. Early multi-user \gls{ofdm} schedulers jointly assign individual subcarriers and select their transmission rates~\cite{wong1999,rheecioffi2000}. These formulations do not map directly to 5G, where scheduling operates at \gls{prb} granularity, one \gls{mcs} applies across a transport block, and \gls{ul} allocations are subject to standardized resource-allocation and waveform constraints. More recent designs use quantized sub-band \gls{cqi} reported by the \gls{ue}. For example, measurements on commercial LTE hardware characterize the gain from sub-band \gls{cqi} reporting~\cite{okvist2011}, while SARA uses sub-band \gls{ri}, \gls{pmi}, and \gls{cqi} reports in 5G-LENA~\cite{sara2025}. Depending on the configured resource-allocation type, the \gls{dl} scheduler may also distribute an allocation across multiple resource-block groups.

A smaller body of work studies frequency-domain scheduling on the \gls{ul}. Earlier work primarily formulates the allocation problem under the contiguity constraint imposed by the \gls{ul} waveform. Krishnamoorthy \textit{et al.} propose frequency-semi-selective scheduling to reduce channel-reporting overhead for \mbox{DFT-S-OFDM}~\cite{krishnamoorthy2008fsss}, while Calabrese \textit{et al.} construct contiguous channel-aware allocations using a search tree~\cite{calabrese2008fdps}. Lee \textit{et al.} further show that proportional-fair frequency-domain scheduling under this constraint is NP-hard~\cite{lee2009proportional}.
Gutierrez \textit{et al.} provide a closer connection to the LTE measurement procedure~\cite{gutierrez2020joint}. They derive a scalar effective \gls{sinr} for each \gls{ue}--\gls{prb} pair in the Vienna LTE-A simulator, use Recursive Maximum Expansion to construct contiguous allocations, and select a common \gls{mcs} from the channel quality across each assigned interval. Because the allocation procedure is shared across their comparisons, the evaluation primarily measures the effect of link adaptation instead of isolating the gain from frequency-selective placement. Their formulation also does not consider how frequency-domain channel quality depends on the selected rank, precoder, or the \gls{ue}'s power distribution across a grant with a variable number of \glspl{prb}.

Taken together, these gaps motivate studying frequency-selective allocation within the complete scheduling loop, which we do using \framework. %\framework exposes frequency-resolved \gls{srs} observations through the shared candidate structure and separates rank and precoder selection, resource allocation, and link adaptation into replaceable stages, while \macemu provides controlled mobility and power-limited operating points.
We formulate and evaluate two real-time allocation frequency-selective policies. The first restricts each \gls{ue} to a fixed target number of \glspl{prb}, while the second retains a bounded set of candidate \gls{prb} counts. We first evaluate these policies offline through simulation, before evaluating whether the properties observed in simulation carry within the real network stack, and how they compare again the proportional fair baseline.

\subsection{Design and Implementation}
\label{sec:freqsel_design}

We now formulate the frequency-selective allocation problem and derive two policies that restrict its search space sufficiently for per-slot execution. We then evaluate their expected allocation quality and computational cost offline in simulation before executing them in \macemu.

\textbf{Sub-band utility.}
Both our algorithms start from constructing a utility profile based on each \gls{ue}'s \gls{srs} feedback.
We read the complex \gls{srs} estimates exposed through the shared candidate structure of \framework. For each candidate \gls{ue} $u$ and \gls{prb} $k$, we apply the rank and precoder selected by the rank/precoder \gls{dapp} (stage~2 in Fig.~\ref{fig:sched_arch}) and estimate the post-\gls{mmse} \gls{sinr} $\gamma_{u,\ell}(k)$ of every scheduled layer $\ell$.
% For multi-layer transmissions, this calculation accounts for spatial conditioning: high channel energy alone does not imply that the layers can be separated reliably.

Since \gls{srs} uses a per-\gls{ue} power-control loop distinct from those of other \gls{ul} channels, the magnitude of the \gls{srs} estimates does not directly represent the \gls{pusch} operating point. We therefore treat the \gls{srs} estimates as a relative frequency profile and combine them with the latest \gls{pusch}-measured \gls{sinr}. We then define the per-\gls{prb} utility as the following rate proxy:
\begin{equation}
  q_u(k) \triangleq
  \sum_{\ell=1}^{L_u}
  \log_2\!\left(1+\gamma_{u,\ell}(k)\right),
  \label{eq:freqsel_prb_score}
\end{equation}
where $L_u$ is the number of layers selected for \gls{ue} $u$. We represent a candidate grant by its first \gls{prb} $s$ and the number $w$ of contiguous \glspl{prb} it contains. Its aggregate utility is
\begin{equation}
  R_u(s,w) \triangleq
  \sum_{k=s}^{s+w-1} q_u(k).
  \label{eq:freqsel_rate}
\end{equation}

\textbf{Allocation problem.}
Let $p_u$ denote the proportional-fair priority of \gls{ue} $u$, $B$ the number of available \glspl{prb}, $U$ the number of candidate \glspl{ue}, and $K$ the number of additional grants permitted by the remaining \gls{pdcch} budget when the dApp is invoked. $\kappa\triangleq\min(U,K)$ is the maximum number of candidates that the allocation can select. Given the interval utilities above, the allocator seeks a set $\mathcal{A}=\{(u,s_u,w_u)\}$ of non-overlapping intervals that maximizes
\begin{equation}
  \max_{\mathcal{A}}
  \sum_{(u,s_u,w_u)\in\mathcal{A}}
  p_u R_u(s_u,w_u),
  \label{eq:freqsel_allocation}
\end{equation}
subject to $|\mathcal{A}|\leq\kappa$, at most one interval per \gls{ue}, and the contiguity and feasibility constraints of the \gls{ul} grant. An exact \gls{dp} can index its state by the next available \gls{prb} and the subset of users that remain available. For a candidate set $\mathcal{W}$ of \gls{prb} counts, with $W=|\mathcal{W}|$, this formulation requires $O(B\cdot 2^U)$ states and $O(BUW\cdot2^U)$ time in the worst case, making \gls{dp} unsuitable for real-time execution. We therefore formulate two real-time policies that restrict different dimensions of this search space.

\textbf{Fixed PRB-count placement.}
Our first policy (\freqselfixed) assumes that up to $\kappa$ candidates will be selected and divides the $B$ available \glspl{prb} equally among them, and enforces a minimum grant size $w_{\min}$. The target number of \glspl{prb} is
\begin{equation}
  w_{\mathrm{tgt}} \triangleq
  \max\!\left(
    w_{\min},
    \left\lfloor\frac{B}{\kappa}\right\rfloor
  \right).
  \label{eq:freqsel_target_prb_count}
\end{equation}
% Within this search, channel information changes the position of the grant.
For each \gls{ue} $u$, the policy examines every contiguous run of available \glspl{prb}. Within a run of length $\ell$, it evaluates all starting positions for an interval containing $w=\min(w_{\mathrm{tgt}},\ell)$ \glspl{prb}. Across all runs, it retains the interval $(s_u,w_u)$ with the largest $R_u(s_u,w_u)$. This produces one candidate interval per \gls{ue} with weight $p_u \times R_u(s_u,w_u)$. Then, since preferred intervals of each \gls{ue} may overlap, a weighted-interval-scheduling \gls{dp} selects the maximum-weight non-overlapping subset containing at most $\kappa$ intervals.

\textbf{Variable PRB-count placement.}
Our second policy, \freqselrefined, separates user ordering from \gls{prb}-count refinement. First, a maximum-weight assignment maps $\kappa$ distinct users to $\kappa$ ordered intervals containing equal numbers of \glspl{prb}. Let $\pi=(u_1,\ldots,u_\kappa)$ denote the resulting user order. Holding $\pi$ fixed, an ordered \gls{dp} chooses $(s_{u_j},w_{u_j})$ such that
\begin{equation}
  s_{u_j}+w_{u_j}\leq s_{u_{j+1}},
  \qquad w_{u_j}\in\mathcal{W},
  \quad 1\leq j<\kappa.
  \label{eq:freqsel_ordered}
\end{equation}
The \gls{dp} maximizes $\sum_{j=1}^{\kappa} p_{u_j} \times R_{u_j}(s_{u_j},w_{u_j})$ over the feasible positions and \gls{prb} counts. %It is therefore exact for the selected user order, while avoiding the subset state of the global optimum.

\textbf{Computational cost.}
Both policies construct $q_u(k)$ and its cumulative values in $O(B)$ time per \gls{ue}. The cumulative representation then allows an interval score to be obtained with two lookups and one subtraction. For \freqselfixed, scanning the starting positions costs $O(UB)$, constructing the interval-compatibility relation costs $O(U^2)$, and selecting at most $\kappa$ intervals costs $O(U\kappa)$. Its overall complexity is therefore $O(UB+U^2+U\kappa)$. For \freqselrefined, assigning $\kappa$ intervals to $U$ users costs $O(U\kappa^2)$, while evaluating $W$ candidate \gls{prb} counts across $B$ positions for each selected user costs $O(\kappa BW)$. Including utility-profile construction, its overall complexity is $O(UB+U\kappa^2+\kappa BW)$.

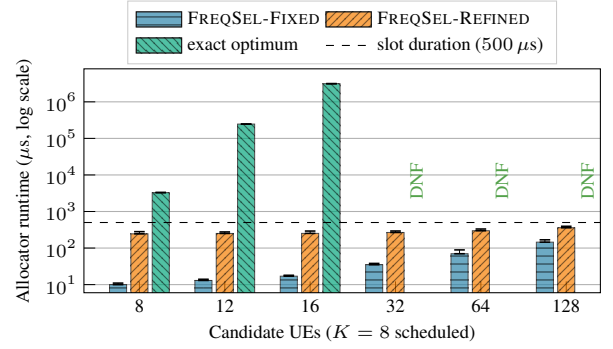
\begin{figure}[t]
  \centering
  \begin{minipage}{0.95\linewidth}
    \centering
    \scriptsize
    \input{figures/freqsel_runtime_preview}
  \end{minipage}
  \caption{Scheduler execution time (offline execution).}
  \label{fig:freqsel_runtime}
\end{figure}

\textbf{Offline policy evaluation.}
We first evaluate whether the two restricted searches can execute within the slot duration. We fix $B=106$, $K=8$ (and hence $\kappa=8$), and $W=30$, vary the number of candidate users from 8 to 128, and measure the execution time outside the scheduler loop. Figure~\ref{fig:freqsel_runtime} shows that both policies remain within the slot budget: with 128 candidates, their 99th-percentile execution times are 168~$\mu$s for \freqselfixed and 394~$\mu$s for \freqselrefined. In contrast, the exact solver already requires 3.3~ms for eight users and 247~ms for twelve users.

We then compare the allocation quality of the two real-time policies. Each channel instance contains eight independent static rank-2 users over a $B=106$-\gls{prb} carrier under a CDL-B channel with 300-ns delay spread. We replay the searches performed by \freqselfixed and \freqselrefined with equal proportional-fair weights and no pre-existing allocations. For every selected interval, we derive the post-\gls{mmse} \glspl{sinr} from the exact channel state and select the \gls{mcs} that maximizes expected rate under \macemu's calibrated \gls{bler} model. We divide the resulting aggregate expected rate by $B$ to obtain the expected cell spectral efficiency in bit/s/Hz.

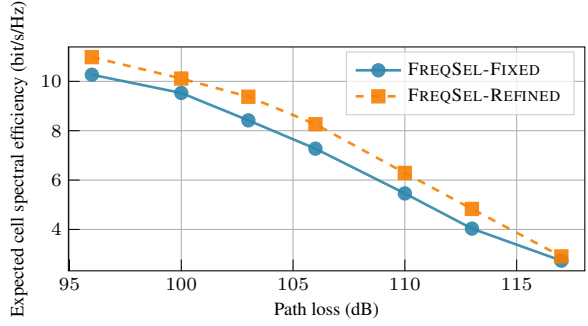
\begin{figure}[t]
  \centering
  \begin{minipage}{0.95\linewidth}
    \centering
    \scriptsize
    \input{figures/f10_bound_vs_pl}
  \end{minipage}
  \caption{Simulation-based expected cell spectral efficiency for 8 static users under CDL-B 300-ns channel.}
  \label{fig:freqsel_ceiling}
\end{figure}

Figure~\ref{fig:freqsel_ceiling} reports this quantity while varying path loss. \freqselrefined outperforms \freqselfixed throughout the sweep, with 10.98 against 10.27~bit/s/Hz at 96~dB path loss, and 110~dB, they achieve 6.29 against 5.46~bit/s/Hz at 110~dB. Overall \freqselrefined's advantage grows from 7.0\% to 15.2\% as the users become power-limited.
Taken together, Figs.~\ref{fig:freqsel_runtime} and~\ref{fig:freqsel_ceiling} identify \freqselrefined as a promising real-time policy in the offline evaluation: the implementation remains within the slot budget, and it outperforms \freqselfixed.

\subsection{\framework-based Evaluation}
\label{sec:freqsel_eval}

We use \framework to evaluate frequency-selective allocation inside the complete \gls{oai} scheduling loop in \macemu. We first select between the two real-time variants introduced above and then compare the retained \freqselfixed policy with proportional fairness across operating conditions. The experiments use saturating \gls{ul} \gls{mgen} traffic, a CDL-B 300~ns channel and two-layer \gls{mimo}. Unless stated otherwise, they use 32 users and periodic \gls{srs} with an interval of 160~slots (80~ms).

\textbf{Mobility models.} For moving users, we set the maximum Doppler according to $f_D=vf_c/c$; at our 3.5-GHz carrier frequency, speeds of 1, 2, and 5~km/h correspond to 3.24, 6.48, and 16.20~Hz, respectively. For static \glspl{ue}, we use 0.5~Hz to represent channel variations caused by moving environmental scatterers, as observed on fixed wireless links~\cite{domazetovic2003doppler}. 

We evaluate with homogeneous scenarios (all \glspl{ue} either moving or static), as well as a mix of static and moving \glspl{ue}. Additionally, we also use a L\'evy-walk mobility model for one scenario, in which users alternate between movement at 5~km/h and quasi-static pauses. Pause durations follow a truncated Pareto distribution with exponent 0.8 over 30--300~s, while movement durations are exponentially distributed with a mean of 60~s. During pauses, the channel continues to evolve at 0.5~Hz. This captures the alternating movement and heavy-tailed pause behavior observed in human-mobility traces~\cite{rhee2011levy}.

\begin{figure}[t]
  \centering
  \begin{minipage}{0.95\linewidth}
    \centering  
    \scriptsize
    \input{figures/f13_pl_sweep}
  \end{minipage}
  \caption{Aggregate \gls{ul} throughput in \macemu.}
  \label{fig:freqsel_pathloss}
\end{figure}
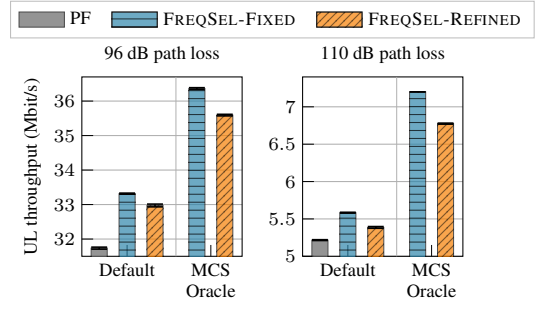

\textbf{Policy selection.} We begin by determining whether superiority of \freqselrefined over \freqselfixed is also observed in end-to-end network performance.
We execute \freqselfixed and \freqselrefined within \framework as \gls{prb} allocation \glspl{dapp}. 
The experiment contains 16 quasi-static users, eight users at 2~km/h, and eight users at 5~km/h. Figure~\ref{fig:freqsel_pathloss} compares aggregate \gls{ul} throughput at 96 and 110~dB path loss. In each panel, the left group compares proportional fairness and our two policies. At 96~dB, \freqselfixed and \freqselrefined achieve 33.32 and 32.97~Mbit/s, corresponding to gains of 5.0\% and 3.9\% over proportional fairness. At 110~dB, they achieve 5.58 and 5.39~Mbit/s, corresponding to gains of 7.0\% and 3.3\%. Thus, \freqselrefined remains below \freqselfixed, and the improvement suggested by the offline optimization does not carry when evaluating with the full scheduler.

The oracle-\gls{mcs} bars on the right of each panel test whether this reversal is caused by the interaction with link adaptation. We replace the default \gls{mcs}-selection \gls{dapp} with an oracle \gls{dapp} that uses the exact \macemu channel state to select the highest \gls{mcs} whose predicted \gls{bler} does not exceed 13\%. At 96~dB, \freqselfixed and \freqselrefined achieve 36.36 and 35.59~Mbit/s, respectively. At 110~dB, they achieve 7.20 and 6.78~Mbit/s. Although the oracle increases throughput for both policies, it does not change their ordering: \freqselrefined remains 2.1\% and 5.9\% below \freqselfixed at the two operating points.

This experiment illustrates how \framework supports policy shaping within the complete system rather than from an isolated optimization objective. By replacing the allocation and link-adaptation stages independently while retaining the rest of the scheduling loop, we find that refining the number of allocated \glspl{prb} increases computational and implementation complexity without improving throughput. We therefore retain \freqselfixed for the remaining experiments, which compare it with proportional fairness under the common CDL-B 300-ns channel at 96~dB path loss, unless stated otherwise.

\begin{figure}[t]
  \centering
  \begin{minipage}{0.95\linewidth}
    \centering
    \scriptsize
    \input{figures/f2_gain_vs_speed}
  \end{minipage}
  \caption{Aggregate \gls{ul} throughput versus user speed.}
  \label{fig:freqsel_gain_speed}
\end{figure}
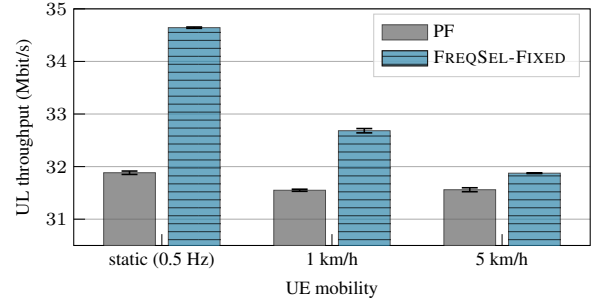

\textbf{Mobility Experiments.} We first isolate the effect of channel evolution by placing all 32 users in the same mobility state. Figure~\ref{fig:freqsel_gain_speed} compares the quasi-static, 1-km/h, and 5-km/h cases. The gain of \freqselfixed is largest in the quasi-static cell and decreases with mobility as the channel evolves between the \gls{srs} observation and the resulting grant. This trend is consistent with the existing studies (e.g., most gain lost at 10~km/h with 20~ms channel sounding intervals ~\cite{okvist2011}).

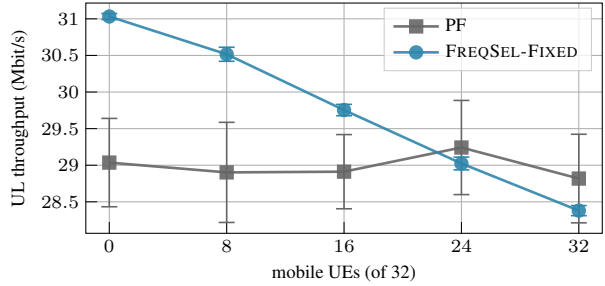
\begin{figure}[t]
  \centering
  \begin{minipage}{0.95\linewidth}
    \centering
    \scriptsize
    \input{figures/f5_proportion_law}
  \end{minipage}
  \caption{Aggregate \gls{ul} throughput.}
  \label{fig:freqsel_proportion}
\end{figure}

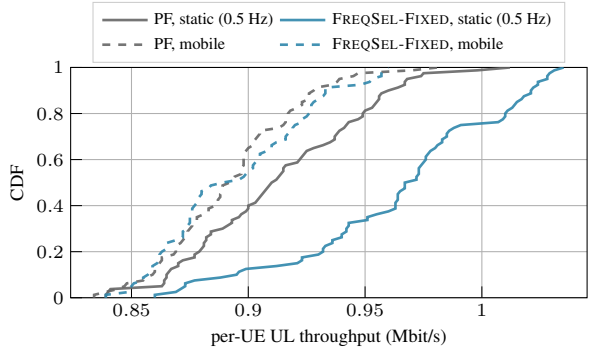
\begin{figure}[t]
  \centering
  \begin{minipage}{0.95\linewidth}
    \centering
    \scriptsize
    \input{figures/f8_class_cdf}
  \end{minipage}
  \caption{Per-user \gls{ul} throughput in a cell with 16 quasi-static users and 16 users moving at 5~km/h.}
  \label{fig:freqsel_class_cdf}
\end{figure}

\textbf{Mixed mobility.} Next, we evaluate \freqselfixed in cells with mixed mobility. In Figure~\ref{fig:freqsel_proportion}, we vary the number of 5-km/h users in a 32-user cell, with the remaining users following the quasi-static model. The gain over proportional fairness decreases as the mobile fraction increases, from 7\%, or 2.0~Mbit/s, when every user is quasi-static to no measurable gain when every user is mobile.
We then show in Figure~\ref{fig:freqsel_class_cdf} how this gain is distributed in a cell containing 16 quasi-static and 16 mobile users. We observe that frequency-selective allocation shifts the throughput distribution of the quasi-static users toward higher rates, while the distribution of the mobile users remains close to that obtained with proportional fairness, as channel evolution and \gls{srs} aging removes most of the benefit of channel-aware \gls{prb} allocation.

\begin{figure}[t]
  \centering
  \begin{minipage}{0.95\linewidth}
    \centering
    \scriptsize
    \input{figures/f9_levy_state_cdf}
  \end{minipage}
  \caption{Per-grant \gls{mcs} conditioned on \gls{ue} state (L\'evy-walk).}
  \label{fig:freqsel_levy}
\end{figure}
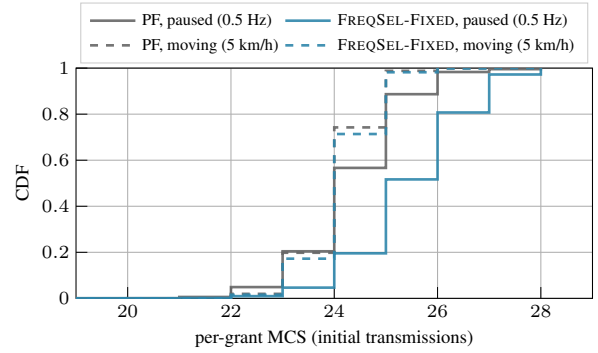

\textbf{L\'evy-walk mobility.} The preceding experiments keep each user in one mobility state throughout the run. We finally examine whether the same behavior holds when individual users alternate between moving and paused periods according to the L\'evy-walk model. Figure~\ref{fig:freqsel_levy} reports the per-grant \gls{mcs} distribution conditioned on the user's state. During quasi-static pauses, \freqselfixed shifts the distribution toward higher \gls{mcs} values relative to proportional fairness. During movement at 5~km/h, the two distributions largely overlap. The benefit therefore follows the user's current mobility state: frequency-selective placement improves link quality while the observed frequency profile remains sufficiently stable, but provides little value after that profile decorrelates.

Overall, \freqselfixed improves aggregate \gls{ul} throughput by up to 7\% in the complete scheduler, with gains concentrated among users whose channels vary slowly. More importantly, this set of experiments illustrates the value of exposing scheduler decisions as separate \gls{dapp} stages. The offline evaluation favors \freqselrefined, but the policy ordering reverses when both algorithms execute inside the complete \gls{oai} scheduling loop. Because resource allocation and \gls{mcs} selection are decoupled, we can replace the default link adaptation with an oracle \gls{dapp} while leaving the algorithms and the rest of the scheduler unchanged. Even under these idealized conditions, the reversal persists. Combined with \macemu's controlled mobility and power conditions, this decomposition lets us identify which the best algorithm within the complete system.

\section{Related Work}
\label{sec:related}

We compare \framework with existing experimentation platforms and with prior work on the two scheduling problems.

\subsection{Experimentation Frameworks, Platforms, and Emulation}
\label{sec:rw_platforms}

Open \gls{ran} stacks such as \gls{oai}~\cite{kaltenberger2019openairinterface} and srsRAN/OCUDU~\cite{gomezmiguelez2016srslte} implement a full protocol stack compliant with 3GPP NR Release 15/16. Their per-slot schedulers --- of which the internals are not standardized, leaving the vendor room to develop their own logic --- however, remain monolithic and represent the motivation for the introduction of \framework. 
To our knowledge, \framework is the first framework to expose the in-\gls{mac} per-slot decision as a pluggable interface in a production \gls{ran} stack.

External controllers provide a complementary control tier. The \gls{oran} architecture hosts control logic in near-\gls{rt} and non-\gls{rt} \glspl{ric}~\cite{polese2023understanding}; FlexRIC connects these controllers to \gls{oai} through an E2 agent and service-oriented \gls{sdk}~\cite{flexric}; and ColO-RAN develops and evaluates machine-learning xApps on programmable platforms~\cite{polese2022colo}. These controllers steer \gls{ue}-level policies over tens of milliseconds or longer, and this is not compatible with real-time scheduling decisions. A \framework dApp applies such a policy directly inside the per-slot scheduler. This design follows the dApp model, which places sub-millisecond inference and control inside the \gls{gnb}~\cite{doro2022dapps}, and its E3 interface for attaching dApps and consuming real-time data~\cite{lacava2025dapps}. \framework specializes this model into synchronous scheduling stages and additionally lets stateful dApps observe transmission outcomes and orchestrate controls across scheduling opportunities.

Existing evaluation platforms trade hardware fidelity for scale and accessibility. Colosseum connects \glspl{sdr} through an \gls{fpga}-based \gls{rf} channel emulator~\cite{bonati2021colosseum}, while POWDER provides a city-scale \gls{ota} testbed~\cite{breen2020powder}. Both retain a real \gls{phy} and radio hardware, which require shared facilities and limit the number of simultaneous \glspl{ue}. ns-3~\cite{henderson2008ns3} and 5G-LENA~\cite{patriciello5glena} scale through simulated time and reimplemented schedulers. Sionna provides differentiable link-level models~\cite{hoydis2022sionna}, and NVIDIA Aerial implements a \gls{cuda}-accelerated \gls{phy}~\cite{nvidia_aerial}.

Closer to \macemu, 5G-EMANE proposes a multi-user proxy over nFAPI for LTE and NR \gls{oai}, providing a \gls{phy}-less Layer~2 proxy~\cite{ryu20225g}. Its 5G evaluation demonstrates up to 15 users. However, the authors report approximately 30~Mbit/s with 40\% packet loss in \gls{dl}. Moreover, its public implementation\footnote{\url{https://github.com/EpiSci/oai-lte-5g-multi-ue-proxy}} has seen no commits since January~2024.

\subsection{Predictive and Proactive \gls{urllc} Scheduling}
\label{sec:rw_urllc}

New \gls{ul} traffic can wait for a periodic \gls{sr}, a small grant, and a \gls{bsr} before the \gls{gnb} issues its data grant. The \gls{tdd} pattern adds slot-alignment delay, and commercial \gls{5g} standalone networks report median \gls{ul} round-trip times of $50$--$70$~ms~\cite{ghoshal2025first}. Configured grants remove the \gls{sr} wait by pre-allocating periodic \gls{ul} resources~\cite{ts38321,le2021urllc}. Their fixed period and phase serve strictly periodic traffic. Event-driven and multi-source arrivals can waste resources or miss reserved opportunities. Configured grants also require the application and \gls{rrc} configuration at the \gls{ue} to remain synchronized. Furthermore, a traffic period that is not aligned with the \gls{tdd} pattern drifts relative to the available \gls{ul} slots.

Learning-based schedulers generally act after demand becomes known. LEASCH uses deep reinforcement learning to allocate resource blocks at the \gls{mac} layer~\cite{altam2020leasch}. Other work jointly allocates resources to \gls{urllc} and \gls{embb} traffic~\cite{anand2020joint}, controls that split through an O-RAN xApp~\cite{sohaib2024drl}, or evaluates data-driven control at scale with ColO-RAN~\cite{polese2022colo}. Our scheduler addresses the preceding access delay. It learns arrivals online from in-\gls{ran} events and issues a grant before the \gls{ue} sends an \gls{sr}. The Layer~2 \gls{dapp} needs no offline dataset or application-to-\gls{rrc} synchronization. To our knowledge, prior work has not demonstrated online arrival prediction and proactive \gls{ul} grants in a production \gls{ran} stack.

\subsection{Frequency-Selective and Channel-Aware Scheduling}
\label{sec:rw_freqsel}

Frequency-selective scheduling first emerged in multi-user \gls{ofdm}. Wong \textit{et al.}~\cite{wong1999} minimize \gls{dl} power under per-user rate constraints, while Rhee and Cioffi~\cite{rheecioffi2000} maximize capacity through dynamic subchannel assignment. Later \gls{dl} work measured a gain of roughly two \gls{cqi} steps in \gls{lte} field trials~\cite{okvist2011} and used per-sub-band \gls{ri}, \gls{pmi}, and \gls{cqi} in the SARA scheduler for 5G-LENA~\cite{sara2025}.

The \gls{ul} adds a contiguity constraint because \mbox{SC-FDMA} (\mbox{DFT-S-OFDM}) assigns each \gls{ue} a contiguous set of \glspl{prb}. This restriction prevents the scheduler from assigning an arbitrary set of strong sub-bands to one \gls{ue}. Lee \textit{et al.}~\cite{lee2009proportional} prove that proportional-fair frequency-domain scheduling is NP-hard under this constraint. Consequently, prior \gls{ul} schedulers use heuristics such as candidate-tree search~\cite{calabrese2008fdps} or a coarse sub-band grid~\cite{krishnamoorthy2008fsss}, primarily evaluated through analysis and link- or system-level simulation.

Our allocator extends this \gls{ul} work in three ways. First, it consumes per-\gls{prb} \gls{srs} observations inside \gls{oai} and places each user's contiguous grant on a favorable sub-band. Second, it derives real-time policies with different restrictions on the exact contiguous-allocation problem. Finally, it uses offline analysis to shape these policies and integrated execution to select between them. To our knowledge, prior work has not combined these properties in a production \gls{ran} stack.

\section{Conclusion}
\label{sec:conc}

We presented \framework, an open \gls{mac}-layer framework for implementing scheduling dApps in the \gls{gnb}. Its staged pipeline exposes self-contained interfaces to \glspl{dapp}, while its software-emulation component, \macemu, executes the same policies with tens of \glspl{ue} at real-time slot pace while emulating realistic channel dynamics. We demonstrate these capabilities with two \gls{dapp}-based use cases that exercise different parts of the scheduler interface. First, proactive scheduling combines observer and candidate-request \glspl{dapp} through persistent context to implement closed-loop control across scheduling opportunities. In \gls{ota} experiments with sparse traffic, it reduces median \gls{rtt} by approximately 50\%.
Second, we develop two real-time frequency-selective schedulers, which we first compare offline through simulations. Although offline analysis favors the more flexible design, executing both in \framework (inside \macemu) shows that the simpler policy performs better within the complete scheduling loop. This use case shows how \framework supports algorithm design: offline and simulation-based analysis identifies potentially useful mechanisms, while execution in the complete scheduler determines which mechanisms affect end-to-end performance and should be retained.
We then evaluate the retained policy further and show that it improves \gls{ul} throughput under favorable channel conditions.
Together, these applications show how \framework supports policy design, controlled cell-scale evaluation, and \gls{ota} execution without reimplementation. The open-source release of \framework and \macemu supports reproducible AI-for-RAN research at the \gls{mac} layer.

\footnotesize
\bibliographystyle{IEEEtran}
\bibliography{refs}

\section*{Acknowledgments}

We thank Robert Schmidt from OpenAirInterface for his precious comments during the review of our open-source contributions.
\end{document}

%% file: acronyms.tex
\newacronym{isac}{ISAC}{Integrated Sensing and Communication}
\newacronym{tpc}{TPC}{Transmit Power Control}
\newacronym{re}{RE}{Resource Element}
\newacronym{ewma}{EWMA}{Exponentially Weighted Moving Average}
\newacronym{gpfb}{GPFB}{Generalized Proportional Fair Buffer}
\newacronym{drb}{DRB}{Data Radio Bearer}
\newacronym{hol}{D-HoL}{Head-of-Line Delay}
\newacronym{bcqi}{BCQI}{Best CQI}
\newacronym{ft}{FT}{Fair Throughput}
\newacronym{rr}{RR}{Round-Robin}
\newacronym{pf}{PF}{Proportional Fair}
\newacronym{x5g}{X5G}{}
\newacronym{ibs}{IBS}{Intent-Based Scheduling}
\newacronym{rapp}{rApp}{RAN Application}
\newacronym{xapp}{xApp}{eXtended Application}
\newacronym{dapp}{dApp}{Distributed Application}
\newacronym{ibn}{IBN}{Intent-Based Networking}
\newacronym{dlsch}{DLSCH}{Downlink Shared Channel}
\newacronym{ocr}{OCR}{Optical Character Recognition}
\newacronym{3gpp}{3GPP}{3rd Generation Partnership Project}
\newacronym{sriov}{SR-IOV}{Single Root I/O Virtualization}
\newacronym{vf}{VF}{Vitual Functions}
\newacronym{4g}{4G}{4th generation}
\newacronym{5g}{5G}{5th generation}
\newacronym{6g}{6G}{6th generation}
\newacronym{5gc}{5GC}{5G Core}
\newacronym{adc}{ADC}{Analog to Digital Converter}
\newacronym{aerpaw}{AERPAW}{Aerial Experimentation and Research Platform for Advanced Wireless}
\newacronym{ai}{AI}{Artificial Intelligence}
\newacronym{aimd}{AIMD}{Additive Increase Multiplicative Decrease}
\newacronym{am}{AM}{Acknowledged Mode}
\newacronym{amc}{AMC}{Adaptive Modulation and Coding}
\newacronym{amf}{AMF}{Access and Mobility Management Function}
\newacronym{aops}{AOPS}{Adaptive Order Prediction Scheduling}
\newacronym{api}{API}{Application Programming Interface}
\newacronym{apn}{APN}{Access Point Name}
\newacronym{ap}{AP}{Application Protocol}
\newacronym{aqm}{AQM}{Active Queue Management}
\newacronym{ausf}{AUSF}{Authentication Server Function}
\newacronym{avc}{AVC}{Advanced Video Coding}
\newacronym{awgn}{AWGN}{Additive White Gaussian Noise}
\newacronym{bbu}{BBU}{Base Band Unit}
\newacronym{bdp}{BDP}{Bandwidth-Delay Product}
\newacronym{ber}{BER}{Bit Error Rate}
\newacronym{bf}{BF}{Beamforming}
\newacronym{bler}{BLER}{Block Error Rate}
\newacronym{brr}{BRR}{Bayesian Ridge Regressor}
\newacronym{bs}{BS}{Base Station}
\newacronym{bsr}{BSR}{Buffer Status Report}
\newacronym{bss}{BSS}{Business Support System}
\newacronym{ca}{CA}{Carrier Aggregation}
\newacronym{caas}{CaaS}{Connectivity-as-a-Service}
\newacronym{cb}{CB}{Code Block}
\newacronym{cc}{CC}{Congestion Control}
\newacronym{ccid}{CCID}{Congestion Control ID}
\newacronym{cco}{CC}{Carrier Component}
\newacronym{cd}{CD}{Continuous Delivery}
\newacronym{cdd}{CDD}{Cyclic Delay Diversity}
\newacronym{cdf}{CDF}{Cumulative Distribution Function}
\newacronym{cg}{CG}{Configured Grant}
\newacronym{cdn}{CDN}{Content Distribution Network}
\newacronym{cli}{CLI}{Command-line Interface}
\newacronym{cn}{CN}{Core Network}
\newacronym{codel}{CoDel}{Controlled Delay Management}
\newacronym{comac}{COMAC}{Converged Multi-Access and Core}
\newacronym{cord}{CORD}{Central Office Re-architected as a Datacenter}
\newacronym{cornet}{CORNET}{COgnitive Radio NETwork}
\newacronym{cosmos}{COSMOS}{Cloud Enhanced Open Software Defined Mobile Wireless Testbed for City-Scale Deployment}
\newacronym{cots}{COTS}{Commercial Off-the-Shelf}
\newacronym{cp}{CP}{Control Plane}
\newacronym{cyp}{CP}{Cyclic Prefix}
\newacronym{up}{UP}{User Plane}
\newacronym{cpu}{CPU}{Central Processing Unit}
\newacronym{cqi}{CQI}{Channel Quality Information}
\newacronym{cr}{CR}{Cognitive Radio}
\newacronym{cran}{CRAN}{Cloud RAN}
\newacronym{crs}{CRS}{Cell Reference Signal}
\newacronym{csi}{CSI}{Channel State Information}
\newacronym{csirs}{CSI-RS}{Channel State Information - Reference Signal}
\newacronym{cu}{CU}{Central Unit}
\newacronym{cubb}{cuBB}{CUDA Baseband}
\newacronym{d2tcp}{D$^2$TCP}{Deadline-aware Data center TCP}
\newacronym{d3}{D$^3$}{Deadline-Driven Delivery}
\newacronym{dac}{DAC}{Digital to Analog Converter}
\newacronym{dag}{DAG}{Directed Acyclic Graph}
\newacronym{das}{DAS}{Distributed Antenna System}
\newacronym{dash}{DASH}{Dynamic Adaptive Streaming over HTTP}
\newacronym{dc}{DC}{Dual Connectivity}
\newacronym{dccp}{DCCP}{Datagram Congestion Control Protocol}
\newacronym{dce}{DCE}{Direct Code Execution}
\newacronym{dci}{DCI}{Downlink Control Information}
\newacronym{dctcp}{DCTCP}{Data Center TCP}
\newacronym{dl}{DL}{Downlink}
\newacronym{dmr}{DMR}{Deadline Miss Ratio}
\newacronym{dmrs}{DMRS}{DeModulation Reference Signal}
\newacronym{drlcc}{DRL-CC}{Deep Reinforcement Learning Congestion Control}
\newacronym{drs}{DRS}{Discovery Reference Signal}
\newacronym{dtx}{DTX}{Discontinuous Transmission}
\newacronym{du}{DU}{Distributed Unit}
\newacronym{e2e}{E2E}{end-to-end}
\newacronym{earfcn}{EARFCN}{E-UTRA Absolute Radio Frequency Channel Number}
\newacronym{ecaas}{ECaaS}{Edge-Cloud-as-a-Service}
\newacronym{ecn}{ECN}{Explicit Congestion Notification}
\newacronym{edf}{EDF}{Earliest Deadline First}
\newacronym{embb}{eMBB}{Enhanced Mobile Broadband}
\newacronym{empower}{EMPOWER}{EMpowering transatlantic PlatfOrms for advanced WirEless Research}
\newacronym{enb}{eNB}{evolved Node Base}
\newacronym{endc}{EN-DC}{E-UTRAN-\gls{nr} \gls{dc}}
\newacronym{epc}{EPC}{Evolved Packet Core}
\newacronym{eps}{EPS}{Evolved Packet System}
\newacronym{es}{ES}{Edge Server}
\newacronym{etsi}{ETSI}{European Telecommunications Standards Institute}
\newacronym[firstplural=Estimated Times of Arrival (ETAs)]{eta}{ETA}{Estimated Time of Arrival}
\newacronym{eutran}{E-UTRAN}{Evolved Universal Terrestrial Access Network}
\newacronym{faas}{FaaS}{Function-as-a-Service}
\newacronym{fapi}{FAPI}{Functional Application Platform Interface}
\newacronym{fdd}{FDD}{Frequency Division Duplexing}
\newacronym{fdm}{FDM}{Frequency Division Multiplexing}
\newacronym{fdma}{FDMA}{Frequency Division Multiple Access}
\newacronym{fed4fire}{FED4FIRE+}{Federation 4 Future Internet Research and Experimentation Plus}
\newacronym{fir}{FIR}{Finite Impulse Response}
\newacronym{fit}{FIT}{Future \acrlong{iot}}
\newacronym{fpga}{FPGA}{Field Programmable Gate Array}
\newacronym{fr2}{FR2}{Frequency Range 2}
\newacronym{fs}{FS}{Fast Switching}
\newacronym{fscc}{FSCC}{Flow Sharing Congestion Control}
\newacronym{ftp}{FTP}{File Transfer Protocol}
\newacronym{fw}{FW}{Flow Window}
\newacronym{ge}{GE}{Gaussian Elimination}
\newacronym{gnb}{gNB}{Next Generation Node Base}
\newacronym{gop}{GOP}{Group of Pictures}
\newacronym{gpr}{GPR}{Gaussian Process Regressor}
\newacronym{gpu}{GPU}{Graphics Processing Unit}
\newacronym{gtp}{GTP}{GPRS Tunneling Protocol}
\newacronym{gtpc}{GTP-C}{GPRS Tunnelling Protocol Control Plane}
\newacronym{gtpu}{GTP-U}{GPRS Tunnelling Protocol User Plane}
\newacronym{gtpv2c}{GTPv2-C}{\gls{gtp} v2 - Control}
\newacronym{gw}{GW}{Gateway}
\newacronym{harq}{HARQ}{Hybrid Automatic Repeat reQuest}
\newacronym{hetnet}{HetNet}{Heterogeneous Network}
\newacronym{hh}{HH}{Hard Handover}
\newacronym{hqf}{HQF}{Highest-quality-first}
\newacronym{hss}{HSS}{Home Subscription Server}
\newacronym{http}{HTTP}{HyperText Transfer Protocol}
\newacronym{ia}{IA}{Initial Access}
\newacronym{iab}{IAB}{Integrated Access and Backhaul}
\newacronym{ic}{IC}{Incident Command}
\newacronym{ietf}{IETF}{Internet Engineering Task Force}
\newacronym{imsi}{IMSI}{International Mobile Subscriber Identity}
\newacronym{imt}{IMT}{International Mobile Telecommunication}
\newacronym{iot}{IoT}{Internet of Things}
\newacronym{ip}{IP}{Internet Protocol}
\newacronym{itu}{ITU}{International Telecommunication Union}
\newacronym{kpi}{KPI}{Key Performance Indicator}
\newacronym{kpm}{KPM}{Key Performance Measurement}
\newacronym{kvm}{KVM}{Kernel-based Virtual Machine}
\newacronym{los}{LOS}{Line-of-Sight}
\newacronym{lsm}{LSM}{Link-to-System Mapping}
\newacronym{lstm}{LSTM}{Long Short Term Memory}
\newacronym{lte}{LTE}{Long Term Evolution}
\newacronym{lxc}{LXC}{Linux Container}
\newacronym{m2m}{M2M}{Machine to Machine}
\newacronym{mac}{MAC}{Medium Access Control}
\newacronym{manet}{MANET}{Mobile Ad Hoc Network}
\newacronym{mano}{MANO}{Management and Orchestration}
\newacronym{mc}{MC}{Multi-Connectivity}
\newacronym{mcc}{MCC}{Mobile Cloud Computing}
\newacronym{mchem}{MCHEM}{Massive Channel Emulator}
\newacronym{mcs}{MCS}{Modulation and Coding Scheme}
\newacronym{tda}{TDA}{Time-Domain Allocation}
\newacronym{mec}{MEC}{Multi-access Edge Computing}
\newacronym{mec2}{MEC}{Mobile Edge Cloud}
\newacronym{mfc}{MFC}{Mobile Fog Computing}
\newacronym{mgen}{MGEN}{Multi-Generator}
\newacronym{mi}{MI}{Mutual Information}
\newacronym{mib}{MIB}{Master Information Block}
\newacronym{miesm}{MIESM}{Mutual Information Based Effective SINR}
\newacronym{mimo}{MIMO}{Multiple Input, Multiple Output}
\newacronym{ml}{ML}{Machine Learning}
\newacronym{mmse}{MMSE}{Minimum Mean Square Error}
\newacronym{mlr}{MLR}{Maximum-local-rate}
\newacronym[plural=\gls{mme}s,firstplural=Mobility Management Entities (MMEs)]{mme}{MME}{Mobility Management Entity}
\newacronym{mmtc}{mMTC}{Massive Machine-Type Communications}
\newacronym{mmwave}{mmWave}{millimeter wave}
\newacronym{mpdccp}{MP-DCCP}{Multipath Datagram Congestion Control Protocol}
\newacronym{mptcp}{MPTCP}{Multipath TCP}
\newacronym{mr}{MR}{Maximum Rate}
\newacronym{mrdc}{MR-DC}{Multi \gls{rat} \gls{dc}}
\newacronym{mse}{MSE}{Mean Square Error}
\newacronym{mss}{MSS}{Maximum Segment Size}
\newacronym{mt}{MT}{Mobile Termination}
\newacronym{mtd}{MTD}{Machine-Type Device}
\newacronym{mtu}{MTU}{Maximum Transmission Unit}
\newacronym{mumimo}{MU-MIMO}{Multi-user \gls{mimo}}
\newacronym{mvno}{MVNO}{Mobile Virtual Network Operator}
\newacronym{nalu}{NALU}{Network Abstraction Layer Unit}
\newacronym{nas}{NAS}{Network Attached Storage}
\newacronym{nat}{NAT}{Network Address Translation}
\newacronym{nbiot}{NB-IoT}{Narrow Band IoT}
\newacronym{nfv}{NFV}{Network Function Virtualization}
\newacronym{nfvi}{NFVI}{Network Function Virtualization Infrastructure}
\newacronym{ni}{NI}{Network Interfaces}
\newacronym{llm}{LLM}{Large Language Model}
\newacronym{dpdk}{DPDK}{Data Plane Development Kit}
\newacronym{cicd}{CI/CD}{Continuous Integration and Continuous Delivery/Deployment}
\newacronym{nic}{NIC}{Network Interface Card}
\newacronym{nlos}{NLOS}{Non-Line-of-Sight}
\newacronym{now}{NOW}{Non Overlapping Window}
\newacronym{nsm}{NSM}{Network Service Mesh}
\newacronym[type=hidden]{nr}{NR}{New Radio}
\newacronym[type=hidden]{ota}{OTA}{Over-the-Air}
\newacronym[type=hidden]{l2}{L2}{Layer 2}
\newacronym[type=hidden]{lcid}{LCID}{Logical Channel Identifier}
\newacronym[type=hidden]{ldpc}{LDPC}{Low-Density Parity-Check}
\newacronym{nrf}{NRF}{Network Repository Function}
\newacronym{nsa}{NSA}{Non Stand Alone}
\newacronym{nse}{NSE}{Network Slicing Engine}
\newacronym{nssf}{NSSF}{Network Slice Selection Function}
\newacronym{o2i}{O2I}{Outdoor to Indoor}
\newacronym{oai}{OAI}{OpenAirInterface}
\newacronym{oal}{OAL}{OpenAirLink}
\newacronym{oaicn}{OAI-CN}{\gls{oai} \acrlong{cn}}
\newacronym{oairan}{OAI-RAN}{\acrlong{oai} \acrlong{ran}}
\newacronym{oam}{OAM}{Operations, Administration and Maintenance}
\newacronym{ofdm}{OFDM}{Orthogonal Frequency Division Multiplexing}
\newacronym{ofdma}{OFDMA}{Orthogonal Frequency Division Multiple Access}
\newacronym{olia}{OLIA}{Opportunistic Linked Increase Algorithm}
\newacronym{omec}{OMEC}{Open Mobile Evolved Core}
\newacronym{onap}{ONAP}{Open Network Automation Platform}
\newacronym{onf}{ONF}{Open Networking Foundation}
\newacronym{onos}{ONOS}{Open Networking Operating System}
\newacronym{oom}{OOM}{\gls{onap} Operations Manager}
\newacronym{opnfv}{OPNFV}{Open Platform for \gls{nfv}}
\newacronym{oran}{O-RAN}{Open RAN}
\newacronym{orbit}{ORBIT}{Open-Access Research Testbed for Next-Generation Wireless Networks}
\newacronym{os}{OS}{Operating System}
\newacronym{oss}{OSS}{Operations Support System}
\newacronym{pa}{PA}{Position-aware}
\newacronym{pase}{PASE}{Prioritization, Arbitration, and Self-adjusting Endpoints}
\newacronym{pawr}{PAWR}{Platforms for Advanced Wireless Research}
\newacronym{pbch}{PBCH}{Physical Broadcast Channel}
\newacronym{pcef}{PCEF}{Policy and Charging Enforcement Function}
\newacronym{pcfich}{PCFICH}{Physical Control Format Indicator Channel}
\newacronym{pcrf}{PCRF}{Policy and Charging Rules Function}
\newacronym{pdcch}{PDCCH}{Physical Downlink Control Channel}
\newacronym{pdcp}{PDCP}{Packet Data Convergence Protocol}
\newacronym{pdf}{PDF}{Probability Density Function}
\newacronym{pdsch}{PDSCH}{Physical Downlink Shared Channel}
\newacronym{pdu}{PDU}{Packet Data Unit}
\newacronym{sdu}{SDU}{Service Data Unit}
\newacronym{pgw}{PGW}{Packet Gateway}
\newacronym{phich}{PHICH}{Physical Hybrid ARQ Indicator Channel}
\newacronym{phy}{PHY}{Physical}
\newacronym{pmch}{PMCH}{Physical Multicast Channel}
\newacronym{pmi}{PMI}{Precoding Matrix Indicators}
\newacronym{powder}{POWDER}{Platform for Open Wireless Data-driven Experimental Research}
\newacronym{ppo}{PPO}{Proximal Policy Optimization}
\newacronym{ppp}{PPP}{Poisson Point Process}
\newacronym{prach}{PRACH}{Physical Random Access Channel}
\newacronym{prb}{PRB}{Physical Resource Block}
\newacronym{psnr}{PSNR}{Peak Signal to Noise Ratio}
\newacronym{pss}{PSS}{Primary Synchronization Signal}
\newacronym{pucch}{PUCCH}{Physical Uplink Control Channel}
\newacronym{pusch}{PUSCH}{Physical Uplink Shared Channel}
\newacronym{qam}{QAM}{Quadrature Amplitude Modulation}
\newacronym{qci}{QCI}{\gls{qos} Class Identifier}
\newacronym{qoe}{QoE}{Quality of Experience}
\newacronym{qos}{QoS}{Quality of Service}
\newacronym{quic}{QUIC}{Quick UDP Internet Connections}
\newacronym{rach}{RACH}{Random Access Channel}
\newacronym{ran}{RAN}{Radio Access Network}
\newacronym[firstplural=Radio Access Technologies (RATs)]{rat}{RAT}{Radio Access Technology}
\newacronym{rbg}{RBG}{Resource Block Group}
\newacronym{rcn}{RCN}{Research Coordination Network}
\newacronym{rc}{RC}{RAN Control}
\newacronym{rec}{REC}{Radio Edge Cloud}
\newacronym{red}{RED}{Random Early Detection}
\newacronym{renew}{RENEW}{Reconfigurable Eco-system for Next-generation End-to-end Wireless}
\newacronym{rf}{RF}{Radio Frequency}
\newacronym{rfc}{RFC}{Request for Comments}
\newacronym{rfr}{RFR}{Random Forest Regressor}
\newacronym{ric}{RIC}{RAN Intelligent Controller}
\newacronym{nrric}{Near-RT RIC}{Near-Real-Time \gls{ran} Intelligent Controller}
\newacronym{ri}{RI}{Rank Indicator}
\newacronym{rlc}{RLC}{Radio Link Control}
\newacronym{rlf}{RLF}{Radio Link Failure}
\newacronym{rlnc}{RLNC}{Random Linear Network Coding}
\newacronym{rmr}{RMR}{RIC Message Router}
\newacronym{rmse}{RMSE}{Root Mean Squared Error}
\newacronym{rnis}{RNIS}{Radio Network Information Service}
\newacronym{rrc}{RRC}{Radio Resource Control}
\newacronym{rrm}{RRM}{Radio Resource Management}
\newacronym{rru}{RRU}{Remote Radio Unit}
\newacronym{rs}{RS}{Remote Server}
\newacronym{rsrp}{RSRP}{Reference Signal Received Power}
\newacronym{rsrq}{RSRQ}{Reference Signal Received Quality}
\newacronym{rss}{RSS}{Received Signal Strength}
\newacronym{rssi}{RSSI}{Received Signal Strength Indicator}
\newacronym{rtt}{RTT}{Round Trip Time}
\newacronym{ru}{RU}{Radio Unit}
\newacronym{rw}{RW}{Receive Window}
\newacronym{rx}{RX}{Receiver}
\newacronym{s1ap}{S1AP}{S1 Application Protocol}
\newacronym{sa}{SA}{standalone}
\newacronym{sack}{SACK}{Selective Acknowledgment}
\newacronym{sap}{SAP}{Service Access Point}
\newacronym{sc2}{SC2}{Spectrum Collaboration Challenge}
\newacronym{scef}{SCEF}{Service Capability Exposure Function}
\newacronym{sch}{SCH}{Secondary Cell Handover}
\newacronym{scoot}{SCOOT}{Split Cycle Offset Optimization Technique}
\newacronym{sctp}{SCTP}{Stream Control Transmission Protocol}
\newacronym{sdap}{SDAP}{Service Data Adaptation Protocol}
\newacronym{sdk}{SDK}{Software Development Kit}
\newacronym{sdm}{SDM}{Space Division Multiplexing}
\newacronym{sdma}{SDMA}{Spatial Division Multiple Access}
\newacronym{sdl}{SDL}{Shared Data Layer}
\newacronym{sdn}{SDN}{Software-defined Networking}
\newacronym{sdr}{SDR}{Software-defined Radio}
\newacronym{seba}{SEBA}{SDN-Enabled Broadband Access}
\newacronym{sgsn}{SGSN}{Serving GPRS Support Node}
\newacronym{sgw}{SGW}{Service Gateway}
\newacronym{si}{SI}{Study Item}
\newacronym{sib}{SIB}{Secondary Information Block}
\newacronym{sinr}{SINR}{Signal to Interference plus Noise Ratio}
\newacronym{sip}{SIP}{Session Initiation Protocol}
\newacronym{siso}{SISO}{Single Input, Single Output}
\newacronym{sla}{SLA}{Service Level Agreement}
\newacronym{sm}{SM}{Service Model}
\newacronym{e2sm}{E2SM}{E2 Service Model}
\newacronym{e2ap}{E2AP}{E2 Application Protocol}
\newacronym{smf}{SMF}{Session Management Function}
\newacronym{smo}{SMO}{Service Management and Orchestration}
\newacronym{sms}{SMS}{Short Message Service}
\newacronym{smsgmsc}{SMS-GMSC}{\gls{sms}-Gateway}
\newacronym{snr}{SNR}{Signal-to-Noise-Ratio}
\newacronym{son}{SON}{Self-Organizing Network}
\newacronym{sptcp}{SPTCP}{Single Path TCP}
\newacronym{srb}{SRB}{Service Radio Bearer}
\newacronym{srn}{SRN}{Standard Radio Node}
\newacronym{srs}{SRS}{Sounding Reference Signal}
\newacronym{ss}{SS}{Synchronization Signal}
\newacronym{sss}{SSS}{Secondary Synchronization Signal}
\newacronym{st}{ST}{Spanning Tree}
\newacronym{svc}{SVC}{Scalable Video Coding}
\newacronym{tb}{TB}{Transport Block}
\newacronym{tcp}{TCP}{Transmission Control Protocol}
\newacronym{tdd}{TDD}{Time Division Duplexing}
\newacronym{tdm}{TDM}{Time Division Multiplexing}
\newacronym{tdma}{TDMA}{Time Division Multiple Access}
\newacronym{tfl}{TfL}{Transport for London}
\newacronym{tfrc}{TFRC}{TCP-Friendly Rate Control}
\newacronym{tft}{TFT}{Traffic Flow Template}
\newacronym{tgen}{TGEN}{Traffic Generator}
\newacronym{tip}{TIP}{Telecom Infra Project}
\newacronym{tm}{TM}{Transparent Mode}
\newacronym{to}{TO}{Telco Operator}
\newacronym{tr}{TR}{Technical Report}
\newacronym{trp}{TRP}{Transmitter Receiver Pair}
\newacronym{ts}{TS}{Technical Specification}
\newacronym{tti}{TTI}{Transmission Time Interval}
\newacronym{ttt}{TTT}{Time-to-Trigger}
\newacronym{tx}{TX}{Transmitter}
\newacronym{udm}{UDM}{Unified Data Management}
\newacronym{udp}{UDP}{User Datagram Protocol}
\newacronym{udr}{UDR}{Unified Data Repository}
\newacronym{ue}{UE}{User Equipment}
\newacronym{uhd}{UHD}{\gls{usrp} Hardware Driver}
\newacronym{ul}{UL}{Uplink}
\newacronym{um}{UM}{Unacknowledged Mode}
\newacronym{uml}{UML}{Unified Modeling Language}
\newacronym{upa}{UPA}{Uniform Planar Array}
\newacronym{upf}{UPF}{User Plane Function}
\newacronym{urllc}{URLLC}{Ultra Reliable and Low Latency Communications}
\newacronym{usa}{U.S.}{United States}
\newacronym{usim}{USIM}{Universal Subscriber Identity Module}
\newacronym{usrp}{USRP}{Universal Software Radio Peripheral}
\newacronym{utc}{UTC}{Urban Traffic Control}
\newacronym{vim}{VIM}{Virtualization Infrastructure Manager}
\newacronym{vm}{VM}{Virtual Machine}
\newacronym{vnf}{VNF}{Virtual Network Function}
\newacronym{volte}{VoLTE}{Voice over \gls{lte}}
\newacronym{voltha}{VOLTHA}{Virtual OLT HArdware Abstraction}
\newacronym{vr}{VR}{Virtual Reality}
\newacronym{vran}{vRAN}{Virtualized RAN}
\newacronym{vss}{VSS}{Video Streaming Server}
\newacronym{wbf}{WBF}{Wired Bias Function}
\newacronym{wf}{WF}{Waterfilling}
\newacronym{wg}{WG}{Working Group}
\newacronym{wlan}{WLAN}{Wireless Local Area Network}
\newacronym{osm}{OSM}{Open Source \gls{nfv} Management and Orchestration}
\newacronym{pnf}{PNF}{Physical Network Function}
\newacronym{drl}{DRL}{Deep Reinforcement Learning}
\newacronym{mtc}{MTC}{Machine-type Communications}
\newacronym{osc}{OSC}{O-RAN Software Community}
\newacronym{mns}{MnS}{Management Services}
\newacronym{ves}{VES}{\gls{vnf} Event Stream}
\newacronym{ei}{EI}{Enrichment Information}
\newacronym{fh}{FH}{Fronthaul}
\newacronym{fft}{FFT}{Fast Fourier Transform}
\newacronym{laa}{LAA}{Licensed-Assisted Access}
\newacronym{plfs}{PLFS}{Physical Layer Frequency Signals}
\newacronym{ptp}{PTP}{Precision Time Protocol}
\newacronym{ntp}{NTP}{Network Time Protocol}
\newacronym{cbrs}{CBRS}{Citizen Broadband Radio Service}
\newacronym{rnti}{RNTI}{Radio Network Temporary Identifier}
\newacronym{tbs}{TBS}{Transport Block Size}
\newacronym{nfd}{NFD}{Node Feature Discovery}
\newacronym{mcp}{MCP}{Model Context Protocol}
\newacronym{vpn}{VPN}{Virtual Private Network}
\newacronym{onr}{ONR}{Office of Naval Research}
\newacronym{afosr}{AFOSR}{Air Force Office of Scientific Research}
\newacronym{afrl}{AFRL}{Air Force Research Laboratory}
\newacronym{arl}{ARL}{Army Research Laboratory}

\newacronym{arc}{ARC}{Aerial Research Cloud}

\newacronym{ct}{CT}{Continuous Testing}
\newacronym{mno}{MNO}{Mobile Network Operator}
\newacronym{oci}{OCI}{Open Container Initiative}
\newacronym{macsec}{MACsec}{Media Access Control Security}
\newacronym{pt}{PT}{Plain Text}
\newacronym{cuda}{CUDA}{Compute Unified Device Architecture}
\newacronym{dsp}{DSP}{Digital Signal Processing}

\newacronym{cus}{CUS}{Control, User, Synchronization}
\newacronym{dpd}{DPD}{Digital Pre-Distorsion}
\newacronym{cfr}{CFR}{Crest Factor Reduction}
\newacronym{pci}{PCIe}{Peripheral Component Interconnect Express}
\newacronym{dpu}{DPU}{Data Processing Unit}
\newacronym{rfsoc}{RFSoC}{Radio Frequency System-on-Chip}
\newacronym{if}{IF}{Intermediate Frequency}
\newacronym{nyu}{NYU}{New York University}
\newacronym{gh}{GH}{Grace Hopper}
\newacronym{trl}{TRL}{Technology Readiness Level}
\newacronym{srfa}{SRFA}{Special Research Focus Area}
\newacronym{qsfp}{QSFP}{quad small form factor pluggable}
\newacronym{pse}{PSE}{Performance Specialized Engine}
\newacronym{cae}{CAE}{Cognitive Analysis Engine}
\newacronym{simd}{SIMD}{Single Instruction/Multiple Data}
\newacronym{rt}{RT}{Real-Time}
\newacronym{nrt}{NRT}{Non-Real-Time}
\newacronym{asm}{ASM}{Advanced Sleep Mode}
\newacronym{aoa}{AoA}{Angle of Arrival}
\newacronym{eaxcid}{eAxC\_ID}{extended Antenna-Carrier Identifier}
\newacronym{bwp}{BWP}{Bandwidth Part}
\newacronym{dfe}{DFE}{Digital Front-End}
\newacronym{spi}{SPI}{Serial Peripheral Interface}
\newacronym{gpio}{GPIO}{General Purpose Input/Output}
\newacronym{nco}{NCO}{Numerically Controlled Oscillator}
\newacronym{lo}{LO}{Local Oscillator}
\newacronym{lna}{LNA}{Low-Noise Amplifier}
\newacronym{pll}{PLL}{Phased-Locked Loop}
\newacronym{som}{SOM}{System-on-Module}
\newacronym{papr}{PAPR}{Peak-to-Average Power Ratio}
\newacronym{pcb}{PCB}{Printed Circuit Board}
\newacronym{gcpw}{GCPW}{Grounded Co-Planar Waveguide}
\newacronym{cnn}{CNN}{Convolutional Neural Network}
\newacronym{gmp}{GMP}{Generalized Memory Polynomial}
\newacronym{ngrg}{nGRG}{next Generation Research Group}
\newacronym{mrl}{MRL}{Manufacturing Readiness Level}
\newacronym{fr}{FR}{Frequency Range}

\newacronym{sbom}{SBOM}{Software Bill of Materials}
\newacronym{hbom}{HBOM}{Hardware Bill of Materials}
\newacronym{vex}{VEX}{Vulnerability Exploitability eXchange}
\newacronym{dos}{DoS}{Denial of Service}
\newacronym{sme}{SME}{Small-Medium Enterprise}

\newacronym{ulpi}{ULPI}{Uplink Performance Improvement}
\newacronym{oem}{OEM}{Original Equipment Manufacturer}
\newacronym{nsin}{NSIN}{National Security Innovation Network}
\newacronym{arpu}{ARPU}{Average Revenue per User}
\newacronym{opex}{OPEX}{operational expenses}
\newacronym{txb}{TXB}{Transmit Beam}
\newacronym{cve}{CVE}{Common Vulnerabilities and Exposure}
\newacronym{json}{JSON}{JavaScript Object Notation}
\newacronym{it}{IT}{Information Technology}
\newacronym{ci}{CI}{Continuous Integration}
\newacronym{nlp}{NLP}{Natural Language Processing}
\newacronym{nl}{NL}{Natural Language}
\newacronym{ulsch}{ULSCH}{Uplink Shared Agent}
\newacronym{phr}{PHR}{Power Headroom Report}
\newacronym{fwa}{FWA}{Fixed Wireless Access}
\newacronym{iq}{IQ}{In-Phase/Quadrature}
\newacronym{ram}{RAM}{Random Access Memory}
\newacronym{xr}{XR}{eXtended Reality}
\newacronym{aws}{AWS}{Amazon Web Services}
\newacronym{ra}{RA}{Resource Allocation}
\newacronym{pc}{PC}{Power Control}
\newacronym{a2a}{A2A}{Agent-To-Agent}
\newacronym{ssb}{SSB}{Synchronization Signal Block}
\newacronym{ta}{TA}{Timing Advance}
\newacronym{sr}{SR}{Scheduling Request}
\newacronym{arfcn}{ARFCN}{Absolute Radio Frequency Channel Number}
\newacronym{prg}{PRG}{Precoding Resource Block Group}
\newacronym{nn}{NN}{Neural Network}
\newacronym{ilp}{ILP}{Integer Linear Program}
\newacronym{eesm}{EESM}{Exponential Effective SINR Mapping}
\newacronym{umi}{UMi}{Urban Micro-cell}
\newacronym{uma}{UMa}{Urban Macro-cell}
\newacronym{scs}{SCS}{Subcarrier Spacing}
\newacronym{khz}{kHz}{kilohertz}
\newacronym{mhz}{MHz}{megahertz}
\newacronym{cce}{CCE}{Control Channel Element}
\newacronym{uci}{UCI}{Uplink Control Information}
\newacronym{rfsim}{RFsim}{RF Simulator}
\newacronym{dp}{DP}{Dynamic Program}

%% file: figures/macemu_floor_characterization.tex
% This file was created with tikzplotlib v0.10.1.post13.
\begin{tikzpicture}

\definecolor{darkgrey176}{RGB}{176,176,176}
\definecolor{darkslategrey377976}{RGB}{37,79,76}
\definecolor{lightgrey209}{RGB}{209,209,209}
\definecolor{mediumaquamarine131183179}{RGB}{131,183,179}
\definecolor{seagreen54111106}{RGB}{54,111,106}
\definecolor{sienna1807554}{RGB}{180,75,54}

\begin{groupplot}[group style={group size=2 by 1, horizontal sep=0.25cm}, title style={at={(axis description cs:0.5,-0.36)}, anchor=north}]
\nextgroupplot[
height=0.39\linewidth,
legend cell align={left},
legend style={
  fill opacity=0.8,
  draw opacity=1,
  text opacity=1,
  at={(1.05,1.18)},
  anchor=south,
  draw=none
},
log basis y={10},
tick align=inside,
tick pos=left,
title={(a) MCS},
width=0.55\linewidth,
x grid style={darkgrey176},
xmin=-1.074, xmax=15.074,
xtick style={color=black},
y grid style={lightgrey209},
ylabel={Error-Floor},
ymajorgrids,
ymin=1e-06, ymax=0.0015,
ymode=log,
ytick style={color=black},
ytick={1e-07,1e-06,1e-05,0.0001,0.001,0.01,0.1},
yticklabels={
  \(\displaystyle {10^{-7}}\),
  \(\displaystyle {10^{-6}}\),
  \(\displaystyle {10^{-5}}\),
  \(\displaystyle {10^{-4}}\),
  \(\displaystyle {10^{-3}}\),
  \(\displaystyle {10^{-2}}\),
  \(\displaystyle {10^{-1}}\)
}
]
\draw[draw=seagreen54111106,fill=mediumaquamarine131183179,opacity=0.82,line width=0.26pt,postaction={pattern=north east lines, pattern color=seagreen54111106, fill opacity=0.82}] (axis cs:-0.34,1e-06) rectangle (axis cs:0.34,1.1198098562864e-05);
\draw[draw=seagreen54111106,fill=mediumaquamarine131183179,opacity=0.82,line width=0.26pt,postaction={pattern=north east lines, pattern color=seagreen54111106, fill opacity=0.82}] (axis cs:0.66,1e-06) rectangle (axis cs:1.34,2.5559069138702e-05);
\draw[draw=seagreen54111106,fill=mediumaquamarine131183179,opacity=0.82,line width=0.26pt,postaction={pattern=north east lines, pattern color=seagreen54111106, fill opacity=0.82}] (axis cs:1.66,1e-06) rectangle (axis cs:2.34,2.20772482917729e-05);
\draw[draw=seagreen54111106,fill=mediumaquamarine131183179,opacity=0.82,line width=0.26pt,postaction={pattern=north east lines, pattern color=seagreen54111106, fill opacity=0.82}] (axis cs:2.66,1e-06) rectangle (axis cs:3.34,3.26150123937047e-05);
\draw[draw=seagreen54111106,fill=mediumaquamarine131183179,opacity=0.82,line width=0.26pt,postaction={pattern=north east lines, pattern color=seagreen54111106, fill opacity=0.82}] (axis cs:4.66,1e-06) rectangle (axis cs:5.34,8.27786017314524e-06);
\draw[draw=seagreen54111106,fill=mediumaquamarine131183179,opacity=0.82,line width=0.26pt,postaction={pattern=north east lines, pattern color=seagreen54111106, fill opacity=0.82}] (axis cs:5.66,1e-06) rectangle (axis cs:6.34,6.74149728654734e-06);
\draw[draw=seagreen54111106,fill=mediumaquamarine131183179,opacity=0.82,line width=0.26pt,postaction={pattern=north east lines, pattern color=seagreen54111106, fill opacity=0.82}] (axis cs:6.66,1e-06) rectangle (axis cs:7.34,6.0688449774239e-06);
\draw[draw=seagreen54111106,fill=mediumaquamarine131183179,opacity=0.82,line width=0.26pt,postaction={pattern=north east lines, pattern color=seagreen54111106, fill opacity=0.82}] (axis cs:7.66,1e-06) rectangle (axis cs:8.34,1.09882864866053e-05);
\draw[draw=seagreen54111106,fill=mediumaquamarine131183179,opacity=0.82,line width=0.26pt,postaction={pattern=north east lines, pattern color=seagreen54111106, fill opacity=0.82}] (axis cs:9.66,1e-06) rectangle (axis cs:10.34,9.60066052544415e-06);
\draw[draw=seagreen54111106,fill=mediumaquamarine131183179,opacity=0.82,line width=0.26pt,postaction={pattern=north east lines, pattern color=seagreen54111106, fill opacity=0.82}] (axis cs:10.66,1e-06) rectangle (axis cs:11.34,1.11652314692049e-05);
\draw[draw=seagreen54111106,fill=mediumaquamarine131183179,opacity=0.82,line width=0.26pt,postaction={pattern=north east lines, pattern color=seagreen54111106, fill opacity=0.82}] (axis cs:11.66,1e-06) rectangle (axis cs:12.34,3.41458334400391e-06);
\draw[draw=seagreen54111106,fill=mediumaquamarine131183179,opacity=0.82,line width=0.26pt,postaction={pattern=north east lines, pattern color=seagreen54111106, fill opacity=0.82}] (axis cs:12.66,1e-06) rectangle (axis cs:13.34,1.43869159631465e-05);
\draw[draw=seagreen54111106,fill=mediumaquamarine131183179,opacity=0.82,line width=0.26pt,postaction={pattern=north east lines, pattern color=seagreen54111106, fill opacity=0.82}] (axis cs:13.66,1e-06) rectangle (axis cs:14.34,1.93160730423505e-05);
\path [draw=darkslategrey377976, line width=0.32pt]
(axis cs:0,5.67429741259944e-06)
--(axis cs:0,2.20990772965002e-05);

\path [draw=darkslategrey377976, line width=0.32pt]
(axis cs:1,1.61679675469025e-05)
--(axis cs:1,4.04047354670856e-05);

\path [draw=darkslategrey377976, line width=0.32pt]
(axis cs:2,1.11869973503608e-05)
--(axis cs:2,4.35684133737017e-05);

\path [draw=darkslategrey377976, line width=0.32pt]
(axis cs:3,2.15394430133969e-05)
--(axis cs:3,4.93853520091185e-05);

\path [draw=darkslategrey377976, line width=0.32pt]
(axis cs:5,3.79377112322276e-06)
--(axis cs:5,1.8061871327526e-05);

\path [draw=darkslategrey377976, line width=0.32pt]
(axis cs:6,2.62159727007826e-06)
--(axis cs:6,1.7335802082156e-05);

\path [draw=darkslategrey377976, line width=0.32pt]
(axis cs:7,2.36001957990384e-06)
--(axis cs:7,1.56060843754887e-05);

\path [draw=darkslategrey377976, line width=0.32pt]
(axis cs:8,5.03597093090415e-06)
--(axis cs:8,2.39758314928551e-05);

\path [draw=darkslategrey377976, line width=0.32pt]
(axis cs:10,4.40001595891359e-06)
--(axis cs:10,2.09481321897656e-05);

\path [draw=darkslategrey377976, line width=0.32pt]
(axis cs:11,5.65764297712512e-06)
--(axis cs:11,2.20342156416742e-05);

\path [draw=darkslategrey377976, line width=0.32pt]
(axis cs:12,9.36382812627456e-07)
--(axis cs:12,1.24514277554857e-05);

\path [draw=darkslategrey377976, line width=0.32pt]
(axis cs:13,7.81488582389564e-06)
--(axis cs:13,2.64856341640228e-05);

\path [draw=darkslategrey377976, line width=0.32pt]
(axis cs:14,1.15065898991835e-05)
--(axis cs:14,3.24256537729299e-05);

\addplot [semithick, darkslategrey377976, mark=-, mark size=1.9, mark options={solid}, only marks, forget plot]
table {%
0 5.67429741259944e-06
1 1.61679675469025e-05
2 1.11869973503608e-05
3 2.15394430133969e-05
5 3.79377112322276e-06
6 2.62159727007826e-06
7 2.36001957990384e-06
8 5.03597093090415e-06
10 4.40001595891359e-06
11 5.65764297712512e-06
12 9.36382812627456e-07
13 7.81488582389564e-06
14 1.15065898991835e-05
};
\addplot [semithick, darkslategrey377976, mark=-, mark size=1.9, mark options={solid}, only marks, forget plot]
table {%
0 2.20990772965002e-05
1 4.04047354670856e-05
2 4.35684133737017e-05
3 4.93853520091185e-05
5 1.8061871327526e-05
6 1.7335802082156e-05
7 1.56060843754887e-05
8 2.39758314928551e-05
10 2.09481321897656e-05
11 2.20342156416742e-05
12 1.24514277554857e-05
13 2.64856341640228e-05
14 3.24256537729299e-05
};
\addplot [line width=0.42pt, sienna1807554, dashed]
table {%
-1.074 1.21285957365477e-05
15.074 1.21285957365477e-05
};
\addlegendentry{Average error-floor}

\nextgroupplot[
height=0.39\linewidth,
log basis y={10},
tick align=inside,
title={(b) Allocated PRBs},
width=0.55\linewidth,
x grid style={darkgrey176},
xmin=-0.774, xmax=8.774,
xtick pos=left,
xtick style={color=black},
xtick={0,1,2,3,4,5,6,7,8},
xticklabels={20,,40,,60,,80,,106},
y grid style={lightgrey209},
ymajorgrids,
ymajorticks=false,
yminorticks=false,
ymin=1e-06, ymax=0.0015,
ymode=log,
ytick style={color=black},
yticklabels={}
]
\draw[draw=seagreen54111106,fill=mediumaquamarine131183179,opacity=0.82,line width=0.26pt,postaction={pattern=north east lines, pattern color=seagreen54111106, fill opacity=0.82}] (axis cs:-0.34,1e-06) rectangle (axis cs:0.34,4.14066673015689e-06);
\draw[draw=seagreen54111106,fill=mediumaquamarine131183179,opacity=0.82,line width=0.26pt,postaction={pattern=north east lines, pattern color=seagreen54111106, fill opacity=0.82}] (axis cs:1.66,1e-06) rectangle (axis cs:2.34,1.11912372612245e-05);
\draw[draw=seagreen54111106,fill=mediumaquamarine131183179,opacity=0.82,line width=0.26pt,postaction={pattern=north east lines, pattern color=seagreen54111106, fill opacity=0.82}] (axis cs:2.66,1e-06) rectangle (axis cs:3.34,1.65602669515033e-05);
\draw[draw=seagreen54111106,fill=mediumaquamarine131183179,opacity=0.82,line width=0.26pt,postaction={pattern=north east lines, pattern color=seagreen54111106, fill opacity=0.82}] (axis cs:3.66,1e-06) rectangle (axis cs:4.34,1.93237515131188e-05);
\draw[draw=seagreen54111106,fill=mediumaquamarine131183179,opacity=0.82,line width=0.26pt,postaction={pattern=north east lines, pattern color=seagreen54111106, fill opacity=0.82}] (axis cs:4.66,1e-06) rectangle (axis cs:5.34,4.13948119882135e-06);
\draw[draw=seagreen54111106,fill=mediumaquamarine131183179,opacity=0.82,line width=0.26pt,postaction={pattern=north east lines, pattern color=seagreen54111106, fill opacity=0.82}] (axis cs:5.66,1e-06) rectangle (axis cs:6.34,2.62228472767573e-05);
\draw[draw=seagreen54111106,fill=mediumaquamarine131183179,opacity=0.82,line width=0.26pt,postaction={pattern=north east lines, pattern color=seagreen54111106, fill opacity=0.82}] (axis cs:6.66,1e-06) rectangle (axis cs:7.34,1.93144741289517e-05);
\draw[draw=seagreen54111106,fill=mediumaquamarine131183179,opacity=0.82,line width=0.26pt,postaction={pattern=north east lines, pattern color=seagreen54111106, fill opacity=0.82}] (axis cs:7.66,1e-06) rectangle (axis cs:8.34,1.10329761867e-05);
\path [draw=darkslategrey377976, line width=0.32pt]
(axis cs:0,1.40817584527727e-06)
--(axis cs:0,1.21753473618309e-05);

\path [draw=darkslategrey377976, line width=0.32pt]
(axis cs:2,5.67082064927184e-06)
--(axis cs:2,2.208553684246e-05);

\path [draw=darkslategrey377976, line width=0.32pt]
(axis cs:3,9.4734508767102e-06)
--(axis cs:3,2.8948372793324e-05);

\path [draw=darkslategrey377976, line width=0.32pt]
(axis cs:4,1.15111639796232e-05)
--(axis cs:4,3.24385434212412e-05);

\path [draw=darkslategrey377976, line width=0.32pt]
(axis cs:5,1.89713680163503e-06)
--(axis cs:5,9.03216845818358e-06);

\path [draw=darkslategrey377976, line width=0.32pt]
(axis cs:6,1.67883365685355e-05)
--(axis cs:6,4.09590354987433e-05);

\path [draw=darkslategrey377976, line width=0.32pt]
(axis cs:7,1.15056374232658e-05)
--(axis cs:7,3.24229697187615e-05);

\path [draw=darkslategrey377976, line width=0.32pt]
(axis cs:8,5.5906264945872e-06)
--(axis cs:8,2.17732160690661e-05);

\addplot [semithick, darkslategrey377976, mark=-, mark size=1.9, mark options={solid}, only marks]
table {%
0 1.40817584527727e-06
2 5.67082064927184e-06
3 9.4734508767102e-06
4 1.15111639796232e-05
5 1.89713680163503e-06
6 1.67883365685355e-05
7 1.15056374232658e-05
8 5.5906264945872e-06
};
\addplot [semithick, darkslategrey377976, mark=-, mark size=1.9, mark options={solid}, only marks]
table {%
0 1.21753473618309e-05
2 2.208553684246e-05
3 2.8948372793324e-05
4 3.24385434212412e-05
5 9.03216845818358e-06
6 4.09590354987433e-05
7 3.24229697187615e-05
8 2.17732160690661e-05
};
\addplot [line width=0.42pt, sienna1807554, dashed]
table {%
-0.773999999999997 1.16069809910815e-05
8.774 1.16069809910815e-05
};
\end{groupplot}

\end{tikzpicture}

%% file: figures/olla_mcs.tex
% This file was created with tikzplotlib v0.10.1.post13.
\begin{tikzpicture}

\definecolor{crimson2143940}{RGB}{214,39,40}
\definecolor{darkgrey176}{RGB}{176,176,176}
\definecolor{dimgrey102}{RGB}{102,102,102}
\definecolor{gainsboro216}{RGB}{216,216,216}
\definecolor{steelblue31119180}{RGB}{31,119,180}

\begin{axis}[
height=0.62\linewidth,
tick align=inside,
tick pos=left,
width=\linewidth,
x grid style={darkgrey176},
xlabel={Channel attenuation (dB)},
xmin=-0.8, xmax=13.6,
xtick style={color=black},
xtick={0,1,2,3,4.9,5.9,6.9,7.9,9.8,10.8,11.8,12.8},
xticklabels={16,22,28,34,16,22,28,34,16,22,28,34},
y grid style={darkgrey176},
ylabel={MCS (initial TX)},
ymajorgrids,
ymin=-0.8, ymax=24.5,
ytick style={color=black},
legend to name=ollaLegend,
legend columns=2,
legend style={
  draw=darkgrey176,
  fill opacity=0.8,
  text opacity=1,
  font=\scriptsize,
  /tikz/every even column/.append style={column sep=0.5cm}
},
]
\addlegendimage{area legend, fill=steelblue31119180, fill opacity=0.75}
\addlegendentry{\gls{oal} (hardware)}
\addlegendimage{area legend, fill=crimson2143940, fill opacity=0.55}
\addlegendentry{\macemu}
\path [fill=steelblue31119180, fill opacity=0.75, line width=0pt]
(axis cs:-0.00175941949420568,0)
--(axis cs:0,0)
--(axis cs:0,0.5)
--(axis cs:0,0.5)
--(axis cs:0,1.5)
--(axis cs:0,1.5)
--(axis cs:0,2.5)
--(axis cs:0,2.5)
--(axis cs:0,3.5)
--(axis cs:0,3.5)
--(axis cs:0,4.5)
--(axis cs:0,4.5)
--(axis cs:0,5.5)
--(axis cs:0,5.5)
--(axis cs:0,6.5)
--(axis cs:0,6.5)
--(axis cs:0,7.5)
--(axis cs:0,7.5)
--(axis cs:0,8.5)
--(axis cs:0,8.5)
--(axis cs:0,9.5)
--(axis cs:0,9.5)
--(axis cs:0,10.5)
--(axis cs:0,10.5)
--(axis cs:0,11.5)
--(axis cs:0,11.5)
--(axis cs:0,12.5)
--(axis cs:0,12.5)
--(axis cs:0,13.5)
--(axis cs:0,13.5)
--(axis cs:0,14.5)
--(axis cs:0,14.5)
--(axis cs:0,15.5)
--(axis cs:0,15.5)
--(axis cs:0,16.5)
--(axis cs:0,16.5)
--(axis cs:0,17.5)
--(axis cs:0,17.5)
--(axis cs:0,18.5)
--(axis cs:0,18.5)
--(axis cs:0,19.5)
--(axis cs:0,19.5)
--(axis cs:0,20.5)
--(axis cs:0,20.5)
--(axis cs:0,21.5)
--(axis cs:0,21.5)
--(axis cs:0,22)
--(axis cs:-0.0016341047154161,22)
--(axis cs:-0.0016341047154161,22)
--(axis cs:-0.0016341047154161,21.5)
--(axis cs:-0.0157846495363353,21.5)
--(axis cs:-0.0157846495363353,20.5)
--(axis cs:-0.057875377436179,20.5)
--(axis cs:-0.057875377436179,19.5)
--(axis cs:-0.123013999451002,19.5)
--(axis cs:-0.123013999451002,18.5)
--(axis cs:-0.143465371349461,18.5)
--(axis cs:-0.143465371349461,17.5)
--(axis cs:-0.203771855494158,17.5)
--(axis cs:-0.203771855494158,16.5)
--(axis cs:-0.335242096217881,16.5)
--(axis cs:-0.335242096217881,15.5)
--(axis cs:-0.42,15.5)
--(axis cs:-0.42,14.5)
--(axis cs:-0.266263829380945,14.5)
--(axis cs:-0.266263829380945,13.5)
--(axis cs:-0.166793970568929,13.5)
--(axis cs:-0.166793970568929,12.5)
--(axis cs:-0.127690746995429,12.5)
--(axis cs:-0.127690746995429,11.5)
--(axis cs:-0.0997756268722625,11.5)
--(axis cs:-0.0997756268722625,10.5)
--(axis cs:-0.0421007530821468,10.5)
--(axis cs:-0.0421007530821468,9.5)
--(axis cs:-0.0120051558080416,9.5)
--(axis cs:-0.0120051558080416,8.5)
--(axis cs:-0.0179751518695772,8.5)
--(axis cs:-0.0179751518695772,7.5)
--(axis cs:-0.00240103116160833,7.5)
--(axis cs:-0.00240103116160833,6.5)
--(axis cs:-0.00241606893506308,6.5)
--(axis cs:-0.00241606893506308,5.5)
--(axis cs:-0.00219551492439342,5.5)
--(axis cs:-0.00219551492439342,4.5)
--(axis cs:-0.00180954540572151,4.5)
--(axis cs:-0.00180954540572151,3.5)
--(axis cs:-0.00181957058802468,3.5)
--(axis cs:-0.00181957058802468,2.5)
--(axis cs:-0.0018145579968731,2.5)
--(axis cs:-0.0018145579968731,1.5)
--(axis cs:-0.00180954540572151,1.5)
--(axis cs:-0.00180954540572151,0.5)
--(axis cs:-0.00175941949420568,0.5)
--(axis cs:-0.00175941949420568,0)
--cycle;

\addplot [very thin, dimgrey102]
table {%
0 -0.5
0 22.5
};
\path [fill=crimson2143940, fill opacity=0.55, line width=0pt]
(axis cs:0.01831652547676,5)
--(axis cs:0,5)
--(axis cs:0,5.5)
--(axis cs:0,5.5)
--(axis cs:0,6.5)
--(axis cs:0,6.5)
--(axis cs:0,7.5)
--(axis cs:0,7.5)
--(axis cs:0,8.5)
--(axis cs:0,8.5)
--(axis cs:0,9.5)
--(axis cs:0,9.5)
--(axis cs:0,10.5)
--(axis cs:0,10.5)
--(axis cs:0,11.5)
--(axis cs:0,11.5)
--(axis cs:0,12.5)
--(axis cs:0,12.5)
--(axis cs:0,13.5)
--(axis cs:0,13.5)
--(axis cs:0,14.5)
--(axis cs:0,14.5)
--(axis cs:0,15.5)
--(axis cs:0,15.5)
--(axis cs:0,16.5)
--(axis cs:0,16.5)
--(axis cs:0,17.5)
--(axis cs:0,17.5)
--(axis cs:0,18.5)
--(axis cs:0,18.5)
--(axis cs:0,19.5)
--(axis cs:0,19.5)
--(axis cs:0,20.5)
--(axis cs:0,20.5)
--(axis cs:0,21.5)
--(axis cs:0,21.5)
--(axis cs:0,22.5)
--(axis cs:0,22.5)
--(axis cs:0,23)
--(axis cs:0.000243949284929101,23)
--(axis cs:0.000243949284929101,23)
--(axis cs:0.000243949284929101,22.5)
--(axis cs:0.00485188022247878,22.5)
--(axis cs:0.00485188022247878,21.5)
--(axis cs:0.0136645481405425,21.5)
--(axis cs:0.0136645481405425,20.5)
--(axis cs:0.0466993471413583,20.5)
--(axis cs:0.0466993471413583,19.5)
--(axis cs:0.125247628704016,19.5)
--(axis cs:0.125247628704016,18.5)
--(axis cs:0.122757313087031,18.5)
--(axis cs:0.122757313087031,17.5)
--(axis cs:0.250613843865983,17.5)
--(axis cs:0.250613843865983,16.5)
--(axis cs:0.300511637187523,16.5)
--(axis cs:0.300511637187523,15.5)
--(axis cs:0.377897771462254,15.5)
--(axis cs:0.377897771462254,14.5)
--(axis cs:0.220106630511794,14.5)
--(axis cs:0.220106630511794,13.5)
--(axis cs:0.205252829607222,13.5)
--(axis cs:0.205252829607222,12.5)
--(axis cs:0.148487186278025,12.5)
--(axis cs:0.148487186278025,11.5)
--(axis cs:0.135595144206425,11.5)
--(axis cs:0.135595144206425,10.5)
--(axis cs:0.0286945346397855,10.5)
--(axis cs:0.0286945346397855,9.5)
--(axis cs:0.0199835122571088,9.5)
--(axis cs:0.0199835122571088,8.5)
--(axis cs:0.0183402427683503,8.5)
--(axis cs:0.0183402427683503,7.5)
--(axis cs:0.00398450498717531,7.5)
--(axis cs:0.00398450498717531,6.5)
--(axis cs:0.00816891286061197,6.5)
--(axis cs:0.00816891286061197,5.5)
--(axis cs:0.01831652547676,5.5)
--(axis cs:0.01831652547676,5)
--cycle;

\addplot [very thin, dimgrey102]
table {%
0 4.5
0 23.5
};
\path [fill=steelblue31119180, fill opacity=0.75, line width=0pt]
(axis cs:0.997487567522578,0)
--(axis cs:1,0)
--(axis cs:1,0.5)
--(axis cs:1,0.5)
--(axis cs:1,1.5)
--(axis cs:1,1.5)
--(axis cs:1,2.5)
--(axis cs:1,2.5)
--(axis cs:1,3.5)
--(axis cs:1,3.5)
--(axis cs:1,4.5)
--(axis cs:1,4.5)
--(axis cs:1,5.5)
--(axis cs:1,5.5)
--(axis cs:1,6.5)
--(axis cs:1,6.5)
--(axis cs:1,7.5)
--(axis cs:1,7.5)
--(axis cs:1,8.5)
--(axis cs:1,8.5)
--(axis cs:1,9.5)
--(axis cs:1,9.5)
--(axis cs:1,10.5)
--(axis cs:1,10.5)
--(axis cs:1,11.5)
--(axis cs:1,11.5)
--(axis cs:1,12.5)
--(axis cs:1,12.5)
--(axis cs:1,13.5)
--(axis cs:1,13.5)
--(axis cs:1,14.5)
--(axis cs:1,14.5)
--(axis cs:1,15)
--(axis cs:0.99029655843764,15)
--(axis cs:0.99029655843764,15)
--(axis cs:0.99029655843764,14.5)
--(axis cs:0.963492366787019,14.5)
--(axis cs:0.963492366787019,13.5)
--(axis cs:0.912234323545302,13.5)
--(axis cs:0.912234323545302,12.5)
--(axis cs:0.828675682118779,12.5)
--(axis cs:0.828675682118779,11.5)
--(axis cs:0.743105621142883,11.5)
--(axis cs:0.743105621142883,10.5)
--(axis cs:0.626273827024167,10.5)
--(axis cs:0.626273827024167,9.5)
--(axis cs:0.60968882553831,9.5)
--(axis cs:0.60968882553831,8.5)
--(axis cs:0.681458927481327,8.5)
--(axis cs:0.681458927481327,7.5)
--(axis cs:0.798519124552537,7.5)
--(axis cs:0.798519124552537,6.5)
--(axis cs:0.869861891939276,6.5)
--(axis cs:0.869861891939276,5.5)
--(axis cs:0.887574172513243,5.5)
--(axis cs:0.887574172513243,4.5)
--(axis cs:0.956500289305865,4.5)
--(axis cs:0.956500289305865,3.5)
--(axis cs:0.962696640371882,3.5)
--(axis cs:0.962696640371882,2.5)
--(axis cs:0.979598458856358,2.5)
--(axis cs:0.979598458856358,1.5)
--(axis cs:0.987563090844902,1.5)
--(axis cs:0.987563090844902,0.5)
--(axis cs:0.997487567522578,0.5)
--(axis cs:0.997487567522578,0)
--cycle;

\addplot [very thin, dimgrey102]
table {%
1 -0.5
1 15.5
};
\path [fill=crimson2143940, fill opacity=0.55, line width=0pt]
(axis cs:1.0145123286243,1)
--(axis cs:1,1)
--(axis cs:1,1.5)
--(axis cs:1,1.5)
--(axis cs:1,2.5)
--(axis cs:1,2.5)
--(axis cs:1,3.5)
--(axis cs:1,3.5)
--(axis cs:1,4.5)
--(axis cs:1,4.5)
--(axis cs:1,5.5)
--(axis cs:1,5.5)
--(axis cs:1,6.5)
--(axis cs:1,6.5)
--(axis cs:1,7.5)
--(axis cs:1,7.5)
--(axis cs:1,8.5)
--(axis cs:1,8.5)
--(axis cs:1,9.5)
--(axis cs:1,9.5)
--(axis cs:1,10.5)
--(axis cs:1,10.5)
--(axis cs:1,11.5)
--(axis cs:1,11.5)
--(axis cs:1,12.5)
--(axis cs:1,12.5)
--(axis cs:1,13.5)
--(axis cs:1,13.5)
--(axis cs:1,14.5)
--(axis cs:1,14.5)
--(axis cs:1,15.5)
--(axis cs:1,15.5)
--(axis cs:1,16.5)
--(axis cs:1,16.5)
--(axis cs:1,17)
--(axis cs:1.0003904498775,17)
--(axis cs:1.0003904498775,17)
--(axis cs:1.0003904498775,16.5)
--(axis cs:1.00127352343221,16.5)
--(axis cs:1.00127352343221,15.5)
--(axis cs:1.00340092790492,15.5)
--(axis cs:1.00340092790492,14.5)
--(axis cs:1.04437627065631,14.5)
--(axis cs:1.04437627065631,13.5)
--(axis cs:1.11187666162748,13.5)
--(axis cs:1.11187666162748,12.5)
--(axis cs:1.12392952092999,12.5)
--(axis cs:1.12392952092999,11.5)
--(axis cs:1.3870015117552,11.5)
--(axis cs:1.3870015117552,10.5)
--(axis cs:1.42,10.5)
--(axis cs:1.42,9.5)
--(axis cs:1.27436584475838,9.5)
--(axis cs:1.27436584475838,8.5)
--(axis cs:1.21389720064641,8.5)
--(axis cs:1.21389720064641,7.5)
--(axis cs:1.20966793515091,7.5)
--(axis cs:1.20966793515091,6.5)
--(axis cs:1.11494552468331,6.5)
--(axis cs:1.11494552468331,5.5)
--(axis cs:1.16284679142991,5.5)
--(axis cs:1.16284679142991,4.5)
--(axis cs:1.04355888025856,4.5)
--(axis cs:1.04355888025856,3.5)
--(axis cs:1.0411249543867,3.5)
--(axis cs:1.0411249543867,2.5)
--(axis cs:1.03780430589585,2.5)
--(axis cs:1.03780430589585,1.5)
--(axis cs:1.0145123286243,1.5)
--(axis cs:1.0145123286243,1)
--cycle;

\addplot [very thin, dimgrey102]
table {%
1 0.5
1 17.5
};
\path [fill=steelblue31119180, fill opacity=0.75, line width=0pt]
(axis cs:1.95604467415196,0)
--(axis cs:2,0)
--(axis cs:2,0.5)
--(axis cs:2,0.5)
--(axis cs:2,1.5)
--(axis cs:2,1.5)
--(axis cs:2,2.5)
--(axis cs:2,2.5)
--(axis cs:2,3.5)
--(axis cs:2,3.5)
--(axis cs:2,4.5)
--(axis cs:2,4.5)
--(axis cs:2,5.5)
--(axis cs:2,5.5)
--(axis cs:2,6.5)
--(axis cs:2,6.5)
--(axis cs:2,7.5)
--(axis cs:2,7.5)
--(axis cs:2,8.5)
--(axis cs:2,8.5)
--(axis cs:2,9.5)
--(axis cs:2,9.5)
--(axis cs:2,10.5)
--(axis cs:2,10.5)
--(axis cs:2,11)
--(axis cs:1.99846967266679,11)
--(axis cs:1.99846967266679,11)
--(axis cs:1.99846967266679,10.5)
--(axis cs:1.99520941008733,10.5)
--(axis cs:1.99520941008733,9.5)
--(axis cs:1.97081565971603,9.5)
--(axis cs:1.97081565971603,8.5)
--(axis cs:1.86729816432009,8.5)
--(axis cs:1.86729816432009,7.5)
--(axis cs:1.58,7.5)
--(axis cs:1.58,6.5)
--(axis cs:1.70443058278382,6.5)
--(axis cs:1.70443058278382,5.5)
--(axis cs:1.8371448939583,5.5)
--(axis cs:1.8371448939583,4.5)
--(axis cs:1.84018475613378,4.5)
--(axis cs:1.84018475613378,3.5)
--(axis cs:1.89041525574764,3.5)
--(axis cs:1.89041525574764,2.5)
--(axis cs:1.93278203528783,2.5)
--(axis cs:1.93278203528783,1.5)
--(axis cs:1.96223251945583,1.5)
--(axis cs:1.96223251945583,0.5)
--(axis cs:1.95604467415196,0.5)
--(axis cs:1.95604467415196,0)
--cycle;

\addplot [very thin, dimgrey102]
table {%
2 -0.5
2 11.5
};
\path [fill=crimson2143940, fill opacity=0.55, line width=0pt]
(axis cs:2.0930786986336,0)
--(axis cs:2,0)
--(axis cs:2,0.5)
--(axis cs:2,0.5)
--(axis cs:2,1.5)
--(axis cs:2,1.5)
--(axis cs:2,2.5)
--(axis cs:2,2.5)
--(axis cs:2,3.5)
--(axis cs:2,3.5)
--(axis cs:2,4.5)
--(axis cs:2,4.5)
--(axis cs:2,5.5)
--(axis cs:2,5.5)
--(axis cs:2,6.5)
--(axis cs:2,6.5)
--(axis cs:2,7.5)
--(axis cs:2,7.5)
--(axis cs:2,8.5)
--(axis cs:2,8.5)
--(axis cs:2,9.5)
--(axis cs:2,9.5)
--(axis cs:2,10.5)
--(axis cs:2,10.5)
--(axis cs:2,11.5)
--(axis cs:2,11.5)
--(axis cs:2,12.5)
--(axis cs:2,12.5)
--(axis cs:2,13)
--(axis cs:2.00023472179738,13)
--(axis cs:2.00023472179738,13)
--(axis cs:2.00023472179738,12.5)
--(axis cs:2.00101252540048,12.5)
--(axis cs:2.00101252540048,11.5)
--(axis cs:2.00612807986702,11.5)
--(axis cs:2.00612807986702,10.5)
--(axis cs:2.02329268659932,10.5)
--(axis cs:2.02329268659932,9.5)
--(axis cs:2.06912556932988,9.5)
--(axis cs:2.06912556932988,8.5)
--(axis cs:2.1412611005321,8.5)
--(axis cs:2.1412611005321,7.5)
--(axis cs:2.35453576897955,7.5)
--(axis cs:2.35453576897955,6.5)
--(axis cs:2.28241634613597,6.5)
--(axis cs:2.28241634613597,5.5)
--(axis cs:2.1679020245353,5.5)
--(axis cs:2.1679020245353,4.5)
--(axis cs:2.1221957074798,4.5)
--(axis cs:2.1221957074798,3.5)
--(axis cs:2.09482070256126,3.5)
--(axis cs:2.09482070256126,2.5)
--(axis cs:2.08793322864387,2.5)
--(axis cs:2.08793322864387,1.5)
--(axis cs:2.02103521519506,1.5)
--(axis cs:2.02103521519506,0.5)
--(axis cs:2.0930786986336,0.5)
--(axis cs:2.0930786986336,0)
--cycle;

\addplot [very thin, dimgrey102]
table {%
2 -0.5
2 13.5
};
\path [fill=steelblue31119180, fill opacity=0.75, line width=0pt]
(axis cs:2.99039029451672,0)
--(axis cs:3,0)
--(axis cs:3,0.5)
--(axis cs:3,0.5)
--(axis cs:3,1.5)
--(axis cs:3,1.5)
--(axis cs:3,2.5)
--(axis cs:3,2.5)
--(axis cs:3,3.5)
--(axis cs:3,3.5)
--(axis cs:3,4.5)
--(axis cs:3,4.5)
--(axis cs:3,5.5)
--(axis cs:3,5.5)
--(axis cs:3,6.5)
--(axis cs:3,6.5)
--(axis cs:3,7.5)
--(axis cs:3,7.5)
--(axis cs:3,8.5)
--(axis cs:3,8.5)
--(axis cs:3,9)
--(axis cs:2.99641684625003,9)
--(axis cs:2.99641684625003,9)
--(axis cs:2.99641684625003,8.5)
--(axis cs:2.96734190012504,8.5)
--(axis cs:2.96734190012504,7.5)
--(axis cs:2.75822651448639,7.5)
--(axis cs:2.75822651448639,6.5)
--(axis cs:2.58,6.5)
--(axis cs:2.58,5.5)
--(axis cs:2.74425519461516,5.5)
--(axis cs:2.74425519461516,4.5)
--(axis cs:2.8477308643946,4.5)
--(axis cs:2.8477308643946,3.5)
--(axis cs:2.85961263202703,3.5)
--(axis cs:2.85961263202703,2.5)
--(axis cs:2.95793705270439,2.5)
--(axis cs:2.95793705270439,1.5)
--(axis cs:2.98609944927768,1.5)
--(axis cs:2.98609944927768,0.5)
--(axis cs:2.99039029451672,0.5)
--(axis cs:2.99039029451672,0)
--cycle;

\addplot [very thin, dimgrey102]
table {%
3 -0.5
3 9.5
};
\path [fill=crimson2143940, fill opacity=0.55, line width=0pt]
(axis cs:3.0708808855136,0)
--(axis cs:3,0)
--(axis cs:3,0.5)
--(axis cs:3,0.5)
--(axis cs:3,1.5)
--(axis cs:3,1.5)
--(axis cs:3,2.5)
--(axis cs:3,2.5)
--(axis cs:3,3.5)
--(axis cs:3,3.5)
--(axis cs:3,4.5)
--(axis cs:3,4.5)
--(axis cs:3,5.5)
--(axis cs:3,5.5)
--(axis cs:3,6.5)
--(axis cs:3,6.5)
--(axis cs:3,7.5)
--(axis cs:3,7.5)
--(axis cs:3,8.5)
--(axis cs:3,8.5)
--(axis cs:3,9.5)
--(axis cs:3,9.5)
--(axis cs:3,10.5)
--(axis cs:3,10.5)
--(axis cs:3,11)
--(axis cs:3.00171103009546,11)
--(axis cs:3.00171103009546,11)
--(axis cs:3.00171103009546,10.5)
--(axis cs:3.00835563872981,10.5)
--(axis cs:3.00835563872981,9.5)
--(axis cs:3.03543935845382,9.5)
--(axis cs:3.03543935845382,8.5)
--(axis cs:3.09745932005099,8.5)
--(axis cs:3.09745932005099,7.5)
--(axis cs:3.19649087941657,7.5)
--(axis cs:3.19649087941657,6.5)
--(axis cs:3.31754463221616,6.5)
--(axis cs:3.31754463221616,5.5)
--(axis cs:3.25953876023458,5.5)
--(axis cs:3.25953876023458,4.5)
--(axis cs:3.12996672325885,4.5)
--(axis cs:3.12996672325885,3.5)
--(axis cs:3.08265424722243,3.5)
--(axis cs:3.08265424722243,2.5)
--(axis cs:3.08013866431909,2.5)
--(axis cs:3.08013866431909,1.5)
--(axis cs:3.03180911209158,1.5)
--(axis cs:3.03180911209158,0.5)
--(axis cs:3.0708808855136,0.5)
--(axis cs:3.0708808855136,0)
--cycle;

\addplot [very thin, dimgrey102]
table {%
3 -0.5
3 11.5
};
\path [fill=steelblue31119180, fill opacity=0.75, line width=0pt]
(axis cs:4.89895696115058,0)
--(axis cs:4.9,0)
--(axis cs:4.9,0.5)
--(axis cs:4.9,0.5)
--(axis cs:4.9,1.5)
--(axis cs:4.9,1.5)
--(axis cs:4.9,2.5)
--(axis cs:4.9,2.5)
--(axis cs:4.9,3.5)
--(axis cs:4.9,3.5)
--(axis cs:4.9,4.5)
--(axis cs:4.9,4.5)
--(axis cs:4.9,5.5)
--(axis cs:4.9,5.5)
--(axis cs:4.9,6.5)
--(axis cs:4.9,6.5)
--(axis cs:4.9,7.5)
--(axis cs:4.9,7.5)
--(axis cs:4.9,8.5)
--(axis cs:4.9,8.5)
--(axis cs:4.9,9.5)
--(axis cs:4.9,9.5)
--(axis cs:4.9,10.5)
--(axis cs:4.9,10.5)
--(axis cs:4.9,11.5)
--(axis cs:4.9,11.5)
--(axis cs:4.9,12.5)
--(axis cs:4.9,12.5)
--(axis cs:4.9,13.5)
--(axis cs:4.9,13.5)
--(axis cs:4.9,14.5)
--(axis cs:4.9,14.5)
--(axis cs:4.9,15.5)
--(axis cs:4.9,15.5)
--(axis cs:4.9,16.5)
--(axis cs:4.9,16.5)
--(axis cs:4.9,17.5)
--(axis cs:4.9,17.5)
--(axis cs:4.9,18.5)
--(axis cs:4.9,18.5)
--(axis cs:4.9,19.5)
--(axis cs:4.9,19.5)
--(axis cs:4.9,20.5)
--(axis cs:4.9,20.5)
--(axis cs:4.9,21.5)
--(axis cs:4.9,21.5)
--(axis cs:4.9,22)
--(axis cs:4.89324292223638,22)
--(axis cs:4.89324292223638,22)
--(axis cs:4.89324292223638,21.5)
--(axis cs:4.8767447686095,21.5)
--(axis cs:4.8767447686095,20.5)
--(axis cs:4.77894400414624,20.5)
--(axis cs:4.77894400414624,19.5)
--(axis cs:4.62217980003023,19.5)
--(axis cs:4.62217980003023,18.5)
--(axis cs:4.48,18.5)
--(axis cs:4.48,17.5)
--(axis cs:4.68775519899799,17.5)
--(axis cs:4.68775519899799,16.5)
--(axis cs:4.52638348413847,16.5)
--(axis cs:4.52638348413847,15.5)
--(axis cs:4.6884173019198,15.5)
--(axis cs:4.6884173019198,14.5)
--(axis cs:4.80404042585354,14.5)
--(axis cs:4.80404042585354,13.5)
--(axis cs:4.84720409441337,13.5)
--(axis cs:4.84720409441337,12.5)
--(axis cs:4.86847301703846,12.5)
--(axis cs:4.86847301703846,11.5)
--(axis cs:4.87966527738787,11.5)
--(axis cs:4.87966527738787,10.5)
--(axis cs:4.89891161163539,10.5)
--(axis cs:4.89891161163539,9.5)
--(axis cs:4.89890254173235,9.5)
--(axis cs:4.89890254173235,8.5)
--(axis cs:4.89891161163539,8.5)
--(axis cs:4.89891161163539,7.5)
--(axis cs:4.89891161163539,7.5)
--(axis cs:4.89891161163539,6.5)
--(axis cs:4.89890254173235,6.5)
--(axis cs:4.89890254173235,5.5)
--(axis cs:4.89890254173235,5.5)
--(axis cs:4.89890254173235,4.5)
--(axis cs:4.89891161163539,4.5)
--(axis cs:4.89891161163539,3.5)
--(axis cs:4.89890254173235,3.5)
--(axis cs:4.89890254173235,2.5)
--(axis cs:4.89891161163539,2.5)
--(axis cs:4.89891161163539,1.5)
--(axis cs:4.89890254173235,1.5)
--(axis cs:4.89890254173235,0.5)
--(axis cs:4.89895696115058,0.5)
--(axis cs:4.89895696115058,0)
--cycle;

\addplot [very thin, dimgrey102]
table {%
4.9 -0.5
4.9 22.5
};
\path [fill=crimson2143940, fill opacity=0.55, line width=0pt]
(axis cs:4.91191120799129,10)
--(axis cs:4.9,10)
--(axis cs:4.9,10.5)
--(axis cs:4.9,10.5)
--(axis cs:4.9,11.5)
--(axis cs:4.9,11.5)
--(axis cs:4.9,12.5)
--(axis cs:4.9,12.5)
--(axis cs:4.9,13.5)
--(axis cs:4.9,13.5)
--(axis cs:4.9,14.5)
--(axis cs:4.9,14.5)
--(axis cs:4.9,15.5)
--(axis cs:4.9,15.5)
--(axis cs:4.9,16.5)
--(axis cs:4.9,16.5)
--(axis cs:4.9,17.5)
--(axis cs:4.9,17.5)
--(axis cs:4.9,18.5)
--(axis cs:4.9,18.5)
--(axis cs:4.9,19.5)
--(axis cs:4.9,19.5)
--(axis cs:4.9,20.5)
--(axis cs:4.9,20.5)
--(axis cs:4.9,21.5)
--(axis cs:4.9,21.5)
--(axis cs:4.9,22)
--(axis cs:4.90712462392057,22)
--(axis cs:4.90712462392057,22)
--(axis cs:4.90712462392057,21.5)
--(axis cs:4.94244531050804,21.5)
--(axis cs:4.94244531050804,20.5)
--(axis cs:5.04384179494173,20.5)
--(axis cs:5.04384179494173,19.5)
--(axis cs:5.13967525669785,19.5)
--(axis cs:5.13967525669785,18.5)
--(axis cs:5.24029821053446,18.5)
--(axis cs:5.24029821053446,17.5)
--(axis cs:5.19765224735274,17.5)
--(axis cs:5.19765224735274,16.5)
--(axis cs:5.2616546342376,16.5)
--(axis cs:5.2616546342376,15.5)
--(axis cs:5.08614752098112,15.5)
--(axis cs:5.08614752098112,14.5)
--(axis cs:5.01296745528266,14.5)
--(axis cs:5.01296745528266,13.5)
--(axis cs:4.9561274964453,13.5)
--(axis cs:4.9561274964453,12.5)
--(axis cs:4.94476299428955,12.5)
--(axis cs:4.94476299428955,11.5)
--(axis cs:4.91431322405594,11.5)
--(axis cs:4.91431322405594,10.5)
--(axis cs:4.91191120799129,10.5)
--(axis cs:4.91191120799129,10)
--cycle;

\addplot [very thin, dimgrey102]
table {%
4.9 9.5
4.9 22.5
};
\path [fill=steelblue31119180, fill opacity=0.75, line width=0pt]
(axis cs:5.89782124479422,0)
--(axis cs:5.9,0)
--(axis cs:5.9,0.5)
--(axis cs:5.9,0.5)
--(axis cs:5.9,1.5)
--(axis cs:5.9,1.5)
--(axis cs:5.9,2.5)
--(axis cs:5.9,2.5)
--(axis cs:5.9,3.5)
--(axis cs:5.9,3.5)
--(axis cs:5.9,4.5)
--(axis cs:5.9,4.5)
--(axis cs:5.9,5.5)
--(axis cs:5.9,5.5)
--(axis cs:5.9,6.5)
--(axis cs:5.9,6.5)
--(axis cs:5.9,7.5)
--(axis cs:5.9,7.5)
--(axis cs:5.9,8.5)
--(axis cs:5.9,8.5)
--(axis cs:5.9,9.5)
--(axis cs:5.9,9.5)
--(axis cs:5.9,10.5)
--(axis cs:5.9,10.5)
--(axis cs:5.9,11.5)
--(axis cs:5.9,11.5)
--(axis cs:5.9,12.5)
--(axis cs:5.9,12.5)
--(axis cs:5.9,13.5)
--(axis cs:5.9,13.5)
--(axis cs:5.9,14.5)
--(axis cs:5.9,14.5)
--(axis cs:5.9,15.5)
--(axis cs:5.9,15.5)
--(axis cs:5.9,16)
--(axis cs:5.89770288183536,16)
--(axis cs:5.89770288183536,16)
--(axis cs:5.89770288183536,15.5)
--(axis cs:5.88388510234117,15.5)
--(axis cs:5.88388510234117,14.5)
--(axis cs:5.82218293026606,14.5)
--(axis cs:5.82218293026606,13.5)
--(axis cs:5.67988873464361,13.5)
--(axis cs:5.67988873464361,12.5)
--(axis cs:5.48,12.5)
--(axis cs:5.48,11.5)
--(axis cs:5.5841067980419,11.5)
--(axis cs:5.5841067980419,10.5)
--(axis cs:5.67941528280815,10.5)
--(axis cs:5.67941528280815,9.5)
--(axis cs:5.69011178723893,9.5)
--(axis cs:5.69011178723893,8.5)
--(axis cs:5.76142327804858,8.5)
--(axis cs:5.76142327804858,7.5)
--(axis cs:5.8276057073074,7.5)
--(axis cs:5.8276057073074,6.5)
--(axis cs:5.82771530263968,6.5)
--(axis cs:5.82771530263968,5.5)
--(axis cs:5.86697235066331,5.5)
--(axis cs:5.86697235066331,4.5)
--(axis cs:5.87887001993591,4.5)
--(axis cs:5.87887001993591,3.5)
--(axis cs:5.89630444539543,3.5)
--(axis cs:5.89630444539543,2.5)
--(axis cs:5.89104386944587,2.5)
--(axis cs:5.89104386944587,1.5)
--(axis cs:5.89736971202522,1.5)
--(axis cs:5.89736971202522,0.5)
--(axis cs:5.89782124479422,0.5)
--(axis cs:5.89782124479422,0)
--cycle;

\addplot [very thin, dimgrey102]
table {%
5.9 -0.5
5.9 16.5
};
\path [fill=crimson2143940, fill opacity=0.55, line width=0pt]
(axis cs:5.91232504402,5)
--(axis cs:5.9,5)
--(axis cs:5.9,5.5)
--(axis cs:5.9,5.5)
--(axis cs:5.9,6.5)
--(axis cs:5.9,6.5)
--(axis cs:5.9,7.5)
--(axis cs:5.9,7.5)
--(axis cs:5.9,8.5)
--(axis cs:5.9,8.5)
--(axis cs:5.9,9.5)
--(axis cs:5.9,9.5)
--(axis cs:5.9,10.5)
--(axis cs:5.9,10.5)
--(axis cs:5.9,11.5)
--(axis cs:5.9,11.5)
--(axis cs:5.9,12.5)
--(axis cs:5.9,12.5)
--(axis cs:5.9,13.5)
--(axis cs:5.9,13.5)
--(axis cs:5.9,14.5)
--(axis cs:5.9,14.5)
--(axis cs:5.9,15.5)
--(axis cs:5.9,15.5)
--(axis cs:5.9,16)
--(axis cs:5.90140240570975,16)
--(axis cs:5.90140240570975,16)
--(axis cs:5.90140240570975,15.5)
--(axis cs:5.91657833608017,15.5)
--(axis cs:5.91657833608017,14.5)
--(axis cs:6.00654539805978,14.5)
--(axis cs:6.00654539805978,13.5)
--(axis cs:6.05638983426121,13.5)
--(axis cs:6.05638983426121,12.5)
--(axis cs:6.25981353393798,12.5)
--(axis cs:6.25981353393798,11.5)
--(axis cs:6.26124761616274,11.5)
--(axis cs:6.26124761616274,10.5)
--(axis cs:6.26167092959053,10.5)
--(axis cs:6.26167092959053,9.5)
--(axis cs:6.09422023222633,9.5)
--(axis cs:6.09422023222633,8.5)
--(axis cs:6.03752215005116,8.5)
--(axis cs:6.03752215005116,7.5)
--(axis cs:5.96509811801777,7.5)
--(axis cs:5.96509811801777,6.5)
--(axis cs:5.96476695445181,6.5)
--(axis cs:5.96476695445181,5.5)
--(axis cs:5.91232504402,5.5)
--(axis cs:5.91232504402,5)
--cycle;

\addplot [very thin, dimgrey102]
table {%
5.9 4.5
5.9 16.5
};
\path [fill=steelblue31119180, fill opacity=0.75, line width=0pt]
(axis cs:6.87197441240776,0)
--(axis cs:6.9,0)
--(axis cs:6.9,0.5)
--(axis cs:6.9,0.5)
--(axis cs:6.9,1.5)
--(axis cs:6.9,1.5)
--(axis cs:6.9,2.5)
--(axis cs:6.9,2.5)
--(axis cs:6.9,3.5)
--(axis cs:6.9,3.5)
--(axis cs:6.9,4.5)
--(axis cs:6.9,4.5)
--(axis cs:6.9,5.5)
--(axis cs:6.9,5.5)
--(axis cs:6.9,6.5)
--(axis cs:6.9,6.5)
--(axis cs:6.9,7.5)
--(axis cs:6.9,7.5)
--(axis cs:6.9,8.5)
--(axis cs:6.9,8.5)
--(axis cs:6.9,9.5)
--(axis cs:6.9,9.5)
--(axis cs:6.9,10)
--(axis cs:6.89832843673135,10)
--(axis cs:6.89832843673135,10)
--(axis cs:6.89832843673135,9.5)
--(axis cs:6.88814481415687,9.5)
--(axis cs:6.88814481415687,8.5)
--(axis cs:6.7981673613009,8.5)
--(axis cs:6.7981673613009,7.5)
--(axis cs:6.48,7.5)
--(axis cs:6.48,6.5)
--(axis cs:6.61540738589779,6.5)
--(axis cs:6.61540738589779,5.5)
--(axis cs:6.69984644028423,5.5)
--(axis cs:6.69984644028423,4.5)
--(axis cs:6.79040858157967,4.5)
--(axis cs:6.79040858157967,3.5)
--(axis cs:6.81556811970484,3.5)
--(axis cs:6.81556811970484,2.5)
--(axis cs:6.86999795025963,2.5)
--(axis cs:6.86999795025963,1.5)
--(axis cs:6.8744781702651,1.5)
--(axis cs:6.8744781702651,0.5)
--(axis cs:6.87197441240776,0.5)
--(axis cs:6.87197441240776,0)
--cycle;

\addplot [very thin, dimgrey102]
table {%
6.9 -0.5
6.9 10.5
};
\path [fill=crimson2143940, fill opacity=0.55, line width=0pt]
(axis cs:6.91096989192465,0)
--(axis cs:6.9,0)
--(axis cs:6.9,0.5)
--(axis cs:6.9,0.5)
--(axis cs:6.9,1.5)
--(axis cs:6.9,1.5)
--(axis cs:6.9,2.5)
--(axis cs:6.9,2.5)
--(axis cs:6.9,3.5)
--(axis cs:6.9,3.5)
--(axis cs:6.9,4.5)
--(axis cs:6.9,4.5)
--(axis cs:6.9,5.5)
--(axis cs:6.9,5.5)
--(axis cs:6.9,6.5)
--(axis cs:6.9,6.5)
--(axis cs:6.9,7.5)
--(axis cs:6.9,7.5)
--(axis cs:6.9,8.5)
--(axis cs:6.9,8.5)
--(axis cs:6.9,9.5)
--(axis cs:6.9,9.5)
--(axis cs:6.9,10.5)
--(axis cs:6.9,10.5)
--(axis cs:6.9,11)
--(axis cs:6.90074656770937,11)
--(axis cs:6.90074656770937,11)
--(axis cs:6.90074656770937,10.5)
--(axis cs:6.90968919447199,10.5)
--(axis cs:6.90968919447199,9.5)
--(axis cs:6.95499108493397,9.5)
--(axis cs:6.95499108493397,8.5)
--(axis cs:7.05174949721809,8.5)
--(axis cs:7.05174949721809,7.5)
--(axis cs:7.19649461121169,7.5)
--(axis cs:7.19649461121169,6.5)
--(axis cs:7.18088143231414,6.5)
--(axis cs:7.18088143231414,5.5)
--(axis cs:7.10951239180473,5.5)
--(axis cs:7.10951239180473,4.5)
--(axis cs:7.02569246326427,4.5)
--(axis cs:7.02569246326427,3.5)
--(axis cs:6.98686891915839,3.5)
--(axis cs:6.98686891915839,2.5)
--(axis cs:6.9483913454544,2.5)
--(axis cs:6.9483913454544,1.5)
--(axis cs:6.92169092794618,1.5)
--(axis cs:6.92169092794618,0.5)
--(axis cs:6.91096989192465,0.5)
--(axis cs:6.91096989192465,0)
--cycle;

\addplot [very thin, dimgrey102]
table {%
6.9 -0.5
6.9 11.5
};
\path [fill=steelblue31119180, fill opacity=0.75, line width=0pt]
(axis cs:7.89674588955304,0)
--(axis cs:7.9,0)
--(axis cs:7.9,0.5)
--(axis cs:7.9,0.5)
--(axis cs:7.9,1.5)
--(axis cs:7.9,1.5)
--(axis cs:7.9,2.5)
--(axis cs:7.9,2.5)
--(axis cs:7.9,3.5)
--(axis cs:7.9,3.5)
--(axis cs:7.9,4.5)
--(axis cs:7.9,4.5)
--(axis cs:7.9,5.5)
--(axis cs:7.9,5.5)
--(axis cs:7.9,6.5)
--(axis cs:7.9,6.5)
--(axis cs:7.9,7.5)
--(axis cs:7.9,7.5)
--(axis cs:7.9,8)
--(axis cs:7.89800447115156,8)
--(axis cs:7.89800447115156,8)
--(axis cs:7.89800447115156,7.5)
--(axis cs:7.82194352396506,7.5)
--(axis cs:7.82194352396506,6.5)
--(axis cs:7.48,6.5)
--(axis cs:7.48,5.5)
--(axis cs:7.76417973395063,5.5)
--(axis cs:7.76417973395063,4.5)
--(axis cs:7.83744827258169,4.5)
--(axis cs:7.83744827258169,3.5)
--(axis cs:7.86244232690143,3.5)
--(axis cs:7.86244232690143,2.5)
--(axis cs:7.88962280604398,2.5)
--(axis cs:7.88962280604398,1.5)
--(axis cs:7.89625755101394,1.5)
--(axis cs:7.89625755101394,0.5)
--(axis cs:7.89674588955304,0.5)
--(axis cs:7.89674588955304,0)
--cycle;

\addplot [very thin, dimgrey102]
table {%
7.9 -0.5
7.9 8.5
};
\path [fill=crimson2143940, fill opacity=0.55, line width=0pt]
(axis cs:7.90434562575967,0)
--(axis cs:7.9,0)
--(axis cs:7.9,0.5)
--(axis cs:7.9,0.5)
--(axis cs:7.9,1.5)
--(axis cs:7.9,1.5)
--(axis cs:7.9,2.5)
--(axis cs:7.9,2.5)
--(axis cs:7.9,3.5)
--(axis cs:7.9,3.5)
--(axis cs:7.9,4.5)
--(axis cs:7.9,4.5)
--(axis cs:7.9,5.5)
--(axis cs:7.9,5.5)
--(axis cs:7.9,6.5)
--(axis cs:7.9,6.5)
--(axis cs:7.9,7.5)
--(axis cs:7.9,7.5)
--(axis cs:7.9,8.5)
--(axis cs:7.9,8.5)
--(axis cs:7.9,9.5)
--(axis cs:7.9,9.5)
--(axis cs:7.9,10)
--(axis cs:7.90126242955528,10)
--(axis cs:7.90126242955528,10)
--(axis cs:7.90126242955528,9.5)
--(axis cs:7.91492346893285,9.5)
--(axis cs:7.91492346893285,8.5)
--(axis cs:7.97240632977273,8.5)
--(axis cs:7.97240632977273,7.5)
--(axis cs:8.08159120550007,7.5)
--(axis cs:8.08159120550007,6.5)
--(axis cs:8.07750417797333,6.5)
--(axis cs:8.07750417797333,5.5)
--(axis cs:8.0203880564456,5.5)
--(axis cs:8.0203880564456,4.5)
--(axis cs:7.98728513173807,4.5)
--(axis cs:7.98728513173807,3.5)
--(axis cs:7.95695626080192,3.5)
--(axis cs:7.95695626080192,2.5)
--(axis cs:7.9296435856189,2.5)
--(axis cs:7.9296435856189,1.5)
--(axis cs:7.90704915274026,1.5)
--(axis cs:7.90704915274026,0.5)
--(axis cs:7.90434562575967,0.5)
--(axis cs:7.90434562575967,0)
--cycle;

\addplot [very thin, dimgrey102]
table {%
7.9 -0.5
7.9 10.5
};
\path [fill=steelblue31119180, fill opacity=0.75, line width=0pt]
(axis cs:9.79813300060962,0)
--(axis cs:9.8,0)
--(axis cs:9.8,0.5)
--(axis cs:9.8,0.5)
--(axis cs:9.8,1.5)
--(axis cs:9.8,1.5)
--(axis cs:9.8,2.5)
--(axis cs:9.8,2.5)
--(axis cs:9.8,3.5)
--(axis cs:9.8,3.5)
--(axis cs:9.8,4.5)
--(axis cs:9.8,4.5)
--(axis cs:9.8,5.5)
--(axis cs:9.8,5.5)
--(axis cs:9.8,6.5)
--(axis cs:9.8,6.5)
--(axis cs:9.8,7.5)
--(axis cs:9.8,7.5)
--(axis cs:9.8,8.5)
--(axis cs:9.8,8.5)
--(axis cs:9.8,9.5)
--(axis cs:9.8,9.5)
--(axis cs:9.8,10.5)
--(axis cs:9.8,10.5)
--(axis cs:9.8,11.5)
--(axis cs:9.8,11.5)
--(axis cs:9.8,12.5)
--(axis cs:9.8,12.5)
--(axis cs:9.8,13.5)
--(axis cs:9.8,13.5)
--(axis cs:9.8,14.5)
--(axis cs:9.8,14.5)
--(axis cs:9.8,15.5)
--(axis cs:9.8,15.5)
--(axis cs:9.8,16.5)
--(axis cs:9.8,16.5)
--(axis cs:9.8,17.5)
--(axis cs:9.8,17.5)
--(axis cs:9.8,18.5)
--(axis cs:9.8,18.5)
--(axis cs:9.8,19.5)
--(axis cs:9.8,19.5)
--(axis cs:9.8,20.5)
--(axis cs:9.8,20.5)
--(axis cs:9.8,21.5)
--(axis cs:9.8,21.5)
--(axis cs:9.8,22)
--(axis cs:9.79698033142077,22)
--(axis cs:9.79698033142077,22)
--(axis cs:9.79698033142077,21.5)
--(axis cs:9.78596503501751,21.5)
--(axis cs:9.78596503501751,20.5)
--(axis cs:9.74100281926389,20.5)
--(axis cs:9.74100281926389,19.5)
--(axis cs:9.65965846756375,19.5)
--(axis cs:9.65965846756375,18.5)
--(axis cs:9.56824043654419,18.5)
--(axis cs:9.56824043654419,17.5)
--(axis cs:9.43985581759505,17.5)
--(axis cs:9.43985581759505,16.5)
--(axis cs:9.38598882648816,16.5)
--(axis cs:9.38598882648816,15.5)
--(axis cs:9.51040404238558,15.5)
--(axis cs:9.51040404238558,14.5)
--(axis cs:9.54916457320767,14.5)
--(axis cs:9.54916457320767,13.5)
--(axis cs:9.61341370440282,13.5)
--(axis cs:9.61341370440282,12.5)
--(axis cs:9.70786763877891,12.5)
--(axis cs:9.70786763877891,11.5)
--(axis cs:9.70894725146987,11.5)
--(axis cs:9.70894725146987,10.5)
--(axis cs:9.77004683586733,10.5)
--(axis cs:9.77004683586733,9.5)
--(axis cs:9.79611177083481,9.5)
--(axis cs:9.79611177083481,8.5)
--(axis cs:9.78641149139347,8.5)
--(axis cs:9.78641149139347,7.5)
--(axis cs:9.79804370933443,7.5)
--(axis cs:9.79804370933443,6.5)
--(axis cs:9.79805182672308,6.5)
--(axis cs:9.79805182672308,5.5)
--(axis cs:9.79804370933443,5.5)
--(axis cs:9.79804370933443,4.5)
--(axis cs:9.79804370933443,4.5)
--(axis cs:9.79804370933443,3.5)
--(axis cs:9.79805182672308,3.5)
--(axis cs:9.79805182672308,2.5)
--(axis cs:9.79804370933443,2.5)
--(axis cs:9.79804370933443,1.5)
--(axis cs:9.79804370933443,1.5)
--(axis cs:9.79804370933443,0.5)
--(axis cs:9.79813300060962,0.5)
--(axis cs:9.79813300060962,0)
--cycle;

\addplot [very thin, dimgrey102]
table {%
9.8 -0.5
9.8 22.5
};
\path [fill=crimson2143940, fill opacity=0.55, line width=0pt]
(axis cs:9.80896724673022,2)
--(axis cs:9.8,2)
--(axis cs:9.8,2.5)
--(axis cs:9.8,2.5)
--(axis cs:9.8,3.5)
--(axis cs:9.8,3.5)
--(axis cs:9.8,4.5)
--(axis cs:9.8,4.5)
--(axis cs:9.8,5.5)
--(axis cs:9.8,5.5)
--(axis cs:9.8,6.5)
--(axis cs:9.8,6.5)
--(axis cs:9.8,7.5)
--(axis cs:9.8,7.5)
--(axis cs:9.8,8.5)
--(axis cs:9.8,8.5)
--(axis cs:9.8,9.5)
--(axis cs:9.8,9.5)
--(axis cs:9.8,10.5)
--(axis cs:9.8,10.5)
--(axis cs:9.8,11.5)
--(axis cs:9.8,11.5)
--(axis cs:9.8,12.5)
--(axis cs:9.8,12.5)
--(axis cs:9.8,13.5)
--(axis cs:9.8,13.5)
--(axis cs:9.8,14.5)
--(axis cs:9.8,14.5)
--(axis cs:9.8,15.5)
--(axis cs:9.8,15.5)
--(axis cs:9.8,16.5)
--(axis cs:9.8,16.5)
--(axis cs:9.8,17.5)
--(axis cs:9.8,17.5)
--(axis cs:9.8,18.5)
--(axis cs:9.8,18.5)
--(axis cs:9.8,19.5)
--(axis cs:9.8,19.5)
--(axis cs:9.8,20.5)
--(axis cs:9.8,20.5)
--(axis cs:9.8,21.5)
--(axis cs:9.8,21.5)
--(axis cs:9.8,22)
--(axis cs:9.80121619374862,22)
--(axis cs:9.80121619374862,22)
--(axis cs:9.80121619374862,21.5)
--(axis cs:9.81704998891598,21.5)
--(axis cs:9.81704998891598,20.5)
--(axis cs:9.8780517069386,20.5)
--(axis cs:9.8780517069386,19.5)
--(axis cs:9.93317612502771,19.5)
--(axis cs:9.93317612502771,18.5)
--(axis cs:10.0140908335181,18.5)
--(axis cs:10.0140908335181,17.5)
--(axis cs:10.22,17.5)
--(axis cs:10.22,16.5)
--(axis cs:10.1954200842385,16.5)
--(axis cs:10.1954200842385,15.5)
--(axis cs:10.0873534138772,15.5)
--(axis cs:10.0873534138772,14.5)
--(axis cs:9.97601058523609,14.5)
--(axis cs:9.97601058523609,13.5)
--(axis cs:9.90786882066061,13.5)
--(axis cs:9.90786882066061,12.5)
--(axis cs:9.90332991576147,12.5)
--(axis cs:9.90332991576147,11.5)
--(axis cs:9.88538378408335,11.5)
--(axis cs:9.88538378408335,10.5)
--(axis cs:9.86746092884061,10.5)
--(axis cs:9.86746092884061,9.5)
--(axis cs:9.83359371536245,9.5)
--(axis cs:9.83359371536245,8.5)
--(axis cs:9.82318914874751,8.5)
--(axis cs:9.82318914874751,7.5)
--(axis cs:9.82054145422301,7.5)
--(axis cs:9.82054145422301,6.5)
--(axis cs:9.8110970405675,6.5)
--(axis cs:9.8110970405675,5.5)
--(axis cs:9.80911272445134,5.5)
--(axis cs:9.80911272445134,4.5)
--(axis cs:9.80116964087786,4.5)
--(axis cs:9.80116964087786,3.5)
--(axis cs:9.80140240523166,3.5)
--(axis cs:9.80140240523166,2.5)
--(axis cs:9.80896724673022,2.5)
--(axis cs:9.80896724673022,2)
--cycle;

\addplot [very thin, dimgrey102]
table {%
9.8 1.5
9.8 22.5
};
\path [fill=steelblue31119180, fill opacity=0.75, line width=0pt]
(axis cs:10.7861552120564,3)
--(axis cs:10.8,3)
--(axis cs:10.8,3.5)
--(axis cs:10.8,3.5)
--(axis cs:10.8,4.5)
--(axis cs:10.8,4.5)
--(axis cs:10.8,5.5)
--(axis cs:10.8,5.5)
--(axis cs:10.8,6.5)
--(axis cs:10.8,6.5)
--(axis cs:10.8,7.5)
--(axis cs:10.8,7.5)
--(axis cs:10.8,8.5)
--(axis cs:10.8,8.5)
--(axis cs:10.8,9.5)
--(axis cs:10.8,9.5)
--(axis cs:10.8,10.5)
--(axis cs:10.8,10.5)
--(axis cs:10.8,11.5)
--(axis cs:10.8,11.5)
--(axis cs:10.8,12.5)
--(axis cs:10.8,12.5)
--(axis cs:10.8,13.5)
--(axis cs:10.8,13.5)
--(axis cs:10.8,14.5)
--(axis cs:10.8,14.5)
--(axis cs:10.8,15)
--(axis cs:10.7985944863068,15)
--(axis cs:10.7985944863068,15)
--(axis cs:10.7985944863068,14.5)
--(axis cs:10.7588767728276,14.5)
--(axis cs:10.7588767728276,13.5)
--(axis cs:10.6542281512509,13.5)
--(axis cs:10.6542281512509,12.5)
--(axis cs:10.5028820543174,12.5)
--(axis cs:10.5028820543174,11.5)
--(axis cs:10.38,11.5)
--(axis cs:10.38,10.5)
--(axis cs:10.5877769936486,10.5)
--(axis cs:10.5877769936486,9.5)
--(axis cs:10.4920203974776,9.5)
--(axis cs:10.4920203974776,8.5)
--(axis cs:10.6143287727366,8.5)
--(axis cs:10.6143287727366,7.5)
--(axis cs:10.6900448471327,7.5)
--(axis cs:10.6900448471327,6.5)
--(axis cs:10.739285632982,6.5)
--(axis cs:10.739285632982,5.5)
--(axis cs:10.7792615020375,5.5)
--(axis cs:10.7792615020375,4.5)
--(axis cs:10.7723964759715,4.5)
--(axis cs:10.7723964759715,3.5)
--(axis cs:10.7861552120564,3.5)
--(axis cs:10.7861552120564,3)
--cycle;

\addplot [very thin, dimgrey102]
table {%
10.8 2.5
10.8 15.5
};
\path [fill=crimson2143940, fill opacity=0.55, line width=0pt]
(axis cs:10.8390451725541,0)
--(axis cs:10.8,0)
--(axis cs:10.8,0.5)
--(axis cs:10.8,0.5)
--(axis cs:10.8,1.5)
--(axis cs:10.8,1.5)
--(axis cs:10.8,2.5)
--(axis cs:10.8,2.5)
--(axis cs:10.8,3.5)
--(axis cs:10.8,3.5)
--(axis cs:10.8,4.5)
--(axis cs:10.8,4.5)
--(axis cs:10.8,5.5)
--(axis cs:10.8,5.5)
--(axis cs:10.8,6.5)
--(axis cs:10.8,6.5)
--(axis cs:10.8,7.5)
--(axis cs:10.8,7.5)
--(axis cs:10.8,8.5)
--(axis cs:10.8,8.5)
--(axis cs:10.8,9.5)
--(axis cs:10.8,9.5)
--(axis cs:10.8,10.5)
--(axis cs:10.8,10.5)
--(axis cs:10.8,11.5)
--(axis cs:10.8,11.5)
--(axis cs:10.8,12.5)
--(axis cs:10.8,12.5)
--(axis cs:10.8,13.5)
--(axis cs:10.8,13.5)
--(axis cs:10.8,14.5)
--(axis cs:10.8,14.5)
--(axis cs:10.8,15.5)
--(axis cs:10.8,15.5)
--(axis cs:10.8,16)
--(axis cs:10.8035872825854,16)
--(axis cs:10.8035872825854,16)
--(axis cs:10.8035872825854,15.5)
--(axis cs:10.8113474664884,15.5)
--(axis cs:10.8113474664884,14.5)
--(axis cs:10.8576937449444,14.5)
--(axis cs:10.8576937449444,13.5)
--(axis cs:10.9430704819159,13.5)
--(axis cs:10.9430704819159,12.5)
--(axis cs:11.1079736541528,12.5)
--(axis cs:11.1079736541528,11.5)
--(axis cs:11.157224483808,11.5)
--(axis cs:11.157224483808,10.5)
--(axis cs:11.0882421285291,10.5)
--(axis cs:11.0882421285291,9.5)
--(axis cs:10.9585349363757,9.5)
--(axis cs:10.9585349363757,8.5)
--(axis cs:10.9660390960088,8.5)
--(axis cs:10.9660390960088,7.5)
--(axis cs:10.9072712262542,7.5)
--(axis cs:10.9072712262542,6.5)
--(axis cs:10.8583794191383,6.5)
--(axis cs:10.8583794191383,5.5)
--(axis cs:10.866681079658,5.5)
--(axis cs:10.866681079658,4.5)
--(axis cs:10.8302402919178,4.5)
--(axis cs:10.8302402919178,3.5)
--(axis cs:10.8214766105894,3.5)
--(axis cs:10.8214766105894,2.5)
--(axis cs:10.8134721736473,2.5)
--(axis cs:10.8134721736473,1.5)
--(axis cs:10.8138694526867,1.5)
--(axis cs:10.8138694526867,0.5)
--(axis cs:10.8390451725541,0.5)
--(axis cs:10.8390451725541,0)
--cycle;

\addplot [very thin, dimgrey102]
table {%
10.8 -0.5
10.8 16.5
};
\path [fill=steelblue31119180, fill opacity=0.75, line width=0pt]
(axis cs:11.768808727997,0)
--(axis cs:11.8,0)
--(axis cs:11.8,0.5)
--(axis cs:11.8,0.5)
--(axis cs:11.8,1.5)
--(axis cs:11.8,1.5)
--(axis cs:11.8,2.5)
--(axis cs:11.8,2.5)
--(axis cs:11.8,3.5)
--(axis cs:11.8,3.5)
--(axis cs:11.8,4.5)
--(axis cs:11.8,4.5)
--(axis cs:11.8,5.5)
--(axis cs:11.8,5.5)
--(axis cs:11.8,6.5)
--(axis cs:11.8,6.5)
--(axis cs:11.8,7.5)
--(axis cs:11.8,7.5)
--(axis cs:11.8,8.5)
--(axis cs:11.8,8.5)
--(axis cs:11.8,9.5)
--(axis cs:11.8,9.5)
--(axis cs:11.8,10)
--(axis cs:11.7989891277284,10)
--(axis cs:11.7989891277284,10)
--(axis cs:11.7989891277284,9.5)
--(axis cs:11.7948949365847,9.5)
--(axis cs:11.7948949365847,8.5)
--(axis cs:11.7358270395883,8.5)
--(axis cs:11.7358270395883,7.5)
--(axis cs:11.38,7.5)
--(axis cs:11.38,6.5)
--(axis cs:11.5205524411683,6.5)
--(axis cs:11.5205524411683,5.5)
--(axis cs:11.5889558545032,5.5)
--(axis cs:11.5889558545032,4.5)
--(axis cs:11.6580279012215,4.5)
--(axis cs:11.6580279012215,3.5)
--(axis cs:11.7400194659685,3.5)
--(axis cs:11.7400194659685,2.5)
--(axis cs:11.740944778518,2.5)
--(axis cs:11.740944778518,1.5)
--(axis cs:11.7753112668724,1.5)
--(axis cs:11.7753112668724,0.5)
--(axis cs:11.768808727997,0.5)
--(axis cs:11.768808727997,0)
--cycle;

\addplot [very thin, dimgrey102]
table {%
11.8 -0.5
11.8 10.5
};
\path [fill=crimson2143940, fill opacity=0.55, line width=0pt]
(axis cs:11.8900618195421,0)
--(axis cs:11.8,0)
--(axis cs:11.8,0.5)
--(axis cs:11.8,0.5)
--(axis cs:11.8,1.5)
--(axis cs:11.8,1.5)
--(axis cs:11.8,2.5)
--(axis cs:11.8,2.5)
--(axis cs:11.8,3.5)
--(axis cs:11.8,3.5)
--(axis cs:11.8,4.5)
--(axis cs:11.8,4.5)
--(axis cs:11.8,5.5)
--(axis cs:11.8,5.5)
--(axis cs:11.8,6.5)
--(axis cs:11.8,6.5)
--(axis cs:11.8,7.5)
--(axis cs:11.8,7.5)
--(axis cs:11.8,8.5)
--(axis cs:11.8,8.5)
--(axis cs:11.8,9.5)
--(axis cs:11.8,9.5)
--(axis cs:11.8,10.5)
--(axis cs:11.8,10.5)
--(axis cs:11.8,11.5)
--(axis cs:11.8,11.5)
--(axis cs:11.8,12)
--(axis cs:11.804769571132,12)
--(axis cs:11.804769571132,12)
--(axis cs:11.804769571132,11.5)
--(axis cs:11.807468470568,11.5)
--(axis cs:11.807468470568,10.5)
--(axis cs:11.8407438754065,10.5)
--(axis cs:11.8407438754065,9.5)
--(axis cs:11.8952178333958,9.5)
--(axis cs:11.8952178333958,8.5)
--(axis cs:11.9537360074618,8.5)
--(axis cs:11.9537360074618,7.5)
--(axis cs:12.0269266057207,7.5)
--(axis cs:12.0269266057207,6.5)
--(axis cs:11.9693177086469,6.5)
--(axis cs:11.9693177086469,5.5)
--(axis cs:11.9788465182178,5.5)
--(axis cs:11.9788465182178,4.5)
--(axis cs:11.9367738584794,4.5)
--(axis cs:11.9367738584794,3.5)
--(axis cs:11.8750256845523,3.5)
--(axis cs:11.8750256845523,2.5)
--(axis cs:11.8759101630811,2.5)
--(axis cs:11.8759101630811,1.5)
--(axis cs:11.8428703436452,1.5)
--(axis cs:11.8428703436452,0.5)
--(axis cs:11.8900618195421,0.5)
--(axis cs:11.8900618195421,0)
--cycle;

\addplot [very thin, dimgrey102]
table {%
11.8 -0.5
11.8 12.5
};
\path [fill=steelblue31119180, fill opacity=0.75, line width=0pt]
(axis cs:12.7986905479156,0)
--(axis cs:12.8,0)
--(axis cs:12.8,0.5)
--(axis cs:12.8,0.5)
--(axis cs:12.8,1.5)
--(axis cs:12.8,1.5)
--(axis cs:12.8,2.5)
--(axis cs:12.8,2.5)
--(axis cs:12.8,3.5)
--(axis cs:12.8,3.5)
--(axis cs:12.8,4.5)
--(axis cs:12.8,4.5)
--(axis cs:12.8,5.5)
--(axis cs:12.8,5.5)
--(axis cs:12.8,6.5)
--(axis cs:12.8,6.5)
--(axis cs:12.8,7.5)
--(axis cs:12.8,7.5)
--(axis cs:12.8,8)
--(axis cs:12.7954869353112,8)
--(axis cs:12.7954869353112,8)
--(axis cs:12.7954869353112,7.5)
--(axis cs:12.7420475324204,7.5)
--(axis cs:12.7420475324204,6.5)
--(axis cs:12.38,6.5)
--(axis cs:12.38,5.5)
--(axis cs:12.6398832454945,5.5)
--(axis cs:12.6398832454945,4.5)
--(axis cs:12.725663412437,4.5)
--(axis cs:12.725663412437,3.5)
--(axis cs:12.7387972414116,3.5)
--(axis cs:12.7387972414116,2.5)
--(axis cs:12.7456196587446,2.5)
--(axis cs:12.7456196587446,1.5)
--(axis cs:12.793067028551,1.5)
--(axis cs:12.793067028551,0.5)
--(axis cs:12.7986905479156,0.5)
--(axis cs:12.7986905479156,0)
--cycle;

\addplot [very thin, dimgrey102]
table {%
12.8 -0.5
12.8 8.5
};
\path [fill=crimson2143940, fill opacity=0.55, line width=0pt]
(axis cs:12.8484036585321,0)
--(axis cs:12.8,0)
--(axis cs:12.8,0.5)
--(axis cs:12.8,0.5)
--(axis cs:12.8,1.5)
--(axis cs:12.8,1.5)
--(axis cs:12.8,2.5)
--(axis cs:12.8,2.5)
--(axis cs:12.8,3.5)
--(axis cs:12.8,3.5)
--(axis cs:12.8,4.5)
--(axis cs:12.8,4.5)
--(axis cs:12.8,5.5)
--(axis cs:12.8,5.5)
--(axis cs:12.8,6.5)
--(axis cs:12.8,6.5)
--(axis cs:12.8,7.5)
--(axis cs:12.8,7.5)
--(axis cs:12.8,8.5)
--(axis cs:12.8,8.5)
--(axis cs:12.8,9.5)
--(axis cs:12.8,9.5)
--(axis cs:12.8,10.5)
--(axis cs:12.8,10.5)
--(axis cs:12.8,11)
--(axis cs:12.8011020893791,11)
--(axis cs:12.8011020893791,11)
--(axis cs:12.8011020893791,10.5)
--(axis cs:12.8096299071962,10.5)
--(axis cs:12.8096299071962,9.5)
--(axis cs:12.8368343950255,9.5)
--(axis cs:12.8368343950255,8.5)
--(axis cs:12.9142989734964,8.5)
--(axis cs:12.9142989734964,7.5)
--(axis cs:12.9777975115586,7.5)
--(axis cs:12.9777975115586,6.5)
--(axis cs:12.9410273159144,6.5)
--(axis cs:12.9410273159144,5.5)
--(axis cs:12.9030520443835,5.5)
--(axis cs:12.9030520443835,4.5)
--(axis cs:12.8773495545794,4.5)
--(axis cs:12.8773495545794,3.5)
--(axis cs:12.8552315302316,3.5)
--(axis cs:12.8552315302316,2.5)
--(axis cs:12.8421054317283,2.5)
--(axis cs:12.8421054317283,1.5)
--(axis cs:12.8339119856889,1.5)
--(axis cs:12.8339119856889,0.5)
--(axis cs:12.8484036585321,0.5)
--(axis cs:12.8484036585321,0)
--cycle;

\addplot [very thin, dimgrey102]
table {%
12.8 -0.5
12.8 11.5
};
\addplot [line width=0.36pt, black]
table {%
-0.3 15
0 15
};
\addplot [line width=0.36pt, black]
table {%
0 15
0.3 15
};
\addplot [line width=0.36pt, black]
table {%
0.7 9
1 9
};
\addplot [line width=0.36pt, black]
table {%
1 9
1.3 9
};
\addplot [line width=0.36pt, black]
table {%
1.7 6
2 6
};
\addplot [line width=0.36pt, black]
table {%
2 6
2.3 6
};
\addplot [line width=0.36pt, black]
table {%
2.7 6
3 6
};
\addplot [line width=0.36pt, black]
table {%
3 6
3.3 6
};
\addplot [line width=0.36pt, black]
table {%
4.6 17
4.9 17
};
\addplot [line width=0.36pt, black]
table {%
4.9 17
5.2 17
};
\addplot [line width=0.36pt, black]
table {%
5.6 11
5.9 11
};
\addplot [line width=0.36pt, black]
table {%
5.9 11
6.2 11
};
\addplot [line width=0.36pt, black]
table {%
6.6 6
6.9 6
};
\addplot [line width=0.36pt, black]
table {%
6.9 6
7.2 6
};
\addplot [line width=0.36pt, black]
table {%
7.6 6
7.9 6
};
\addplot [line width=0.36pt, black]
table {%
7.9 6
8.2 6
};
\addplot [line width=0.36pt, black]
table {%
9.5 16
9.8 16
};
\addplot [line width=0.36pt, black]
table {%
9.8 16
10.1 16
};
\addplot [line width=0.36pt, black]
table {%
10.5 10
10.8 10
};
\addplot [line width=0.36pt, black]
table {%
10.8 10
11.1 10
};
\addplot [line width=0.36pt, black]
table {%
11.5 6
11.8 6
};
\addplot [line width=0.36pt, black]
table {%
11.8 6
12.1 6
};
\addplot [line width=0.36pt, black]
table {%
12.5 6
12.8 6
};
\addplot [line width=0.36pt, black]
table {%
12.8 6
13.1 6
};
\addplot [line width=0.32pt, gainsboro216]
table {%
3.95 -0.8
3.95 24.5
};
\addplot [line width=0.32pt, gainsboro216]
table {%
8.85 -0.8
8.85 24.5
};
\draw (axis cs:1.5,23.2) node[
  scale=0.45,
  anchor=base,
  text=black,
  rotate=0.0
]{TDL-A30};
\draw (axis cs:6.4,23.2) node[
  scale=0.45,
  anchor=base,
  text=black,
  rotate=0.0
]{TDL-B100};
\draw (axis cs:11.3,23.2) node[
  scale=0.45,
  anchor=base,
  text=black,
  rotate=0.0
]{TDL-C300};
\end{axis}

\end{tikzpicture}

%% file: figures/bler_grid.tex
% Auto-generated by export_olla_validation_figures.py
\definecolor{steelblue}{RGB}{31,119,180}
\definecolor{crimson}{RGB}{214,39,40}
\begin{tikzpicture}
\begin{groupplot}[
  group style={group size=4 by 3, horizontal sep=0.18cm,
    vertical sep=0.28cm, xlabels at=edge bottom,
    ylabels at=edge left, xticklabels at=edge bottom,
    yticklabels at=edge left},
  width=0.22\linewidth, height=0.14\linewidth,
  scale only axis,
  ymode=log, ymin=1e-3, ymax=1, xmin=-1, xmax=24,
  xtick={0,10,20}, ytick={1e-3,1e-2,1e-1,1},
  ymajorgrids, xmajorgrids, grid style={lightgray!60, line width=0.3pt},
  minor y tick num=0, tick align=inside, tick pos=left,
  xlabel={MCS (initial TX)},
  label style={font=\footnotesize}, title style={font=\footnotesize},
  tick label style={font=\scriptsize},
]
\nextgroupplot[title={16 dB}, ylabel={TDL-A30 BLER}, legend columns=2, legend to name=blerlegend, legend style={draw=lightgray, font=\footnotesize, /tikz/every even column/.append style={column sep=0.35cm}}]
\addplot[steelblue, opacity=0.9, mark=*, mark size=1.3pt, line width=1pt] coordinates {(8,0.0092025) (9,0.006263) (10,0.0051197) (11,0.0055765) (12,0.011149) (13,0.020856) (14,0.034338) (15,0.073351) (16,0.10112) (17,0.094706) (18,0.12166) (19,0.18113) (20,0.25567) (21,0.34519)};
\addplot[crimson, densely dashed, mark=square, opacity=0.9, mark size=1.2pt, line width=1pt] coordinates {(5,0.0016648) (6,0.0012) (7,0.02551) (8,0.0020321) (9,0.0094947) (10,0.0066124) (11,0.0056222) (12,0.016566) (13,0.016458) (14,0.032603) (15,0.075573) (16,0.10764) (17,0.088445) (18,0.14286) (19,0.17167) (20,0.19822) (21,0.25986) (22,0.55517)};
\addlegendimage{steelblue, mark=*, mark size=1.3pt, line width=1pt}
\addlegendentry{\gls{oal} (hardware)}
\addlegendimage{crimson, densely dashed, mark=square, mark size=1.2pt, line width=1pt}
\addlegendentry{\macemu twin}
\nextgroupplot[title={22 dB}]
\addplot[steelblue, opacity=0.9, mark=*, mark size=1.3pt, line width=1pt] coordinates {(1,0.0012) (2,0.0012) (3,0.0012) (4,0.0012) (5,0.0033423) (6,0.0034535) (7,0.013165) (8,0.049868) (9,0.10416) (10,0.050765) (11,0.07847) (12,0.14901) (13,0.23993) (14,0.40727) (15,0.55353)};
\addplot[crimson, densely dashed, mark=square, opacity=0.9, mark size=1.2pt, line width=1pt] coordinates {(1,0.0012) (2,0.0012) (3,0.002307) (4,0.0023456) (5,0.0020391) (6,0.0055238) (7,0.017056) (8,0.035246) (9,0.06396) (10,0.094936) (11,0.098524) (12,0.15753) (13,0.19655) (14,0.27917) (15,0.29936)};
\nextgroupplot[title={28 dB}]
\addplot[steelblue, opacity=0.9, mark=*, mark size=1.3pt, line width=1pt] coordinates {(0,0.01088) (1,0.0070469) (2,0.0045781) (3,0.0059199) (4,0.0080144) (5,0.013763) (6,0.033471) (7,0.096497) (8,0.27567) (9,0.49145) (10,0.35417)};
\addplot[crimson, densely dashed, mark=square, opacity=0.9, mark size=1.2pt, line width=1pt] coordinates {(0,0.0014587) (1,0.0025161) (2,0.0047629) (3,0.0085184) (4,0.017099) (5,0.030413) (6,0.055913) (7,0.10434) (8,0.18102) (9,0.19155) (10,0.28502) (11,0.28014)};
\nextgroupplot[title={34 dB}]
\addplot[steelblue, opacity=0.9, mark=*, mark size=1.3pt, line width=1pt] coordinates {(0,0.001938) (1,0.0021436) (2,0.0037191) (3,0.0081717) (4,0.017857) (5,0.029347) (6,0.062468) (7,0.15383) (8,0.60835) (9,0.50312)};
\addplot[crimson, densely dashed, mark=square, opacity=0.9, mark size=1.2pt, line width=1pt] coordinates {(0,0.0021111) (1,0.0025907) (2,0.0051145) (3,0.012463) (4,0.020841) (5,0.047076) (6,0.063007) (7,0.13019) (8,0.19323) (9,0.29323) (10,0.37062) (11,0.28517)};
\nextgroupplot[ylabel={TDL-B100 BLER}]
\addplot[steelblue, opacity=0.9, mark=*, mark size=1.3pt, line width=1pt] coordinates {(11,0.0053524) (12,0.0063291) (13,0.011682) (14,0.019943) (15,0.033822) (16,0.05098) (17,0.049186) (18,0.058177) (19,0.12683) (20,0.21001) (21,0.41108) (22,0.5906)};
\addplot[crimson, densely dashed, mark=square, opacity=0.9, mark size=1.2pt, line width=1pt] coordinates {(10,0.0012) (11,0.0020317) (12,0.0050023) (13,0.0052847) (14,0.0096533) (15,0.031275) (16,0.054163) (17,0.060075) (18,0.072961) (19,0.1374) (20,0.15585) (21,0.27001) (22,0.49633)};
\nextgroupplot[]
\addplot[steelblue, opacity=0.9, mark=*, mark size=1.3pt, line width=1pt] coordinates {(1,0.0012) (2,0.0039158) (3,0.0012) (4,0.0012) (5,0.0049111) (6,0.0020013) (7,0.004481) (8,0.021765) (9,0.054033) (10,0.014269) (11,0.02856) (12,0.089795) (13,0.16991) (14,0.31666) (15,0.54271) (16,0.83397)};
\addplot[crimson, densely dashed, mark=square, opacity=0.9, mark size=1.2pt, line width=1pt] coordinates {(5,0.0053738) (6,0.0019563) (7,0.0026099) (8,0.0089413) (9,0.032708) (10,0.048689) (11,0.06755) (12,0.097256) (13,0.14998) (14,0.2182) (15,0.44728)};
\nextgroupplot[]
\addplot[steelblue, opacity=0.9, mark=*, mark size=1.3pt, line width=1pt] coordinates {(0,0.0047357) (1,0.0012) (2,0.0051411) (3,0.0012745) (4,0.0026512) (5,0.0074016) (6,0.01791) (7,0.088079) (8,0.40498) (9,0.6472)};
\addplot[crimson, densely dashed, mark=square, opacity=0.9, mark size=1.2pt, line width=1pt] coordinates {(0,0.0012) (1,0.0012) (2,0.0028012) (3,0.0042622) (4,0.0056982) (5,0.018213) (6,0.033156) (7,0.090886) (8,0.18414) (9,0.34161) (10,0.46795)};
\nextgroupplot[]
\addplot[steelblue, opacity=0.9, mark=*, mark size=1.3pt, line width=1pt] coordinates {(0,0.0012) (1,0.0012) (2,0.0027807) (3,0.0048463) (4,0.014265) (5,0.028633) (6,0.062142) (7,0.28483) (8,0.89766)};
\addplot[crimson, densely dashed, mark=square, opacity=0.9, mark size=1.2pt, line width=1pt] coordinates {(0,0.0013524) (1,0.0012) (2,0.0017447) (3,0.0068311) (4,0.0079454) (5,0.017458) (6,0.049123) (7,0.10606) (8,0.22635) (9,0.42604) (10,0.39944)};
\nextgroupplot[ylabel={TDL-C300 BLER}]
\addplot[steelblue, opacity=0.9, mark=*, mark size=1.3pt, line width=1pt] coordinates {(8,0.0083632) (10,0.01084) (11,0.013194) (12,0.017974) (13,0.023232) (14,0.031682) (15,0.05962) (16,0.071486) (17,0.064372) (18,0.11145) (19,0.187) (20,0.28564) (21,0.38751)};
\addplot[crimson, densely dashed, mark=square, opacity=0.9, mark size=1.2pt, line width=1pt] coordinates {(2,0.011681) (5,0.0038314) (6,0.0083901) (7,0.035127) (8,0.011041) (9,0.0095271) (10,0.018632) (11,0.024058) (12,0.025173) (13,0.032637) (14,0.026515) (15,0.045362) (16,0.060587) (17,0.075704) (18,0.096681) (19,0.13537) (20,0.24827) (21,0.4628)};
\nextgroupplot[]
\addplot[steelblue, opacity=0.9, mark=*, mark size=1.3pt, line width=1pt] coordinates {(3,0.0012) (4,0.0017319) (5,0.0018442) (6,0.015433) (7,0.013826) (8,0.024924) (9,0.076061) (10,0.030771) (11,0.052405) (12,0.10584) (13,0.21606) (14,0.36108)};
\addplot[crimson, densely dashed, mark=square, opacity=0.9, mark size=1.2pt, line width=1pt] coordinates {(0,0.022008) (1,0.011033) (2,0.010485) (3,0.02165) (4,0.019657) (5,0.024405) (6,0.020768) (7,0.021453) (8,0.025947) (9,0.036271) (10,0.056959) (11,0.065187) (12,0.10256) (13,0.15445) (14,0.22102) (15,0.4707) (16,0.39869)};
\nextgroupplot[]
\addplot[steelblue, opacity=0.9, mark=*, mark size=1.3pt, line width=1pt] coordinates {(0,0.013106) (1,0.0097548) (2,0.0059026) (3,0.0089286) (4,0.0089728) (5,0.015571) (6,0.030266) (7,0.11942) (8,0.40146) (9,0.6648)};
\addplot[crimson, densely dashed, mark=square, opacity=0.9, mark size=1.2pt, line width=1pt] coordinates {(0,0.017186) (1,0.016389) (2,0.022813) (3,0.022779) (4,0.024703) (5,0.021688) (6,0.030232) (7,0.054175) (8,0.12722) (9,0.22339) (10,0.30315) (11,0.24045) (12,0.474)};
\nextgroupplot[]
\addplot[steelblue, opacity=0.9, mark=*, mark size=1.3pt, line width=1pt] coordinates {(0,0.0018762) (1,0.020907) (2,0.007635) (3,0.011962) (4,0.026373) (5,0.041888) (6,0.064285) (7,0.31616) (8,0.87534)};
\addplot[crimson, densely dashed, mark=square, opacity=0.9, mark size=1.2pt, line width=1pt] coordinates {(0,0.022299) (1,0.023388) (2,0.018551) (3,0.026371) (4,0.018554) (5,0.02488) (6,0.037253) (7,0.070629) (8,0.1533) (9,0.27386) (10,0.38028) (11,0.51456)};
\end{groupplot}
\node[anchor=south] at ($(group c1r1.north west)!0.5!(group c4r1.north east)+(0,0.55cm)$) {\ref{blerlegend}};
\end{tikzpicture}

%% file: figures/macemu_throughput.tex
% Auto-generated by figures/scripts/plot_macemu_throughput.py
% Per-bin aggregate throughput (DL and UL), mean +/- 95% CI over 300 bins.
\begin{tikzpicture}
\definecolor{macgdl}{RGB}{46,134,171}
\definecolor{macgul}{RGB}{247,127,0}
\begin{axis}[
  width=\fwidth, height=\fheight,
  xlabel={Number of UEs},
  ylabel={Cell throughput (Mbps)},
  xmode=log, log basis x=10,
  xmin=0.85, xmax=105,
  xtick={1,2,5,10,20,40,90},
  xticklabels={1,2,5,10,20,40,90},
  ymin=0, ymax=130,
  ytick={0,30,60,90,120},
  yticklabel={\pgfmathprintnumber[fixed,precision=0]{\tick}},
  scaled y ticks=false,
  grid=major, grid style={dotted, gray!40},
  legend style={at={(0.97,0.5)}, anchor=east, draw=gray!40, fill=white, fill opacity=0.9, text opacity=1, font=\small},
  legend cell align=left,
  tick align=inside, tick pos=left,
]
\addplot[
  color=macgdl, opacity=0.9, mark=*, mark size=1.6pt,
  line width=1pt,
  error bars/.cd, y dir=both, y explicit,
  error bar style={line width=0.65pt, macgdl},
  error mark options={mark size=1.8pt, line width=0.65pt},
]
table[x=N, y=dl_mean, y error=dl_ci, col sep=comma] {
N,dl_mean,dl_ci
1,84.817,2.567
2,115.111,2.912
5,115.639,1.038
10,115.359,0.456
20,115.623,0.265
30,115.798,0.262
40,115.449,0.238
50,115.331,0.242
60,115.515,0.239
70,115.810,0.176
80,115.917,0.162
90,116.010,0.165
};
\addlegendentry{DL aggregate}
\addplot[
  color=macgul, opacity=0.9, mark=square*, mark size=1.6pt,
  line width=1pt,
  error bars/.cd, y dir=both, y explicit,
  error bar style={line width=0.65pt, macgul},
  error mark options={mark size=1.8pt, line width=0.65pt},
]
table[x=N, y=ul_mean, y error=ul_ci, col sep=comma] {
N,ul_mean,ul_ci
1,32.084,0.808
2,32.287,0.467
5,32.158,0.365
10,31.224,0.324
20,31.682,0.233
30,31.486,0.220
40,31.073,0.208
50,30.471,0.202
60,30.236,0.206
70,30.437,0.232
80,30.769,0.215
90,30.587,0.212
};
\addlegendentry{UL aggregate}
\end{axis}
\end{tikzpicture}

%% file: figures/scaling_stage.tex
% This file was created with tikzplotlib v0.10.1.post13.
\begin{tikzpicture}

\definecolor{cadetblue104169194}{RGB}{104,169,194}
\definecolor{darkorange2411431}{RGB}{241,143,1}
\definecolor{grey142}{RGB}{142,142,142}
\definecolor{lightslategrey111138183}{RGB}{111,138,183}
\definecolor{mediumvioletred16259114}{RGB}{162,59,114}
\definecolor{steelblue46134171}{RGB}{46,134,171}

\begin{axis}[
height=0.5\linewidth,
legend cell align={left},
legend columns=2,
legend style={
  fill opacity=0.8,
  draw opacity=1,
  text opacity=1,
  at={(0.03,0.97)},
  anchor=north west,
  draw=none
},
log basis x={10},
tick align=inside,
tick pos=left,
width=\linewidth,
x grid style={lightgray},
xlabel={Concurrent UEs},
xmajorgrids,
xmin=1, xmax=90,
xmode=log,
xtick={1,2,5,10,20,40,90},
xticklabels={1,2,5,10,20,40,90},
log ticks with fixed point,
minor xtick={},
xtick style={color=black},
y grid style={lightgray},
ylabel={Mean exec.\ time (\(\displaystyle \mu\)s/slot)},
ymajorgrids,
ymin=0, ymax=394.13535,
ytick style={color=black}
]
\path [fill=steelblue46134171, fill opacity=0.78]
(axis cs:1,12.233)
--(axis cs:1,0)
--(axis cs:2,0)
--(axis cs:5,0)
--(axis cs:10,0)
--(axis cs:20,0)
--(axis cs:30,0)
--(axis cs:40,0)
--(axis cs:50,0)
--(axis cs:60,0)
--(axis cs:70,0)
--(axis cs:80,0)
--(axis cs:90,0)
--(axis cs:90,67.967)
--(axis cs:90,67.967)
--(axis cs:80,66)
--(axis cs:70,64.367)
--(axis cs:60,61.767)
--(axis cs:50,61)
--(axis cs:40,58.467)
--(axis cs:30,54.4)
--(axis cs:20,52.067)
--(axis cs:10,47.967)
--(axis cs:5,42.867)
--(axis cs:2,25.667)
--(axis cs:1,12.233)
--cycle;
\addlegendimage{area legend, fill=steelblue46134171, fill opacity=0.78}
\addlegendentry{UL receive}

\path [fill=cadetblue104169194, fill opacity=0.78]
(axis cs:1,12.233)
--(axis cs:1,12.233)
--(axis cs:2,25.667)
--(axis cs:5,42.867)
--(axis cs:10,47.967)
--(axis cs:20,52.067)
--(axis cs:30,54.4)
--(axis cs:40,58.467)
--(axis cs:50,61)
--(axis cs:60,61.767)
--(axis cs:70,64.367)
--(axis cs:80,66)
--(axis cs:90,67.967)
--(axis cs:90,67.967)
--(axis cs:90,67.967)
--(axis cs:80,66)
--(axis cs:70,64.367)
--(axis cs:60,61.767)
--(axis cs:50,61)
--(axis cs:40,58.467)
--(axis cs:30,54.4)
--(axis cs:20,52.067)
--(axis cs:10,47.967)
--(axis cs:5,42.867)
--(axis cs:2,25.667)
--(axis cs:1,12.233)
--cycle;
\addlegendimage{area legend, fill=cadetblue104169194, fill opacity=0.78}
\addlegendentry{HARQ cleanup}

\path [fill=mediumvioletred16259114, fill opacity=0.78]
(axis cs:1,12.233)
--(axis cs:1,12.233)
--(axis cs:2,25.667)
--(axis cs:5,42.867)
--(axis cs:10,47.967)
--(axis cs:20,52.067)
--(axis cs:30,54.4)
--(axis cs:40,58.467)
--(axis cs:50,61)
--(axis cs:60,61.767)
--(axis cs:70,64.367)
--(axis cs:80,66)
--(axis cs:90,67.967)
--(axis cs:90,68.967)
--(axis cs:90,68.967)
--(axis cs:80,67)
--(axis cs:70,65.367)
--(axis cs:60,62.767)
--(axis cs:50,61)
--(axis cs:40,58.467)
--(axis cs:30,54.4)
--(axis cs:20,52.067)
--(axis cs:10,47.967)
--(axis cs:5,42.867)
--(axis cs:2,25.667)
--(axis cs:1,12.233)
--cycle;
\addlegendimage{area legend, fill=mediumvioletred16259114, fill opacity=0.78}
\addlegendentry{ACK/NACK}

\path [fill=darkorange2411431, fill opacity=0.78]
(axis cs:1,14.8)
--(axis cs:1,12.233)
--(axis cs:2,25.667)
--(axis cs:5,42.867)
--(axis cs:10,47.967)
--(axis cs:20,52.067)
--(axis cs:30,54.4)
--(axis cs:40,58.467)
--(axis cs:50,61)
--(axis cs:60,62.767)
--(axis cs:70,65.367)
--(axis cs:80,67)
--(axis cs:90,68.967)
--(axis cs:90,193)
--(axis cs:90,193)
--(axis cs:80,177.1)
--(axis cs:70,160.6)
--(axis cs:60,142.967)
--(axis cs:50,128.1)
--(axis cs:40,112.534)
--(axis cs:30,95.533)
--(axis cs:20,80.8)
--(axis cs:10,63.567)
--(axis cs:5,52.934)
--(axis cs:2,31.667)
--(axis cs:1,14.8)
--cycle;
\addlegendimage{area legend, fill=darkorange2411431, fill opacity=0.78}
\addlegendentry{Scheduler}

\path [fill=lightslategrey111138183, fill opacity=0.78]
(axis cs:1,17.8)
--(axis cs:1,14.8)
--(axis cs:2,31.667)
--(axis cs:5,52.934)
--(axis cs:10,63.567)
--(axis cs:20,80.8)
--(axis cs:30,95.533)
--(axis cs:40,112.534)
--(axis cs:50,128.1)
--(axis cs:60,142.967)
--(axis cs:70,160.6)
--(axis cs:80,177.1)
--(axis cs:90,193)
--(axis cs:90,229)
--(axis cs:90,229)
--(axis cs:80,211.267)
--(axis cs:70,192.567)
--(axis cs:60,173.534)
--(axis cs:50,155.1)
--(axis cs:40,135.534)
--(axis cs:30,114.566)
--(axis cs:20,95.8)
--(axis cs:10,73.567)
--(axis cs:5,60.634)
--(axis cs:2,36.667)
--(axis cs:1,17.8)
--cycle;
\addlegendimage{area legend, fill=lightslategrey111138183, fill opacity=0.78}
\addlegendentry{DL send}

\path [fill=grey142, fill opacity=0.78]
(axis cs:1,19.8)
--(axis cs:1,17.8)
--(axis cs:2,36.667)
--(axis cs:5,60.634)
--(axis cs:10,73.567)
--(axis cs:20,95.8)
--(axis cs:30,114.566)
--(axis cs:40,135.534)
--(axis cs:50,155.1)
--(axis cs:60,173.534)
--(axis cs:70,192.567)
--(axis cs:80,211.267)
--(axis cs:90,229)
--(axis cs:90,375.367)
--(axis cs:90,375.367)
--(axis cs:80,341.2)
--(axis cs:70,307.5)
--(axis cs:60,268.601)
--(axis cs:50,233.267)
--(axis cs:40,196.534)
--(axis cs:30,158.699)
--(axis cs:20,123.533)
--(axis cs:10,85.567)
--(axis cs:5,65.667)
--(axis cs:2,39.3)
--(axis cs:1,19.8)
--cycle;
\addlegendimage{area legend, fill=grey142, fill opacity=0.78}
\addlegendentry{Slot tick}

\end{axis}

\end{tikzpicture}

%% file: figures/scaling_violation.tex
% This file was created with tikzplotlib v0.10.1.post13.
\begin{tikzpicture}

\definecolor{mediumvioletred16259114}{RGB}{162,59,114}

\begin{axis}[
height=0.5\linewidth,
log basis x={10},
log basis y={10},
tick align=inside,
tick pos=left,
width=\linewidth,
x grid style={lightgray},
xlabel={Concurrent UEs},
xmajorgrids,
xmin=0.85, xmax=105,
xmode=log,
xtick={1,2,5,10,20,40,90},
xticklabels={1,2,5,10,20,40,90},
log ticks with fixed point,
minor xtick={},
xtick style={color=black},
y grid style={lightgray},
ylabel={Deadline-violation prob.},
ymajorgrids,
ymin=1e-05, ymax=0.005,
ymode=log,
ytick style={color=black},
ytick={1e-05,0.0001,0.001},
yticklabels={
  \(\displaystyle {10^{-5}}\),
  \(\displaystyle {10^{-4}}\),
  \(\displaystyle {10^{-3}}\)
}
]
\draw[draw=black,fill=mediumvioletred16259114,opacity=0.72,line width=0.18pt,postaction={pattern=north east lines, fill opacity=0.72}] (axis cs:1.82,1e-05) rectangle (axis cs:2.18,1.667e-05);
\draw[draw=black,fill=mediumvioletred16259114,opacity=0.72,line width=0.18pt,postaction={pattern=north east lines, fill opacity=0.72}] (axis cs:4.55,1e-05) rectangle (axis cs:5.45,8.333e-05);
\draw[draw=black,fill=mediumvioletred16259114,opacity=0.72,line width=0.18pt,postaction={pattern=north east lines, fill opacity=0.72}] (axis cs:9.1,1e-05) rectangle (axis cs:10.9,5e-05);
\draw[draw=black,fill=mediumvioletred16259114,opacity=0.72,line width=0.18pt,postaction={pattern=north east lines, fill opacity=0.72}] (axis cs:18.2,1e-05) rectangle (axis cs:21.8,0.00011667);
\draw[draw=black,fill=mediumvioletred16259114,opacity=0.72,line width=0.18pt,postaction={pattern=north east lines, fill opacity=0.72}] (axis cs:27.3,1e-05) rectangle (axis cs:32.7,0.00036667);
\draw[draw=black,fill=mediumvioletred16259114,opacity=0.72,line width=0.18pt,postaction={pattern=north east lines, fill opacity=0.72}] (axis cs:36.4,1e-05) rectangle (axis cs:43.6,3.333e-05);
\draw[draw=black,fill=mediumvioletred16259114,opacity=0.72,line width=0.18pt,postaction={pattern=north east lines, fill opacity=0.72}] (axis cs:45.5,1e-05) rectangle (axis cs:54.5,0.00266667);
\draw[draw=black,fill=mediumvioletred16259114,opacity=0.72,line width=0.18pt,postaction={pattern=north east lines, fill opacity=0.72}] (axis cs:54.6,1e-05) rectangle (axis cs:65.4,0.0002);
\draw[draw=black,fill=mediumvioletred16259114,opacity=0.72,line width=0.18pt,postaction={pattern=north east lines, fill opacity=0.72}] (axis cs:81.9,1e-05) rectangle (axis cs:98.1,0.00091667);
\end{axis}

\end{tikzpicture}

%% file: figures/schedReq.tex
\begin{tikzpicture}[
    >={Stealth[length=2.5mm]},
    lifeline/.style={dashed, gray!70},
    msg/.style={->, thick},
    font=\small
  ]
  % --- geometry ---
  \def\ue{0}      % UE lifeline x
  \def\gnb{.8\linewidth}     % gNB lifeline x
  \def\ytop{0}    % top (actor boxes)
  \def\ybot{-2.9} % bottom of lifelines

  % --- actor heads ---
  \node[draw, fill=blue!8, minimum width=1.4cm, minimum height=6mm] (UE)  at (\ue,\ytop)  {UE};
  \node[draw, fill=blue!8, minimum width=1.4cm, minimum height=6mm] (GNB) at (\gnb,\ytop) {gNB};

  % --- lifelines ---
  \draw[lifeline] (UE.south)  -- (\ue,\ybot);
  \draw[lifeline] (GNB.south) -- (\gnb,\ybot);

  % --- messages (y decreases with time) ---
  % 1: UE -> gNB
  \draw[msg] (\ue,-0.6) -- node[above, midway] {SR tx (1 bit), predefined periodicity} (\gnb,-0.6);
  % 2: gNB -> UE
  \draw[msg] (\gnb,-1.3) -- node[above, midway] {UL min grant} (\ue,-1.3);
  % 3: UE -> gNB
  \draw[msg] (\ue,-2.0) -- node[above, midway] {BSR (+ optional small SDU if capacity)} (\gnb,-2.0);
  % 4: gNB -> UE
  \draw[msg] (\gnb,-2.7) -- node[above, midway] {Actual UL grant} (\ue,-2.7);

\end{tikzpicture}

%% file: figures/baseline.tex
% This file was created with tikzplotlib v0.10.1.post13.
\begin{tikzpicture}[
scale=0.9
]

\definecolor{darkgrey176}{RGB}{176,176,176}
\definecolor{dimgrey102}{RGB}{102,102,102}
\definecolor{lightgrey204}{RGB}{204,204,204}
\definecolor{steelblue46134171}{RGB}{46,134,171}

\begin{axis}[
width=1.1\linewidth,
height=0.4\linewidth,
grid=major,
legend cell align={left},
legend style={
  fill opacity=0.8,
  draw opacity=1,
  text opacity=1,
  at={(0.97,0.03)},
  anchor=south east,
  draw=lightgrey204
},
thick,
tick align=inside,
tick pos=left,
x grid style={darkgrey176},
xlabel={RTT (ms)},
xmajorgrids,
xmin=0, xmax=40,
xtick style={color=black},
y grid style={darkgrey176},
ylabel={},
ymajorgrids,
ymin=0, ymax=1,
ytick style={color=black}
]
\addplot [line width=1pt, dimgrey102, opacity=0.9]
table {%
6.88 0
6.88 0.001
6.92 0.014
6.94 0.028
6.96 0.042
7.38 0.056
7.88 0.07
7.9 0.084
7.91 0.098
7.93 0.111
7.99 0.125
8.36 0.139
8.4 0.153
8.91 0.167
8.93 0.181
8.95 0.195
9.34 0.209
9.37 0.222
9.39 0.236
9.4 0.25
9.44 0.264
15 0.278
15.5 0.292
15.9 0.306
16 0.32
16.9 0.333
17.1 0.347
17.9 0.361
18.1 0.375
18.9 0.389
19.9 0.403
19.9 0.417
19.9 0.431
19.9 0.444
19.9 0.458
19.9 0.472
20 0.486
20.1 0.5
20.4 0.514
20.5 0.528
20.9 0.542
20.9 0.555
20.9 0.569
20.9 0.583
20.9 0.597
21 0.611
21.4 0.625
21.9 0.639
21.9 0.653
21.9 0.666
21.9 0.68
21.9 0.694
22.7 0.708
22.9 0.722
22.9 0.736
22.9 0.75
22.9 0.764
23 0.777
23.6 0.791
23.9 0.805
23.9 0.819
23.9 0.833
23.9 0.847
24 0.861
24.8 0.875
24.9 0.888
24.9 0.902
24.9 0.916
25.8 0.93
25.9 0.944
26.9 0.958
28.4 0.972
28.9 0.986
32.9 0.999
33.5 1
};
\addlegendentry{Default SR-based scheduling}
\addplot [line width=1pt, steelblue46134171, opacity=0.9, dashed]
table {%
6.4 0
6.4 0.001
6.9 0.014
6.91 0.028
6.92 0.042
6.93 0.056
6.94 0.07
7.37 0.084
7.41 0.098
7.43 0.111
7.47 0.125
7.87 0.139
7.88 0.153
7.89 0.167
7.9 0.181
7.9 0.195
7.91 0.209
7.91 0.222
7.92 0.236
7.92 0.25
7.93 0.264
7.94 0.278
7.95 0.292
8.44 0.306
8.89 0.32
8.9 0.333
8.9 0.347
8.91 0.361
8.92 0.375
8.92 0.389
8.93 0.403
8.94 0.417
8.96 0.431
9.86 0.444
9.88 0.458
9.89 0.472
9.9 0.486
9.9 0.5
9.9 0.514
9.91 0.528
9.91 0.542
9.92 0.555
9.92 0.569
9.93 0.583
9.93 0.597
9.94 0.611
9.98 0.625
10.9 0.639
10.9 0.653
10.9 0.666
10.9 0.68
10.9 0.694
10.9 0.708
10.9 0.722
10.9 0.736
10.9 0.75
10.9 0.764
10.9 0.777
10.9 0.791
11.2 0.805
11.4 0.819
11.4 0.833
11.9 0.847
11.9 0.861
11.9 0.875
11.9 0.888
11.9 0.902
13.9 0.916
13.9 0.93
13.9 0.944
13.9 0.958
15.7 0.972
16 0.986
18.9 0.999
18.9 1
};
\addlegendentry{Always scheduling in UL}
\end{axis}

\end{tikzpicture}

%% file: figures/backward_lookup.tex
% This file was created with tikzplotlib v0.10.1.post13.
\begin{tikzpicture}[
scale=0.9
]

\definecolor{darkgrey176}{RGB}{176,176,176}
\definecolor{darkcyan6167125}{RGB}{6,167,125}
\definecolor{darkorange2471270}{RGB}{247,127,0}
\definecolor{lightgrey204}{RGB}{204,204,204}
\definecolor{mediumpurple148103189}{RGB}{148,103,189}
\definecolor{steelblue46134171}{RGB}{46,134,171}

\begin{axis}[
grid=major,
height=.6\linewidth,
width=\linewidth,
legend cell align={left},
legend style={
  fill opacity=0.8,
  draw opacity=1,
  text opacity=1,
  at={(0.97,0.03)},
  anchor=south east,
  draw=lightgrey204
},
thick,
tick align=inside,
tick pos=left,
x grid style={darkgrey176},
xlabel={RTT (ms)},
xmajorgrids,
xmin=0, xmax=40,
xtick style={color=black},
y grid style={darkgrey176},
ylabel={CDF},
ymajorgrids,
ymin=0, ymax=1,
ytick style={color=black}
]
\addplot [line width=1pt, steelblue46134171, opacity=0.9]
table {%
6.88 0
6.88 0.001
6.9 0.014
6.92 0.028
6.93 0.042
6.94 0.056
6.95 0.07
7.38 0.084
7.42 0.098
7.44 0.111
7.85 0.125
7.88 0.139
7.89 0.153
7.9 0.167
7.91 0.181
7.92 0.195
7.92 0.209
7.93 0.222
7.93 0.236
7.94 0.25
7.95 0.264
8.29 0.278
8.33 0.292
8.35 0.306
8.36 0.32
8.37 0.333
8.39 0.347
8.89 0.361
8.91 0.375
8.93 0.389
8.94 0.403
9.08 0.417
9.4 0.431
9.86 0.444
9.89 0.458
9.91 0.472
9.92 0.486
9.94 0.5
10.3 0.514
10.4 0.528
10.4 0.542
10.4 0.555
10.9 0.569
13.4 0.583
19.9 0.597
20 0.611
20.4 0.625
20.5 0.639
20.5 0.653
20.5 0.666
20.5 0.68
20.9 0.694
20.9 0.708
20.9 0.722
20.9 0.736
21 0.75
21.1 0.764
21.4 0.777
21.5 0.791
23.9 0.805
23.9 0.819
23.9 0.833
23.9 0.847
23.9 0.861
24 0.875
24.8 0.888
24.9 0.902
24.9 0.916
24.9 0.93
24.9 0.944
25.5 0.958
26.5 0.972
28.4 0.986
56.9 0.999
83.9 1
};
\addlegendentry{1 subframe backward}
\addplot [line width=1pt, darkorange2471270, opacity=0.9, dashed]
table {%
6.46 0
6.46 0.001
6.9 0.014
6.91 0.028
6.92 0.042
6.93 0.056
6.95 0.07
7.04 0.084
7.39 0.098
7.4 0.111
7.42 0.125
7.44 0.139
7.47 0.153
7.88 0.167
7.89 0.181
7.9 0.195
7.9 0.209
7.91 0.222
7.92 0.236
7.92 0.25
7.93 0.264
7.94 0.278
8.4 0.292
8.45 0.306
8.89 0.32
8.9 0.333
8.9 0.347
8.91 0.361
8.91 0.375
8.92 0.389
8.93 0.403
8.94 0.417
9.32 0.431
9.86 0.444
9.89 0.458
9.89 0.472
9.9 0.486
9.91 0.5
9.91 0.514
9.91 0.528
9.92 0.542
9.92 0.555
9.93 0.569
9.94 0.583
9.95 0.597
10.4 0.611
10.4 0.625
10.9 0.639
10.9 0.653
10.9 0.666
10.9 0.68
10.9 0.694
10.9 0.708
10.9 0.722
10.9 0.736
10.9 0.75
10.9 0.764
11 0.777
11.4 0.791
11.4 0.805
11.8 0.819
11.9 0.833
11.9 0.847
11.9 0.861
12.9 0.875
12.9 0.888
13.9 0.902
13.9 0.916
13.9 0.93
13.9 0.944
14.5 0.958
15.9 0.972
16.9 0.986
57.9 0.999
61.3 1
};
\addlegendentry{2 subframes backward}
\addplot [line width=1pt, darkcyan6167125, opacity=0.9, dash pattern=on 1pt off 3pt on 3pt off 3pt]
table {%
6.42 0
6.42 0.001
6.9 0.014
6.92 0.028
6.92 0.042
6.93 0.056
6.93 0.07
6.94 0.084
6.94 0.098
6.96 0.111
7.42 0.125
7.43 0.139
7.87 0.153
7.89 0.167
7.9 0.181
7.91 0.195
7.92 0.209
7.92 0.222
7.92 0.236
7.93 0.25
7.93 0.264
7.94 0.278
8.15 0.292
8.44 0.306
8.89 0.32
8.9 0.333
8.91 0.347
8.92 0.361
8.92 0.375
8.93 0.389
8.93 0.403
8.93 0.417
8.94 0.431
8.94 0.444
8.95 0.458
8.96 0.472
9.88 0.486
9.9 0.5
9.91 0.514
9.91 0.528
9.92 0.542
9.92 0.555
9.92 0.569
9.93 0.583
9.93 0.597
9.93 0.611
9.94 0.625
9.94 0.639
9.95 0.653
10.7 0.666
10.9 0.68
10.9 0.694
10.9 0.708
10.9 0.722
10.9 0.736
10.9 0.75
10.9 0.764
10.9 0.777
10.9 0.791
10.9 0.805
11 0.819
11 0.833
11.4 0.847
11.9 0.861
11.9 0.875
11.9 0.888
12.4 0.902
12.9 0.916
12.9 0.93
13.9 0.944
14.9 0.958
15.9 0.972
16.9 0.986
18 0.999
18.9 1
};
\addlegendentry{3 subframes backward}
\addplot [line width=1pt, mediumpurple148103189, opacity=0.9, dotted]
table {%
6.41 0
6.41 0.001
6.9 0.014
6.91 0.028
6.92 0.042
6.92 0.056
6.93 0.07
6.93 0.084
6.94 0.098
6.95 0.111
7.41 0.125
7.43 0.139
7.88 0.153
7.9 0.167
7.91 0.181
7.91 0.195
7.92 0.209
7.92 0.222
7.93 0.236
7.93 0.25
7.94 0.264
7.96 0.278
8.43 0.292
8.89 0.306
8.91 0.32
8.91 0.333
8.92 0.347
8.92 0.361
8.93 0.375
8.93 0.389
8.93 0.403
8.94 0.417
8.94 0.431
8.95 0.444
8.96 0.458
8.98 0.472
9.88 0.486
9.9 0.5
9.91 0.514
9.91 0.528
9.92 0.542
9.92 0.555
9.92 0.569
9.93 0.583
9.93 0.597
9.93 0.611
9.94 0.625
9.95 0.639
9.96 0.653
10.4 0.666
10.9 0.68
10.9 0.694
10.9 0.708
10.9 0.722
10.9 0.736
10.9 0.75
10.9 0.764
10.9 0.777
10.9 0.791
10.9 0.805
11 0.819
11 0.833
11.4 0.847
11.9 0.861
11.9 0.875
11.9 0.888
12.9 0.902
12.9 0.916
13.9 0.93
14.9 0.944
16.4 0.958
16.9 0.972
17.9 0.986
21.9 0.999
32.4 1
};
\addlegendentry{4 subframes backward}
\end{axis}

\end{tikzpicture}

%% file: figures/ewma_comparison_all.tex
% This file was created with tikzplotlib v0.10.1.post13.
\begin{tikzpicture}[
scale=0.9
]

\definecolor{darkcyan6167125}{RGB}{6,167,125}
\definecolor{darkgrey176}{RGB}{176,176,176}
\definecolor{darkorange2471270}{RGB}{247,127,0}
\definecolor{dimgrey102}{RGB}{102,102,102}
\definecolor{lightgrey204}{RGB}{204,204,204}
\definecolor{saddlebrown}{RGB}{139,69,19}
\definecolor{steelblue46134171}{RGB}{46,134,171}

\begin{axis}[
width=1.1\linewidth,
height=0.4\linewidth,
grid=major,
legend cell align={left},
legend style={
  fill opacity=0.8,
  draw opacity=1,
  text opacity=1,
  at={(0.5,1.02)},
  anchor=south,
  draw=lightgrey204,
  legend columns=3,    % Move this INSIDE the legend style block
  column sep=0.12cm,
  row sep=-0.5pt,
  font=\footnotesize,
  inner xsep=3pt,
  inner ysep=1pt
},
thick,
tick align=inside,
tick pos=left,
x grid style={darkgrey176},
xlabel={RTT (ms)},
xmajorgrids,
xmin=0, xmax=40,
xtick style={color=black},
y grid style={darkgrey176},
ylabel={},
ymajorgrids,
ymin=0, ymax=1,
ytick style={color=black}
]
\addplot [line width=1pt, dimgrey102, opacity=0.9]
table {%
8.89 0
8.89 0.00100502512562814
14.8 0.014070351758794
14.9 0.0281407035175879
14.9 0.0422110552763819
14.9 0.0562814070351759
14.9 0.0703517587939698
14.9 0.0834170854271357
14.9 0.0974874371859296
14.9 0.111557788944724
14.9 0.125628140703518
14.9 0.139698492462312
15 0.152763819095477
15.8 0.166834170854271
15.8 0.180904522613065
15.9 0.194974874371859
15.9 0.209045226130653
15.9 0.222110552763819
15.9 0.236180904522613
15.9 0.250251256281407
15.9 0.264321608040201
15.9 0.278391959798995
15.9 0.291457286432161
16 0.305527638190955
16.6 0.319597989949749
16.8 0.333668341708543
16.9 0.347738693467337
16.9 0.360804020100502
16.9 0.374874371859297
16.9 0.38894472361809
16.9 0.403015075376884
16.9 0.417085427135678
16.9 0.430150753768844
16.9 0.444221105527638
17.1 0.458291457286432
17.8 0.472361809045226
17.9 0.48643216080402
17.9 0.499497487437186
17.9 0.51356783919598
17.9 0.527638190954774
17.9 0.541708542713568
17.9 0.555778894472362
17.9 0.569849246231156
17.9 0.582914572864322
17.9 0.596984924623116
18.8 0.61105527638191
18.8 0.625125628140703
18.9 0.639195979899498
18.9 0.652261306532663
18.9 0.666331658291457
18.9 0.680402010050251
18.9 0.694472361809045
18.9 0.708542713567839
18.9 0.721608040201005
18.9 0.735678391959799
19 0.749748743718593
19.9 0.763819095477387
19.9 0.777889447236181
20.8 0.790954773869347
20.9 0.805025125628141
21 0.819095477386935
21.9 0.833165829145729
21.9 0.847236180904523
21.9 0.860301507537688
22.8 0.874371859296482
22.9 0.888442211055276
22.9 0.90251256281407
23.8 0.916582914572864
23.9 0.92964824120603
23.9 0.943718592964824
27 0.957788944723618
148 0.971859296482412
1238 0.985929648241206
1915 0.998994974874372
2008 1
};
\addlegendentry{Baseline (Reactive)}
\addplot [thick, saddlebrown, opacity=0.8]
table {%
7.25 0
7.25 0.001
7.8 0.014
8.15 0.028
8.77 0.042
8.94 0.056
9.54 0.07
9.58 0.084
9.61 0.098
9.63 0.111
9.65 0.125
9.68 0.139
9.69 0.153
9.71 0.167
9.74 0.181
9.82 0.195
9.83 0.209
9.84 0.222
9.84 0.236
9.85 0.25
9.85 0.264
9.86 0.278
9.88 0.292
9.89 0.306
9.89 0.32
9.9 0.333
9.9 0.347
9.9 0.361
9.9 0.375
9.91 0.389
9.91 0.403
9.92 0.417
9.92 0.431
9.92 0.444
9.92 0.458
9.93 0.472
9.93 0.486
9.93 0.5
9.94 0.514
9.94 0.528
9.94 0.542
9.94 0.555
9.94 0.569
9.94 0.583
9.95 0.597
9.95 0.611
9.95 0.625
9.95 0.639
9.96 0.653
9.96 0.666
9.96 0.68
9.96 0.694
9.96 0.708
9.96 0.722
9.97 0.736
9.97 0.75
9.97 0.764
9.97 0.777
9.97 0.791
9.98 0.805
9.98 0.819
9.98 0.833
9.98 0.847
9.98 0.861
9.99 0.875
9.99 0.888
10 0.902
10.8 0.916
11.9 0.93
14 0.944
14.9 0.958
15 0.972
16.9 0.986
17.9 0.999
18.1 1
};
\addlegendentry{TW=0 N=1}
\addplot [thick, saddlebrown, opacity=0.8, dashed]
table {%
6.51 0
6.51 0.001
7.2 0.014
7.66 0.028
7.98 0.042
8.39 0.056
8.8 0.07
8.98 0.084
9.63 0.098
9.64 0.111
9.67 0.125
9.68 0.139
9.69 0.153
9.72 0.167
9.72 0.181
9.73 0.195
9.74 0.209
9.75 0.222
9.76 0.236
9.76 0.25
9.77 0.264
9.78 0.278
9.79 0.292
9.79 0.306
9.79 0.32
9.79 0.333
9.8 0.347
9.8 0.361
9.8 0.375
9.8 0.389
9.81 0.403
9.81 0.417
9.82 0.431
9.82 0.444
9.82 0.458
9.83 0.472
9.83 0.486
9.83 0.5
9.83 0.514
9.84 0.528
9.84 0.542
9.84 0.555
9.85 0.569
9.85 0.583
9.86 0.597
9.86 0.611
9.86 0.625
9.87 0.639
9.87 0.653
9.87 0.666
9.88 0.68
9.88 0.694
9.89 0.708
9.9 0.722
9.91 0.736
9.92 0.75
9.93 0.764
9.93 0.777
9.95 0.791
9.96 0.805
9.96 0.819
9.98 0.833
9.99 0.847
10 0.861
10.1 0.875
10.6 0.888
11 0.902
11.8 0.916
13.8 0.93
14.7 0.944
14.9 0.958
15.7 0.972
17 0.986
19.8 0.999
21.6 1
};
\addlegendentry{TW=0 N=2}
\addplot [thick, saddlebrown, opacity=0.8, dotted]
table {%
6.91 0
6.91 0.001
7.96 0.014
8.94 0.028
9.51 0.042
9.53 0.056
9.63 0.07
9.64 0.084
9.65 0.098
9.65 0.111
9.66 0.125
9.67 0.139
9.68 0.153
9.69 0.167
9.7 0.181
9.73 0.195
9.75 0.209
9.76 0.222
9.77 0.236
9.78 0.25
9.78 0.264
9.79 0.278
9.8 0.292
9.81 0.306
9.83 0.32
9.84 0.333
9.84 0.347
9.85 0.361
9.86 0.375
9.86 0.389
9.87 0.403
9.87 0.417
9.88 0.431
9.88 0.444
9.88 0.458
9.89 0.472
9.89 0.486
9.9 0.5
9.9 0.514
9.91 0.528
9.91 0.542
9.91 0.555
9.91 0.569
9.92 0.583
9.92 0.597
9.93 0.611
9.93 0.625
9.93 0.639
9.94 0.653
9.94 0.666
9.94 0.68
9.95 0.694
9.95 0.708
9.95 0.722
9.95 0.736
9.96 0.75
9.96 0.764
9.96 0.777
9.96 0.791
9.96 0.805
9.96 0.819
9.97 0.833
9.97 0.847
9.97 0.861
9.97 0.875
9.98 0.888
9.98 0.902
9.98 0.916
9.99 0.93
9.99 0.944
10 0.958
10.1 0.972
13.8 0.986
15.9 0.999
16.1 1
};
\addlegendentry{TW=0 N=3}
\addplot [thick, steelblue46134171, opacity=0.8]
table {%
6.93 0
6.93 0.001
8 0.014
8.19 0.028
8.89 0.042
8.93 0.056
9.18 0.07
9.7 0.084
9.71 0.098
9.8 0.111
9.81 0.125
9.81 0.139
9.82 0.153
9.82 0.167
9.83 0.181
9.83 0.195
9.83 0.209
9.84 0.222
9.84 0.236
9.85 0.25
9.85 0.264
9.86 0.278
9.87 0.292
9.88 0.306
9.88 0.32
9.88 0.333
9.89 0.347
9.89 0.361
9.89 0.375
9.89 0.389
9.9 0.403
9.9 0.417
9.9 0.431
9.91 0.444
9.91 0.458
9.92 0.472
9.92 0.486
9.92 0.5
9.92 0.514
9.92 0.528
9.93 0.542
9.93 0.555
9.93 0.569
9.93 0.583
9.93 0.597
9.93 0.611
9.93 0.625
9.94 0.639
9.94 0.653
9.94 0.666
9.94 0.68
9.94 0.694
9.95 0.708
9.95 0.722
9.95 0.736
9.95 0.75
9.95 0.764
9.95 0.777
9.95 0.791
9.96 0.805
9.96 0.819
9.96 0.833
9.97 0.847
9.97 0.861
9.97 0.875
9.98 0.888
9.99 0.902
9.99 0.916
10 0.93
10 0.944
11 0.958
14.6 0.972
15 0.986
29.9 0.999
39.9 1
};
\addlegendentry{TW=1 N=1}
\addplot [thick, steelblue46134171, opacity=0.8, dashed]
table {%
6.93 0
6.93 0.001
7.62 0.014
8.19 0.028
8.46 0.042
9.16 0.056
9.56 0.07
9.58 0.084
9.62 0.098
9.63 0.111
9.75 0.125
9.76 0.139
9.77 0.153
9.79 0.167
9.79 0.181
9.8 0.195
9.8 0.209
9.8 0.222
9.81 0.236
9.81 0.25
9.82 0.264
9.82 0.278
9.82 0.292
9.82 0.306
9.83 0.32
9.83 0.333
9.83 0.347
9.83 0.361
9.83 0.375
9.84 0.389
9.84 0.403
9.84 0.417
9.85 0.431
9.85 0.444
9.85 0.458
9.85 0.472
9.85 0.486
9.86 0.5
9.86 0.514
9.86 0.528
9.86 0.542
9.86 0.555
9.87 0.569
9.87 0.583
9.88 0.597
9.88 0.611
9.88 0.625
9.89 0.639
9.89 0.653
9.9 0.666
9.9 0.68
9.91 0.694
9.92 0.708
9.92 0.722
9.93 0.736
9.93 0.75
9.94 0.764
9.94 0.777
9.95 0.791
9.96 0.805
9.97 0.819
9.98 0.833
9.99 0.847
9.99 0.861
10 0.875
10.1 0.888
10.3 0.902
11 0.916
13.9 0.93
14.9 0.944
15 0.958
16.8 0.972
18 0.986
58 0.999
72.9 1
};
\addlegendentry{TW=1 N=2}
\addplot [thick, steelblue46134171, opacity=0.8, dotted]
table {%
7.61 0
7.61 0.001
9.6 0.014
9.63 0.028
9.65 0.042
9.66 0.056
9.7 0.07
9.71 0.084
9.72 0.098
9.73 0.111
9.74 0.125
9.75 0.139
9.76 0.153
9.77 0.167
9.78 0.181
9.8 0.195
9.82 0.209
9.82 0.222
9.83 0.236
9.84 0.25
9.84 0.264
9.84 0.278
9.85 0.292
9.85 0.306
9.86 0.32
9.86 0.333
9.86 0.347
9.87 0.361
9.87 0.375
9.88 0.389
9.88 0.403
9.88 0.417
9.88 0.431
9.89 0.444
9.89 0.458
9.89 0.472
9.89 0.486
9.9 0.5
9.9 0.514
9.9 0.528
9.9 0.542
9.91 0.555
9.91 0.569
9.91 0.583
9.91 0.597
9.92 0.611
9.92 0.625
9.92 0.639
9.92 0.653
9.93 0.666
9.93 0.68
9.93 0.694
9.93 0.708
9.94 0.722
9.94 0.736
9.94 0.75
9.94 0.764
9.94 0.777
9.95 0.791
9.95 0.805
9.95 0.819
9.96 0.833
9.96 0.847
9.96 0.861
9.97 0.875
9.97 0.888
9.97 0.902
9.98 0.916
9.98 0.93
9.98 0.944
9.99 0.958
10 0.972
10 0.986
14.9 0.999
17.7 1
};
\addlegendentry{TW=1 N=3}
\addplot [thick, darkorange2471270, opacity=0.8]
table {%
6.91 0
6.91 0.001
8.18 0.014
8.94 0.028
9.7 0.042
9.8 0.056
9.81 0.07
9.81 0.084
9.82 0.098
9.83 0.111
9.84 0.125
9.84 0.139
9.85 0.153
9.85 0.167
9.85 0.181
9.86 0.195
9.86 0.209
9.86 0.222
9.87 0.236
9.87 0.25
9.88 0.264
9.88 0.278
9.88 0.292
9.88 0.306
9.89 0.32
9.89 0.333
9.89 0.347
9.89 0.361
9.9 0.375
9.9 0.389
9.9 0.403
9.9 0.417
9.91 0.431
9.91 0.444
9.91 0.458
9.91 0.472
9.91 0.486
9.91 0.5
9.91 0.514
9.92 0.528
9.92 0.542
9.92 0.555
9.92 0.569
9.93 0.583
9.93 0.597
9.93 0.611
9.93 0.625
9.94 0.639
9.94 0.653
9.94 0.666
9.94 0.68
9.95 0.694
9.95 0.708
9.95 0.722
9.95 0.736
9.96 0.75
9.96 0.764
9.96 0.777
9.96 0.791
9.96 0.805
9.96 0.819
9.97 0.833
9.97 0.847
9.97 0.861
9.97 0.875
9.97 0.888
9.98 0.902
9.98 0.916
9.99 0.93
10 0.944
10 0.958
10.8 0.972
13.6 0.986
19 0.999
23.5 1
};
\addlegendentry{TW=2 N=1}
\addplot [thick, darkorange2471270, opacity=0.8, dashed]
table {%
7.31 0
7.31 0.001
7.91 0.014
8.43 0.028
8.77 0.042
9.15 0.056
9.62 0.07
9.73 0.084
9.75 0.098
9.77 0.111
9.79 0.125
9.79 0.139
9.8 0.153
9.81 0.167
9.81 0.181
9.81 0.195
9.82 0.209
9.83 0.222
9.83 0.236
9.84 0.25
9.84 0.264
9.85 0.278
9.85 0.292
9.86 0.306
9.86 0.32
9.87 0.333
9.87 0.347
9.87 0.361
9.87 0.375
9.88 0.389
9.89 0.403
9.89 0.417
9.89 0.431
9.9 0.444
9.9 0.458
9.9 0.472
9.91 0.486
9.91 0.5
9.91 0.514
9.91 0.528
9.92 0.542
9.92 0.555
9.92 0.569
9.92 0.583
9.93 0.597
9.93 0.611
9.93 0.625
9.93 0.639
9.93 0.653
9.94 0.666
9.94 0.68
9.94 0.694
9.94 0.708
9.94 0.722
9.95 0.736
9.95 0.75
9.95 0.764
9.96 0.777
9.96 0.791
9.96 0.805
9.97 0.819
9.97 0.833
9.97 0.847
9.97 0.861
9.98 0.875
9.98 0.888
9.99 0.902
9.99 0.916
10 0.93
10.7 0.944
11.9 0.958
14.4 0.972
15.8 0.986
18.7 0.999
19.2 1
};
\addlegendentry{TW=2 N=2}
\addplot [thick, darkorange2471270, opacity=0.8, dotted]
table {%
5.93 0
5.93 0.001
7.53 0.014
7.94 0.028
8.5 0.042
8.91 0.056
9.46 0.07
9.67 0.084
9.68 0.098
9.7 0.111
9.71 0.125
9.72 0.139
9.73 0.153
9.74 0.167
9.74 0.181
9.75 0.195
9.75 0.209
9.78 0.222
9.79 0.236
9.8 0.25
9.8 0.264
9.81 0.278
9.81 0.292
9.82 0.306
9.82 0.32
9.83 0.333
9.83 0.347
9.83 0.361
9.84 0.375
9.84 0.389
9.84 0.403
9.84 0.417
9.84 0.431
9.84 0.444
9.85 0.458
9.85 0.472
9.85 0.486
9.85 0.5
9.85 0.514
9.85 0.528
9.86 0.542
9.86 0.555
9.86 0.569
9.86 0.583
9.87 0.597
9.87 0.611
9.87 0.625
9.88 0.639
9.89 0.653
9.89 0.666
9.9 0.68
9.9 0.694
9.91 0.708
9.91 0.722
9.92 0.736
9.92 0.75
9.93 0.764
9.93 0.777
9.94 0.791
9.96 0.805
9.96 0.819
9.97 0.833
9.98 0.847
9.99 0.861
10 0.875
10.1 0.888
10.2 0.902
11 0.916
13.8 0.93
14.9 0.944
15 0.958
16.1 0.972
17.9 0.986
18.9 0.999
18.9 1
};
\addlegendentry{TW=2 N=3}
\addplot [thick, darkcyan6167125, opacity=0.8]
table {%
6.69 0
6.69 0.001
7.51 0.014
7.79 0.028
8.25 0.042
8.57 0.056
8.9 0.07
9.06 0.084
9.57 0.098
9.69 0.111
9.69 0.125
9.71 0.139
9.72 0.153
9.73 0.167
9.73 0.181
9.74 0.195
9.74 0.209
9.75 0.222
9.75 0.236
9.76 0.25
9.76 0.264
9.77 0.278
9.78 0.292
9.8 0.306
9.81 0.32
9.81 0.333
9.82 0.347
9.82 0.361
9.83 0.375
9.83 0.389
9.83 0.403
9.84 0.417
9.85 0.431
9.86 0.444
9.87 0.458
9.89 0.472
9.9 0.486
9.9 0.5
9.91 0.514
9.91 0.528
9.91 0.542
9.92 0.555
9.92 0.569
9.92 0.583
9.93 0.597
9.93 0.611
9.93 0.625
9.93 0.639
9.94 0.653
9.94 0.666
9.94 0.68
9.94 0.694
9.95 0.708
9.95 0.722
9.96 0.736
9.96 0.75
9.96 0.764
9.96 0.777
9.96 0.791
9.97 0.805
9.97 0.819
9.97 0.833
9.98 0.847
9.98 0.861
9.98 0.875
9.99 0.888
10 0.902
10 0.916
10.8 0.93
13.6 0.944
14.7 0.958
15 0.972
17 0.986
19.7 0.999
20.9 1
};
\addlegendentry{TW=3 N=1}
\addplot [thick, darkcyan6167125, opacity=0.8, dashed]
table {%
6.91 0
6.91 0.001
7.83 0.014
8.24 0.028
8.45 0.042
8.94 0.056
9.39 0.07
9.44 0.084
9.63 0.098
9.66 0.111
9.66 0.125
9.67 0.139
9.68 0.153
9.69 0.167
9.7 0.181
9.71 0.195
9.72 0.209
9.75 0.222
9.83 0.236
9.85 0.25
9.85 0.264
9.86 0.278
9.87 0.292
9.87 0.306
9.88 0.32
9.88 0.333
9.88 0.347
9.89 0.361
9.89 0.375
9.9 0.389
9.9 0.403
9.9 0.417
9.9 0.431
9.91 0.444
9.91 0.458
9.91 0.472
9.91 0.486
9.92 0.5
9.92 0.514
9.92 0.528
9.92 0.542
9.93 0.555
9.93 0.569
9.93 0.583
9.94 0.597
9.94 0.611
9.94 0.625
9.95 0.639
9.95 0.653
9.95 0.666
9.95 0.68
9.95 0.694
9.95 0.708
9.96 0.722
9.96 0.736
9.96 0.75
9.97 0.764
9.97 0.777
9.97 0.791
9.97 0.805
9.97 0.819
9.98 0.833
9.98 0.847
9.98 0.861
9.99 0.875
10 0.888
10 0.902
10.5 0.916
11.7 0.93
13.7 0.944
14.5 0.958
15 0.972
16.9 0.986
18.9 0.999
22.3 1
};
\addlegendentry{TW=3 N=2}
\addplot [thick, darkcyan6167125, opacity=0.8, dotted]
table {%
6.87 0
6.87 0.001
7.49 0.014
7.85 0.028
8.33 0.042
8.8 0.056
8.96 0.07
9.18 0.084
9.72 0.098
9.72 0.111
9.74 0.125
9.75 0.139
9.76 0.153
9.77 0.167
9.78 0.181
9.79 0.195
9.81 0.209
9.82 0.222
9.86 0.236
9.86 0.25
9.88 0.264
9.88 0.278
9.88 0.292
9.89 0.306
9.89 0.32
9.9 0.333
9.9 0.347
9.9 0.361
9.9 0.375
9.91 0.389
9.91 0.403
9.91 0.417
9.92 0.431
9.92 0.444
9.92 0.458
9.92 0.472
9.92 0.486
9.92 0.5
9.93 0.514
9.93 0.528
9.93 0.542
9.93 0.555
9.93 0.569
9.94 0.583
9.94 0.597
9.94 0.611
9.94 0.625
9.94 0.639
9.94 0.653
9.95 0.666
9.95 0.68
9.95 0.694
9.95 0.708
9.96 0.722
9.96 0.736
9.96 0.75
9.96 0.764
9.96 0.777
9.97 0.791
9.97 0.805
9.97 0.819
9.98 0.833
9.98 0.847
9.98 0.861
9.99 0.875
10 0.888
10 0.902
10.8 0.916
11.8 0.93
13.7 0.944
14.9 0.958
15 0.972
16.1 0.986
18.1 0.999
18.6 1
};
\addlegendentry{TW=3 N=3}
\end{axis}

\end{tikzpicture}

%% file: figures/rtt_overhead.tex
% This file was created with tikzplotlib v0.10.1.post13.
\begin{tikzpicture}
\usetikzlibrary{patterns}
\definecolor{darkcyan6167125}{RGB}{6,167,125}
\definecolor{darkgrey176}{RGB}{176,176,176}
\definecolor{darkorange2471270}{RGB}{247,127,0}
\definecolor{dimgrey102}{RGB}{102,102,102}
\definecolor{grey}{RGB}{128,128,128}
\definecolor{lightgrey204}{RGB}{204,204,204}
\definecolor{saddlebrown}{RGB}{139,69,19}
\definecolor{steelblue46134171}{RGB}{46,134,171}

\begin{axis}[
width=0.99\linewidth,
height=0.4\linewidth,
legend cell align={left},
legend style={
  fill opacity=0.95, 
  draw opacity=1, 
  text opacity=1, 
  draw=lightgrey204,
  at={(0.5,1.02)},
  anchor=south,
  legend columns=3,  % 3 columns (will create 2 rows with 5 entries)
  column sep=0.3cm,  % Tighter column spacing
  font=\small,       % Smaller font
  inner sep=2pt,     % Reduce internal padding
  nodes={inner sep=1pt}  % Reduce padding around text
},
tick align=inside,
tick pos=left,
x grid style={darkgrey176},
xmin=-1.185, xmax=16.085,
xtick style={color=black},
xtick={0,1.5,2.5,3.5,5.3,6.3,7.3,9.1,10.1,11.1,12.9,13.9,14.9},
xticklabels={BL
 ,1,2,3,1,2,3,1,2,3,1,2,3},
y grid style={darkgrey176},
ylabel={\% Scheduled UL Slots},
ymajorgrids,
ymin=0, ymax=27.2791046724093,
ytick style={color=black}
]
\draw[draw=black,fill=dimgrey102,opacity=0.8,line width=0.28pt] (axis cs:-0.4,0) rectangle (axis cs:0.4,4.27142857142857);
\addlegendimage{ybar,ybar legend,draw=black,fill=dimgrey102,opacity=0.8,line width=0.28pt}
\addlegendentry{Baseline (Reactive)}

\draw[draw=black,fill=saddlebrown,opacity=0.7,line width=0.28pt,postaction={pattern=vertical lines, fill opacity=0.7}] (axis cs:1.1,0) rectangle (axis cs:1.9,5.52090330422544);
\addlegendimage{ybar,ybar legend,draw=black,fill=saddlebrown,opacity=0.7,line width=0.28pt,postaction={pattern=vertical lines, fill opacity=0.7}}
\addlegendentry{TW = 0}

\draw[draw=black,fill=saddlebrown,opacity=0.7,line width=0.28pt,postaction={pattern=vertical lines, fill opacity=0.7}] (axis cs:2.1,0) rectangle (axis cs:2.9,7.59710840162569);
\draw[draw=black,fill=saddlebrown,opacity=0.7,line width=0.28pt,postaction={pattern=vertical lines, fill opacity=0.7}] (axis cs:3.1,0) rectangle (axis cs:3.9,8.53325336532054);
\draw[draw=black,fill=steelblue46134171,opacity=0.7,line width=0.28pt,postaction={pattern=horizontal lines, fill opacity=0.7}] (axis cs:4.9,0) rectangle (axis cs:5.7,7.59897205686217);
\addlegendimage{ybar,ybar legend,draw=black,fill=steelblue46134171,opacity=0.7,line width=0.28pt,postaction={pattern=horizontal lines, fill opacity=0.7}}
\addlegendentry{TW = 1}

\draw[draw=black,fill=steelblue46134171,opacity=0.7,line width=0.28pt,postaction={pattern=horizontal lines, fill opacity=0.7}] (axis cs:5.9,0) rectangle (axis cs:6.7,11.3582114641116);
\draw[draw=black,fill=steelblue46134171,opacity=0.7,line width=0.28pt,postaction={pattern=horizontal lines, fill opacity=0.7}] (axis cs:6.9,0) rectangle (axis cs:7.7,13.1414141414141);
\draw[draw=black,fill=darkorange2471270,opacity=0.7,line width=0.28pt,postaction={pattern=north east lines, fill opacity=0.7}] (axis cs:8.7,0) rectangle (axis cs:9.5,9.73569023569024);
\addlegendimage{ybar,ybar legend,draw=black,fill=darkorange2471270,opacity=0.7,line width=0.28pt,postaction={pattern=north east lines, fill opacity=0.7}}
\addlegendentry{TW = 2}

\draw[draw=black,fill=darkorange2471270,opacity=0.7,line width=0.28pt,postaction={pattern=north east lines, fill opacity=0.7}] (axis cs:9.7,0) rectangle (axis cs:10.5,15.7040731504572);
\draw[draw=black,fill=darkorange2471270,opacity=0.7,line width=0.28pt,postaction={pattern=north east lines, fill opacity=0.7}] (axis cs:10.7,0) rectangle (axis cs:11.5,18.9655172413793);
\draw[draw=black,fill=darkcyan6167125,opacity=0.7,line width=0.28pt,postaction={pattern=north west lines, fill opacity=0.7}] (axis cs:12.5,0) rectangle (axis cs:13.3,11.3036303630363);
\addlegendimage{ybar,ybar legend,draw=black,fill=darkcyan6167125,opacity=0.7,line width=0.28pt,postaction={pattern=north west lines, fill opacity=0.7}}
\addlegendentry{TW = 3}

\draw[draw=black,fill=darkcyan6167125,opacity=0.7,line width=0.28pt,postaction={pattern=north west lines, fill opacity=0.7}] (axis cs:13.5,0) rectangle (axis cs:14.3,18.3694374475231);
\draw[draw=black,fill=darkcyan6167125,opacity=0.7,line width=0.28pt,postaction={pattern=north west lines, fill opacity=0.7}] (axis cs:14.5,0) rectangle (axis cs:15.3,23.7209605847037);
\addplot [semithick, grey, opacity=0.3, dashed, forget plot]
table {%
0.75 0
0.75 27.2791046724093
};
\addplot [semithick, grey, opacity=0.3, dashed, forget plot]
table {%
4.4 0
4.4 27.2791046724093
};
\addplot [semithick, grey, opacity=0.3, dashed, forget plot]
table {%
8.2 0
8.2 27.2791046724093
};
\addplot [semithick, grey, opacity=0.3, dashed, forget plot]
table {%
12 0
12 27.2791046724093
};
\end{axis}

\end{tikzpicture}

%% file: figures/ml_inference.tex
% This file was created with tikzplotlib v0.10.1.post13.
\begin{tikzpicture}

\definecolor{darkgrey176}{RGB}{176,176,176}
\definecolor{darkorange2471270}{RGB}{247,127,0}
\definecolor{dimgrey102}{RGB}{102,102,102}
\definecolor{lightgrey204}{RGB}{204,204,204}
\definecolor{steelblue46134171}{RGB}{46,134,171}

\begin{axis}[
height=0.25\textwidth,
legend cell align={left},
legend style={
  fill opacity=0.95,
  draw opacity=1,
  text opacity=1,
  at={(0.03,0.97)},
  anchor=north west,
  draw=lightgrey204
},
tick align=inside,
tick pos=left,
width=0.5\textwidth,
x grid style={darkgrey176},
xlabel={Number of Input Words},
xmajorgrids,
xmin=47.5, xmax=102.5,
xtick style={color=black},
y grid style={darkgrey176},
ylabel={\scriptsize Avg RTT (ms) - network + inference},
ymajorgrids,
ymin=35.4653670953519, ymax=73.9353457841816,
ytick style={color=black}
]
\path [draw=steelblue46134171, fill=steelblue46134171, opacity=0.18, line width=0pt]
(axis cs:50,39.1836230973302)
--(axis cs:50,37.2140024902987)
--(axis cs:51,38.2999783767163)
--(axis cs:52,37.9473153643746)
--(axis cs:53,37.5060750754884)
--(axis cs:54,38.0667894843573)
--(axis cs:55,38.7966348680611)
--(axis cs:56,37.7484128751609)
--(axis cs:57,38.2687372267453)
--(axis cs:58,38.8950609939111)
--(axis cs:59,39.9159707454149)
--(axis cs:60,39.8939840509134)
--(axis cs:61,41.77397662538)
--(axis cs:62,42.0920033932024)
--(axis cs:63,42.3257634367741)
--(axis cs:64,41.5399435277986)
--(axis cs:65,41.3865037701699)
--(axis cs:66,42.4688271718643)
--(axis cs:67,42.6425325893519)
--(axis cs:68,41.8683622221975)
--(axis cs:69,42.960973791864)
--(axis cs:70,42.7521737462427)
--(axis cs:71,43.6688767626716)
--(axis cs:72,44.0009294418939)
--(axis cs:73,43.1019391494855)
--(axis cs:74,46.1231914355798)
--(axis cs:75,46.2148301520456)
--(axis cs:76,46.8090289103077)
--(axis cs:77,46.975307284675)
--(axis cs:78,46.7954075021546)
--(axis cs:79,47.7483551333466)
--(axis cs:80,47.3071268244592)
--(axis cs:81,47.1592210066478)
--(axis cs:82,48.2202419232727)
--(axis cs:83,47.4766688588727)
--(axis cs:84,48.5629146019602)
--(axis cs:85,48.3107695208481)
--(axis cs:86,50.148136210266)
--(axis cs:87,49.8531027216693)
--(axis cs:88,51.7312419363338)
--(axis cs:89,51.2721401459759)
--(axis cs:90,51.3956771751542)
--(axis cs:91,52.0325857224387)
--(axis cs:92,52.649716663973)
--(axis cs:93,51.8365150517122)
--(axis cs:94,52.6262039305104)
--(axis cs:95,52.7338223806227)
--(axis cs:96,51.7924492686307)
--(axis cs:97,52.4097509143641)
--(axis cs:98,53.8155174067587)
--(axis cs:99,53.9033593044764)
--(axis cs:100,57.3397596941464)
--(axis cs:100,60.4353417467138)
--(axis cs:100,60.4353417467138)
--(axis cs:99,57.9792924711333)
--(axis cs:98,57.5145187893198)
--(axis cs:97,54.9213617943737)
--(axis cs:96,55.3775821777979)
--(axis cs:95,55.4340970203434)
--(axis cs:94,55.6685558290524)
--(axis cs:93,55.3813021665619)
--(axis cs:92,57.0445859289919)
--(axis cs:91,54.7534615781748)
--(axis cs:90,54.108489429974)
--(axis cs:89,53.5591596731196)
--(axis cs:88,55.128664364637)
--(axis cs:87,51.9478319908307)
--(axis cs:86,52.6981684956163)
--(axis cs:85,52.76970912283)
--(axis cs:84,52.3784508553262)
--(axis cs:83,50.5393463627489)
--(axis cs:82,52.5698130608543)
--(axis cs:81,49.893152595625)
--(axis cs:80,51.1362572052438)
--(axis cs:79,51.2309145074301)
--(axis cs:78,50.1414664089565)
--(axis cs:77,50.766867715325)
--(axis cs:76,49.1852585323153)
--(axis cs:75,49.3051661194476)
--(axis cs:74,49.3662648459736)
--(axis cs:73,45.9527461141961)
--(axis cs:72,47.4779906949482)
--(axis cs:71,47.0706361219438)
--(axis cs:70,45.0241870984175)
--(axis cs:69,46.3065189306029)
--(axis cs:68,45.2301113757924)
--(axis cs:67,45.412797256802)
--(axis cs:66,46.8575281819818)
--(axis cs:65,43.4944303877248)
--(axis cs:64,44.0778320028136)
--(axis cs:63,44.6045623753807)
--(axis cs:62,45.3253800776971)
--(axis cs:61,46.1036550557795)
--(axis cs:60,43.4500834090866)
--(axis cs:59,44.3038830124799)
--(axis cs:58,42.5913087804478)
--(axis cs:57,41.7445193100968)
--(axis cs:56,41.0189105876962)
--(axis cs:55,43.4974515128913)
--(axis cs:54,40.933607348976)
--(axis cs:53,41.2212689791564)
--(axis cs:52,41.7176771726624)
--(axis cs:51,41.8490263532837)
--(axis cs:50,39.1836230973302)
--cycle;

\path [draw=dimgrey102, fill=dimgrey102, opacity=0.18, line width=0pt]
(axis cs:50,48.1167790974305)
--(axis cs:50,46.6942962221571)
--(axis cs:51,48.3335258814526)
--(axis cs:52,47.8353111879544)
--(axis cs:53,47.5924094958656)
--(axis cs:54,48.1972962919578)
--(axis cs:55,48.2249922174634)
--(axis cs:56,49.173989726093)
--(axis cs:57,47.388471321136)
--(axis cs:58,48.7706932800825)
--(axis cs:59,49.1043238113034)
--(axis cs:60,49.0375645797246)
--(axis cs:61,51.348972638649)
--(axis cs:62,52.3182166158816)
--(axis cs:63,52.7805537445817)
--(axis cs:64,52.4481647407179)
--(axis cs:65,52.8000928958674)
--(axis cs:66,52.3927080762689)
--(axis cs:67,53.3850053716109)
--(axis cs:68,53.3958973344071)
--(axis cs:69,53.2427424706308)
--(axis cs:70,52.9060024008551)
--(axis cs:71,53.4193416224013)
--(axis cs:72,54.093117001661)
--(axis cs:73,54.4456736185705)
--(axis cs:74,56.3107347891286)
--(axis cs:75,56.3286517866846)
--(axis cs:76,57.4736805990185)
--(axis cs:77,57.8033928280368)
--(axis cs:78,57.8687505416233)
--(axis cs:79,58.2121919878551)
--(axis cs:80,58.0739444735218)
--(axis cs:81,59.0167240898364)
--(axis cs:82,58.9075210784058)
--(axis cs:83,58.405857765608)
--(axis cs:84,59.1547051400755)
--(axis cs:85,58.911690499405)
--(axis cs:86,60.0100279098108)
--(axis cs:87,60.5826170956675)
--(axis cs:88,62.044192104725)
--(axis cs:89,60.6900313033556)
--(axis cs:90,64.3511104816498)
--(axis cs:91,64.6843319180084)
--(axis cs:92,62.1400120140073)
--(axis cs:93,64.1566956314038)
--(axis cs:94,65.5454166648645)
--(axis cs:95,66.2266554625842)
--(axis cs:96,65.1669367237464)
--(axis cs:97,67.3419225719302)
--(axis cs:98,67.9575185756632)
--(axis cs:99,66.236854785315)
--(axis cs:100,67.9186027946505)
--(axis cs:100,71.1757421300807)
--(axis cs:100,71.1757421300807)
--(axis cs:99,68.8372911951728)
--(axis cs:98,71.6543992968858)
--(axis cs:97,72.1867103892348)
--(axis cs:96,68.7901302316107)
--(axis cs:95,68.2249747499762)
--(axis cs:94,67.8781357856849)
--(axis cs:93,67.9276675564135)
--(axis cs:92,64.1381530714199)
--(axis cs:91,67.1666273457953)
--(axis cs:90,67.1500738978374)
--(axis cs:89,63.3268136865434)
--(axis cs:88,65.4490256137216)
--(axis cs:87,63.3777112918325)
--(axis cs:86,62.8743272078362)
--(axis cs:85,60.5565827419743)
--(axis cs:84,60.9135996073993)
--(axis cs:83,61.8029951749325)
--(axis cs:82,61.1154932813826)
--(axis cs:81,62.7474487738)
--(axis cs:80,60.8261011997456)
--(axis cs:79,61.3980120898148)
--(axis cs:78,60.3552290805989)
--(axis cs:77,61.6308288700764)
--(axis cs:76,60.6522327671017)
--(axis cs:75,58.5920251769517)
--(axis cs:74,58.1277186380558)
--(axis cs:73,57.5538653114295)
--(axis cs:72,55.7577988404443)
--(axis cs:71,56.3121979160602)
--(axis cs:70,54.9885797156498)
--(axis cs:69,56.5625310976512)
--(axis cs:68,56.1056026857949)
--(axis cs:67,55.6079014222035)
--(axis cs:66,54.3458513186028)
--(axis cs:65,55.2318924936063)
--(axis cs:64,56.3319793817311)
--(axis cs:63,56.3558435703354)
--(axis cs:62,54.0846928232719)
--(axis cs:61,54.8554985690805)
--(axis cs:60,51.2872115308281)
--(axis cs:59,51.3700911044861)
--(axis cs:58,52.1026340429944)
--(axis cs:57,49.6516532683377)
--(axis cs:56,51.9977510624785)
--(axis cs:55,50.6224249349176)
--(axis cs:54,50.8732068179898)
--(axis cs:53,49.9232425590794)
--(axis cs:52,49.5701463305641)
--(axis cs:51,50.8196889185474)
--(axis cs:50,48.1167790974305)
--cycle;

\addplot [line width=1.1pt, steelblue46134171, opacity=0.9, dash pattern=on 5pt off 3pt, mark=*, mark size=1.6, mark options={solid,draw=steelblue46134171,fill=steelblue46134171}]
table {%
50 38.1988127938144
51 40.074502365
52 39.8324962685185
53 39.3636720273224
54 39.5001984166667
55 41.1470431904762
56 39.3836617314286
57 40.0066282684211
58 40.7431848871795
59 42.1099268789474
60 41.67203373
61 43.9388158405797
62 43.7086917354497
63 43.4651629060774
64 42.8088877653061
65 42.4404670789474
66 44.6631776769231
67 44.0276649230769
68 43.549236798995
69 44.6337463612335
70 43.8881804223301
71 45.3697564423077
72 45.739460068421
73 44.5273426318408
74 47.7447281407767
75 47.7599981357466
76 47.9971437213115
77 48.8710875
78 48.4684369555556
79 49.4896348203883
80 49.2216920148515
81 48.5261868011364
82 50.3950274920635
83 49.0080076108108
84 50.4706827286432
85 50.5402393218391
86 51.4231523529412
87 50.90046735625
88 53.4299531504854
89 52.4156499095477
90 52.7520833025641
91 53.3930236503068
92 54.8471512964824
93 53.6089086091371
94 54.1473798797814
95 54.0839597004831
96 53.5850157232143
97 53.6655563543689
98 55.6650180980392
99 55.9413258878049
100 58.8875507204301
};
\addlegendentry{ML-Aware UL Scheduler}
\addplot [line width=1.1pt, dimgrey102, opacity=0.9, mark=square*, mark size=1.6, mark options={solid,draw=dimgrey102,fill=dimgrey102}]
table {%
50 47.4055376597938
51 49.5766074
52 48.7027287592593
53 48.7578260274725
54 49.5352515549738
55 49.4237085761905
56 50.5858703942857
57 48.5200622947368
58 50.4366636615385
59 50.2372074578947
60 50.1623880552764
61 53.1022356038647
62 53.2014547195767
63 54.5681986574586
64 54.3900720612245
65 54.0159926947368
66 53.3692796974359
67 54.4964533969072
68 54.750750010101
69 54.902636784141
70 53.9472910582524
71 54.8657697692308
72 54.9254579210526
73 55.999769465
74 57.2192267135922
75 57.4603384818182
76 59.0629566830601
77 59.7171108490566
78 59.1119898111111
79 59.805102038835
80 59.4500228366337
81 60.8820864318182
82 60.0115071798942
83 60.1044264702703
84 60.0341523737374
85 59.7341366206897
86 61.4421775588235
87 61.98016419375
88 63.7466088592233
89 62.0084224949495
90 65.7505921897436
91 65.9254796319018
92 63.1390825427136
93 66.0421815939086
94 66.7117762252747
95 67.2258151062802
96 66.9785334776786
97 69.7643164805825
98 69.8059589362745
99 67.5370729902439
100 69.5471724623656
};
\addlegendentry{Baseline}
\end{axis}

\end{tikzpicture}

%% file: figures/mgen_comparison_all.tex
% This file was created with tikzplotlib v0.10.1.post13.
\begin{tikzpicture}[
scale=0.9
]

\definecolor{darkcyan6167125}{RGB}{6,167,125}
\definecolor{darkgrey176}{RGB}{176,176,176}
\definecolor{darkorange2471270}{RGB}{247,127,0}
\definecolor{dimgrey102}{RGB}{102,102,102}
\definecolor{lightgrey204}{RGB}{204,204,204}
\definecolor{saddlebrown}{RGB}{139,69,19}
\definecolor{steelblue46134171}{RGB}{46,134,171}

\begin{axis}[
width=1.\linewidth,
height=0.4\linewidth,
width=\columnwidth,
grid=major,
legend cell align={left},
legend style={
  fill opacity=0.8,
  draw opacity=1,
  text opacity=1,
  at={(0.5,1.02)},
  anchor=south,
  draw=lightgrey204,
  legend columns=3,    % Move this INSIDE the legend style block
  column sep=0.12cm,
  row sep=-0.5pt,
  font=\footnotesize,
  inner xsep=3pt,
  inner ysep=1pt
},
thick,
tick align=inside,
tick pos=left,
x grid style={darkgrey176},
xlabel={Latency (ms)},
xmajorgrids,
xmin=0, xmax=40,
xtick style={color=black},
y grid style={darkgrey176},
ylabel={},
ymajorgrids,
ymin=0, ymax=1,
ytick style={color=black}
]
\addplot [line width=1pt, dimgrey102, opacity=0.9]
table {%
5.54199999896809 0
5.54199999896809 0.000103082156478714
6.29000000481028 0.0139160911246263
6.62300000840332 0.0278321822492527
7.0209999976214 0.041748273373879
7.3990000091726 0.0555612823420266
7.76399999449495 0.0694773734666529
8.08300000790041 0.0833934645912792
8.46599999931641 0.0973095557159056
8.94800000241958 0.111122564684053
9.43700000061654 0.12503865580868
9.94099999661557 0.138954746933306
10.4629999987083 0.152767755901453
11.4960000064457 0.16668384702608
12.4999999970896 0.180599938150706
13.4369999868795 0.194516029275332
14.2590000032214 0.20832903824348
14.8300000000745 0.222245129368106
15.3629999986151 0.236161220492733
15.7409999956144 0.250077311617359
16.2430000054883 0.263890320585507
16.5490000072168 0.277806411710133
16.7990000045393 0.291722502834759
17.0190000062576 0.305535511802907
17.2980000061216 0.319451602927533
17.6510000019334 0.33336769405216
17.9560000105994 0.347283785176786
18.2039999926928 0.361096794144934
18.4440000011818 0.37501288526956
18.7359999981709 0.388928976394186
19.0719999955036 0.402741985362334
19.4190000038361 0.41665807648696
19.7190000035334 0.430574167611586
20.0539999932516 0.444490258736213
20.3529999998864 0.45830326770436
20.6529999995837 0.472219358828987
20.9340000001248 0.486135449953613
21.3470000016969 0.500051541078239
22.1339999989141 0.513864550046387
23.019000000204 0.527780641171013
23.9029999938793 0.54169673229564
24.6029999980237 0.555509741263787
25.1360000111163 0.569425832388414
25.6260000023758 0.58334192351304
26.5599999984261 0.597258014637666
27.4890000000596 0.611071023605814
28.442000009818 0.62498711473044
29.5089999999618 0.638903205855066
32.8250000020489 0.652716214823214
35.5929999932414 0.66663230594784
37.5570000032894 0.680548397072467
38.9279999944847 0.694464488197093
45.2070000028471 0.708277497165241
47.048000007635 0.722193588289867
48.5390000103507 0.736109679414493
50.8030000055442 0.749922688382641
56.0510000068462 0.763838779507267
57.9310000030091 0.777754870631894
59.4050000072457 0.79167096175652
65.227999992203 0.805483970724668
67.3509999905946 0.819400061849294
68.965999991633 0.83331615297392
70.9399999905145 0.847232244098547
76.5800000081072 0.861045253066694
78.4500000008848 0.87496134419132
79.9819999956526 0.888877435315947
85.3460000071209 0.902690444284094
87.3399999982212 0.916606535408721
88.768000001437 0.930522626533347
89.9289999942994 0.944438717657973
96.6110000008484 0.958251726626121
98.4910000115633 0.972167817750747
99.8519999993732 0.986083908875374
109.937999994145 0.999896917843521
112.173000001349 1
};
\addlegendentry{Baseline (Reactive)}
\addplot [thick, saddlebrown, opacity=0.8]
table {%
5.37899999471847 0
5.37899999471847 0.000103082156478714
5.76700000965502 0.0139160911246263
5.90000000374857 0.0278321822492527
5.99200000579003 0.041748273373879
6.083000000217 0.0555612823420266
6.18599999870639 0.0694773734666529
6.27600000007078 0.0833934645912792
6.36800000211224 0.0973095557159056
6.46499999857042 0.111122564684053
6.55099999858066 0.12503865580868
6.63499999791384 0.138954746933306
6.71699999656994 0.152767755901453
6.81099999928847 0.16668384702608
6.90699998813216 0.180599938150706
6.99999999778811 0.194516029275332
7.08199999644421 0.20832903824348
7.18400000187103 0.222245129368106
7.27699999697506 0.236161220492733
7.36300001153722 0.250077311617359
7.46199999412056 0.263890320585507
7.56600000022445 0.277806411710133
7.66400000429712 0.291722502834759
7.73899999330752 0.305535511802907
7.8489999868907 0.319451602927533
7.96400000399444 0.33336769405216
8.05399999080691 0.347283785176786
8.14600000740029 0.361096794144934
8.2469999906607 0.37501288526956
8.34300000860821 0.388928976394186
8.43800000438932 0.402741985362334
8.5419999959413 0.41665807648696
8.64000000001397 0.430574167611586
8.73600000340957 0.444490258736213
8.86100000934675 0.45830326770436
8.96900000225287 0.472219358828987
9.08000000345055 0.486135449953613
9.2000000004191 0.500051541078239
9.31700000364799 0.513864550046387
9.43200000619981 0.527780641171013
9.54000001365785 0.54169673229564
9.64699999894947 0.555509741263787
9.75299999117851 0.569425832388414
9.86499999999069 0.58334192351304
9.9740000005113 0.597258014637666
10.1079999876674 0.611071023605814
10.3720000042813 0.62498711473044
11.0010000062175 0.638903205855066
11.331000001519 0.652716214823214
11.7019999888726 0.66663230594784
12.2400000109337 0.680548397072467
12.6120000059018 0.694464488197093
13.026999993599 0.708277497165241
13.2920000032755 0.722193588289867
13.6560000100872 0.736109679414493
14.2049999994924 0.749922688382641
14.4769999897107 0.763838779507267
14.8539999936474 0.777754870631894
15.2059999963967 0.79167096175652
16.2030000064988 0.805483970724668
16.6509999980917 0.819400061849294
17.6750000100583 0.83331615297392
18.7029999942752 0.847232244098547
19.4360000023153 0.861045253066694
19.6760000108043 0.87496134419132
20.3739999997197 0.888877435315947
21.2249999894993 0.902690444284094
21.9039999938104 0.916606535408721
22.6679999905173 0.930522626533347
23.0489999958081 0.944438717657973
23.5549999924842 0.958251726626121
24.0860000048997 0.972167817750747
24.6610000031069 0.986083908875374
30.2280000032624 0.999896917843521
30.6540000019595 1
};
\addlegendentry{TW=0 N=1}
\addplot [thick, saddlebrown, opacity=0.8, dashed]
table {%
5.28399999893736 0
5.28399999893736 0.000103114044132811
5.63799998781178 0.0139203959579295
5.8019999996759 0.0278407919158589
5.88100000459235 0.0417611878737884
5.94000000273809 0.0555784697875851
6.01199999800883 0.0694988657455145
6.08999999531079 0.083419261703444
6.1649999988731 0.0972365436172407
6.24899999820627 0.11115693957517
6.33699999889359 0.1250773355331
6.43899998976849 0.138894617446896
6.56300000264309 0.152815013404826
6.68399999267422 0.166735409362755
6.80599999031983 0.180552691276552
6.92600000184029 0.194473087234481
7.05000000016298 0.208393483192411
7.17500000610016 0.222210765106207
7.31499999528751 0.236131161064137
7.45800000731833 0.250051557022066
7.58700000005774 0.263868838935863
7.73300000582822 0.277789234893793
7.8769999963697 0.291709630851722
8.0279999965569 0.305526912765519
8.14599999284837 0.319447308723448
8.31799999286886 0.333367704681378
8.45900000422262 0.347288100639307
8.6099999898579 0.361105382553104
8.77099999343045 0.375025778511033
8.931999997003 0.388946174468963
9.05699998838827 0.402763456382759
9.19199999771081 0.416683852340689
9.32100000500213 0.430604248298618
9.43900000129361 0.444421530212415
9.55099999555387 0.458341926170344
9.69399999303278 0.472262322128274
9.92299999052193 0.486079604042071
10.1489999942714 0.5
10.5000000039581 0.513920395957929
10.9220000013011 0.527737677871726
11.2590000062482 0.541658073829656
11.5729999961331 0.555578469787585
11.9049999921117 0.569395751701382
12.2950000077253 0.583316147659311
12.5740000075893 0.597236543617241
12.7869999996619 0.611053825531037
12.9519999900367 0.624974221488967
13.2279999961611 0.638894617446896
13.5129999980563 0.652711899360693
14.0850000025239 0.666632295318622
14.3460000108462 0.680552691276552
14.5480000064708 0.694473087234481
14.7179999912623 0.708290369148278
14.896000007866 0.722210765106208
15.4870000114897 0.736131161064137
16.3470000115922 0.749948442977934
17.2579999925802 0.763868838935863
18.316000001505 0.777789234893792
19.1309999936493 0.791606516807589
19.6750000031898 0.805526912765519
20.0130000011995 0.819447308723448
20.9879999893019 0.833264590637245
21.573999998509 0.847184986595174
21.9819999911124 0.861105382553104
22.436999992351 0.8749226644669
22.8899999929126 0.88884306042483
23.1879999919329 0.902763456382759
23.5929999907967 0.916580738296556
23.9660000079311 0.930501134254485
24.6270000061486 0.944421530212415
25.0519999972312 0.958238812126212
26.1109999992186 0.972159208084141
31.115999998292 0.986079604042071
118.092999997316 0.999896885955867
123.202999995556 1
};
\addlegendentry{TW=0 N=2}
\addplot [thick, saddlebrown, opacity=0.8, dotted]
table {%
5.47800000640564 0
5.47800000640564 0.000103124677735382
5.97600000037346 0.0139218314942766
6.08700000157114 0.0278436629885532
6.18699999176897 0.0416623698050944
6.29299999854993 0.0555842012993709
6.4109999948414 0.0695060327936475
6.5560000075493 0.0833247396101887
6.78300000436138 0.0972465711044653
6.96199999947567 0.111168402598742
7.12400001066271 0.125090234093018
7.36399999004789 0.13890894090956
7.61600000259932 0.152830772403836
7.81799999822397 0.166752603898113
7.98799999756739 0.180571310714654
8.1779999891296 0.194493142208931
8.48199999018107 0.208311849025472
8.73200000205543 0.222233680519748
8.90899999649264 0.236155512014025
9.08899999922141 0.250077343508302
9.32299999112729 0.263896050324843
9.54600000113714 0.277817881819119
9.69500000064727 0.29163658863566
9.8909999942407 0.305558420129937
10.1200000062818 0.319480251624214
10.5730000068434 0.33340208311849
10.8999999938533 0.347220789935031
11.0500000009779 0.361142621429308
11.2709999957588 0.375064452923585
11.5959999966435 0.388883159740126
11.8190000066534 0.402804991234402
12.0629999873927 0.416623698050944
12.3679999960586 0.43054552954522
12.6460000028601 0.444467361039497
12.8709999989951 0.458389192533773
13.1289999990258 0.472207899350315
13.453000006848 0.486129730844591
13.6659999989206 0.499948437661132
13.900999998441 0.513870269155409
14.1989999974612 0.527792100649686
14.4599999912316 0.541610807466227
14.699000006658 0.555532638960503
15.1499999919906 0.56945447045478
15.7060000055935 0.583273177271321
15.956000002916 0.597195008765598
16.1510000034468 0.611116840259874
16.341000009561 0.625038671754151
16.6220000101021 0.638857378570692
17.0270000089658 0.652779210064969
17.3740000027465 0.66659791688151
17.785999996704 0.680519748375786
18.2000000058906 0.694441579870063
18.5979999951087 0.708260286686604
19.0259999944828 0.722182118180881
19.4859999901382 0.736103949675157
19.8240000026999 0.749922656491698
20.3699999983655 0.763844487985975
20.9809999942081 0.777766319480252
21.2219999957597 0.791585026296793
21.4770000020508 0.805506857791069
21.870000011404 0.819428689285346
22.1489999967162 0.833247396101887
22.4250000028405 0.847169227596164
22.7920000033919 0.86109105909044
23.0680000095163 0.875012890584717
23.3419999858597 0.888831597401258
23.6999999906402 0.902753428895535
23.9489999949001 0.916572135712076
24.1850000020349 0.930493967206352
24.5530000102008 0.944415798700629
24.7969999909401 0.95823450551717
25.0430000014603 0.972156337011447
25.6499999959487 0.986078168505723
30.5260000022827 0.999896875322265
31.9770000060089 1
};
\addlegendentry{TW=0 N=3}
\addplot [thick, steelblue46134171, opacity=0.8]
table {%
5.33200000063516 0
5.33200000063516 0.000103082156478714
5.54600000032224 0.0139160911246263
5.65500000084285 0.0278321822492527
5.74300000153016 0.041748273373879
5.82299999950919 0.0555612823420266
5.90099999681115 0.0694773734666529
5.97500000731088 0.0833934645912792
6.03599999158178 0.0973095557159056
6.10699999378994 0.111122564684053
6.18799999938346 0.12503865580868
6.25999999465421 0.138954746933306
6.32200000109151 0.152767755901453
6.38500000059139 0.16668384702608
6.46200000483077 0.180599938150706
6.52999999874737 0.194516029275332
6.5900000045076 0.20832903824348
6.66899999487214 0.222245129368106
6.73900000401773 0.236161220492733
6.81199999235105 0.250077311617359
6.87300000572577 0.263890320585507
6.94300000031944 0.277806411710133
7.01800000388175 0.291722502834759
7.09099999221507 0.305535511802907
7.16799999645445 0.319451602927533
7.24600000830833 0.33336769405216
7.31000000087079 0.347283785176786
7.37099999969359 0.361096794144934
7.44499999564141 0.37501288526956
7.50899998820387 0.388928976394186
7.56800000090152 0.402741985362334
7.64799999888055 0.41665807648696
7.72100000176579 0.430574167611586
7.79000000329688 0.444490258736213
7.85900000482798 0.45830326770436
7.92900001397356 0.472219358828987
8.00500001059845 0.486135449953613
8.07300000451505 0.500051541078239
8.13799999014009 0.513864550046387
8.21400000131689 0.527780641171013
8.28400001046248 0.54169673229564
8.3510000113165 0.555509741263787
8.42100000591017 0.569425832388414
8.48200000473298 0.58334192351304
8.53900000220165 0.597258014637666
8.60399998782668 0.611071023605814
8.68400000035763 0.62498711473044
8.75799999630544 0.638903205855066
8.82299999648239 0.652716214823214
8.90200000139885 0.66663230594784
8.9729999890551 0.680548397072467
9.05100000090897 0.694464488197093
9.10799999837764 0.708277497165241
9.18100000126287 0.722193588289867
9.25099999585655 0.736109679414493
9.32199999806471 0.749922688382641
9.38699999824166 0.763838779507267
9.45200001297053 0.777754870631894
9.51400000485592 0.79167096175652
9.58499999251217 0.805483970724668
9.66799999878276 0.819400061849294
9.72999999066815 0.83331615297392
9.78699998813681 0.847232244098547
9.86299999931362 0.861045253066694
9.940000003553 0.87496134419132
10.0119999988237 0.888877435315947
10.0839999940945 0.902690444284094
10.1480000012089 0.916606535408721
10.2239999978337 0.930522626533347
10.2919999917503 0.944438717657973
10.3680000029271 0.958251726626121
10.4509999946458 0.972167817750747
10.5949999997392 0.986083908875374
20.4420000081882 0.999896917843521
23.1780000030994 1
};
\addlegendentry{TW=1 N=1}
\addplot [thick, steelblue46134171, opacity=0.8, dashed]
table {%
5.34699999843724 0
5.34699999843724 0.000103082156478714
5.62600001285318 0.0139160911246263
5.6990000011865 0.0278321822492527
5.75199999730103 0.041748273373879
5.79500000458211 0.0555612823420266
5.83699998969678 0.0694773734666529
5.87700000323821 0.0833934645912792
5.91500000155065 0.0973095557159056
5.95099999918602 0.111122564684053
5.98500001069624 0.12503865580868
6.02200000139419 0.138954746933306
6.05999999970663 0.152767755901453
6.10600000072736 0.16668384702608
6.15000000107102 0.180599938150706
6.20400000480004 0.194516029275332
6.25600000785198 0.20832903824348
6.31200001225807 0.222245129368106
6.37299999652896 0.236161220492733
6.4400000119349 0.250077311617359
6.50000000314321 0.263890320585507
6.56900000467431 0.277806411710133
6.63200000417419 0.291722502834759
6.72399999166373 0.305535511802907
6.79099999251775 0.319451602927533
6.85999999404885 0.33336769405216
6.93999999202788 0.347283785176786
7.01099999423604 0.361096794144934
7.08299998950679 0.37501288526956
7.14899999729823 0.388928976394186
7.20199999341276 0.402741985362334
7.25400001101661 0.41665807648696
7.32900000002701 0.430574167611586
7.40400000358932 0.444490258736213
7.47200001205783 0.45830326770436
7.54199999209959 0.472219358828987
7.61200000124518 0.486135449953613
7.69100000616163 0.500051541078239
7.76399999449495 0.513864550046387
7.83499999670312 0.527780641171013
7.90799999958836 0.54169673229564
7.97799999418203 0.555509741263787
8.04600000265054 0.569425832388414
8.11200001044199 0.58334192351304
8.18099999742117 0.597258014637666
8.2399999955669 0.611071023605814
8.31099999777507 0.62498711473044
8.38400000066031 0.638903205855066
8.44799999322277 0.652716214823214
8.51600000169128 0.66663230594784
8.58500000322238 0.680548397072467
8.64700000965968 0.694464488197093
8.73099999444094 0.708277497165241
8.80599999800324 0.722193588289867
8.88100000156555 0.736109679414493
8.9529999968363 0.749922688382641
9.02300000598188 0.763838779507267
9.08799999160692 0.777754870631894
9.15800000075251 0.79167096175652
9.22399999399204 0.805483970724668
9.28399999975227 0.819400061849294
9.35399999434594 0.83331615297392
9.42499999655411 0.847232244098547
9.49100000434555 0.861045253066694
9.56200000655372 0.87496134419132
9.63100000808481 0.888877435315947
9.70799999777228 0.902690444284094
9.78100000065751 0.916606535408721
9.85600000421982 0.930522626533347
9.92300000507385 0.944438717657973
10.0039999961155 0.958251726626121
10.0660000025528 0.972167817750747
10.1799999974901 0.986083908875374
62.2869999933755 0.999896917843521
72.5800000072923 1
};
\addlegendentry{TW=1 N=2}
\addplot [thick, steelblue46134171, opacity=0.8, dotted]
table {%
5.3419999894686 0
5.3419999894686 0.000103082156478714
5.57199999457225 0.0139160911246263
5.61500000185333 0.0278321822492527
5.65399999322835 0.041748273373879
5.69600000744686 0.0555612823420266
5.73299999814481 0.0694773734666529
5.76899999578018 0.0833934645912792
5.80600000103004 0.0973095557159056
5.84499999240506 0.111122564684053
5.88499999139458 0.12503865580868
5.91400000848807 0.138954746933306
5.95000000612345 0.152767755901453
5.98799998988397 0.16668384702608
6.03299999784213 0.180599938150706
6.08700000157114 0.194516029275332
6.15000000107102 0.20832903824348
6.20499999786261 0.222245129368106
6.27499999245629 0.236161220492733
6.34799999534152 0.250077311617359
6.42599999264348 0.263890320585507
6.48999999975786 0.277806411710133
6.56099998741411 0.291722502834759
6.62800000282004 0.305535511802907
6.7070000077365 0.319451602927533
6.77700000233017 0.33336769405216
6.85099999827798 0.347283785176786
6.91599999845494 0.361096794144934
6.98900000134017 0.37501288526956
7.05800000287127 0.388928976394186
7.12099998781923 0.402741985362334
7.17799999983981 0.41665807648696
7.24300000001676 0.430574167611586
7.32299999799579 0.444490258736213
7.39100000646431 0.45830326770436
7.45600000664126 0.472219358828987
7.53600000462029 0.486135449953613
7.6029999909224 0.500051541078239
7.67000000632834 0.513864550046387
7.74400000227615 0.527780641171013
7.82100000651553 0.54169673229564
7.89199999417178 0.555509741263787
7.95600000128616 0.569425832388414
8.0279999965569 0.58334192351304
8.10000000637956 0.597258014637666
8.1590000045253 0.611071023605814
8.21899999573361 0.62498711473044
8.29900000826456 0.638903205855066
8.37199999659788 0.652716214823214
8.4389999974519 0.66663230594784
8.49800001014955 0.680548397072467
8.57399999222253 0.694464488197093
8.64299999375362 0.708277497165241
8.72199999867007 0.722193588289867
8.79900000290945 0.736109679414493
8.87200000579469 0.749922688382641
8.93999999971129 0.763838779507267
9.01299998804461 0.777754870631894
9.07599998754449 0.79167096175652
9.14599999669008 0.805483970724668
9.20499999483582 0.819400061849294
9.27200001024175 0.83331615297392
9.35399999434594 0.847232244098547
9.42300001042895 0.861045253066694
9.49000001128297 0.87496134419132
9.55200000316836 0.888877435315947
9.61399999505375 0.902690444284094
9.69500000064727 0.916606535408721
9.77400000556372 0.930522626533347
9.85000000218861 0.944438717657973
9.91799999610521 0.958251726626121
9.99099999899045 0.972167817750747
10.0660000025528 0.986083908875374
19.0460000012536 0.999896917843521
19.5359999925131 1
};
\addlegendentry{TW=1 N=3}
\addplot [thick, darkorange2471270, opacity=0.8]
table {%
5.82799999392591 0
5.82799999392591 0.000643086816720257
6.07099999615457 0.0141479099678457
6.15499999548774 0.0282958199356913
6.20300001173746 0.0418006430868167
6.28300000971649 0.0559485530546624
6.3640000007581 0.0694533762057878
6.43900000432041 0.0836012861736334
6.49200000043493 0.0977491961414791
6.55599999299739 0.111254019292605
6.6439999936847 0.12540192926045
6.69500000367407 0.138906752411576
6.77100000029895 0.153054662379421
6.85799999337178 0.166559485530547
6.933000011486 0.180707395498392
7.00800000049639 0.194855305466238
7.08199999644421 0.208360128617363
7.13199999881908 0.222508038585209
7.19499999831896 0.236012861736334
7.26900000881869 0.25016077170418
7.34900000679772 0.264308681672026
7.40299999597482 0.277813504823151
7.48299999395385 0.291961414790997
7.55500000377651 0.305466237942122
7.64299998991191 0.319614147909968
7.72500000311993 0.333118971061093
7.77799999923445 0.347266881028939
7.86100000550505 0.361414790996785
7.92499999806751 0.37491961414791
7.97900000179652 0.389067524115756
8.04700001026504 0.402572347266881
8.1110000028275 0.416720257234727
8.18000000435859 0.430868167202572
8.24300000385847 0.444372990353698
8.3170000143582 0.458520900321543
8.37700000556652 0.472025723472669
8.44600000709761 0.486173633440514
8.5350000008475 0.49967845659164
8.59800000034738 0.513826366559486
8.66999999561813 0.527974276527331
8.75700000324287 0.541479099678457
8.81899999512825 0.555627009646302
8.90099999378435 0.569131832797428
8.96899998770095 0.583279742765273
9.02899999346118 0.597427652733119
9.09900000260677 0.610932475884244
9.15600000007544 0.62508038585209
9.2199999926379 0.638585209003215
9.28699999349192 0.652733118971061
9.35000000754371 0.666237942122186
9.43400000687689 0.680385852090032
9.53500000468921 0.694533762057878
9.5969999965746 0.708038585209003
9.64599999133497 0.722186495176849
9.72800000454299 0.735691318327974
9.80200000049081 0.74983922829582
9.86499999999069 0.763987138263666
9.93699999526143 0.777491961414791
10.0050000037299 0.791639871382637
10.0519999978133 0.805144694533762
10.1430000067921 0.819292604501608
10.2059999917401 0.832797427652733
10.2570000017295 0.846945337620579
10.3819999931147 0.861093247588424
10.4429999919375 0.87459807073955
10.4969999956666 0.888745980707396
10.5620000103954 0.902250803858521
10.6419999938225 0.916398713826367
10.7370000041556 0.930546623794212
10.8050000126241 0.944051446945338
10.8500000060303 0.958199356913183
10.9280000033323 0.971704180064309
11.0339999955613 0.985852090032154
19.7610000032 0.99935691318328
23.4770000097342 1
};
\addlegendentry{TW=2 N=1}
\addplot [thick, darkorange2471270, opacity=0.8, dashed]
table {%
5.50199999997858 0
5.50199999997858 0.000103103412722961
5.73099999746773 0.0139189607175998
5.77999999222811 0.0278379214351995
5.81699999747798 0.0417568821527993
5.85400000272784 0.055572739457676
5.88799999968614 0.0694917001752758
5.92799999867566 0.0834106608928755
5.96500000392552 0.0972265181977523
6.00199999462347 0.111145478915352
6.03999999293592 0.125064439632952
6.07200000376906 0.138880296937829
6.11300001037307 0.152799257655428
6.14599999971688 0.166718218373028
6.18700000632089 0.180637179090628
6.23800000175834 0.194453036395505
6.29400000616442 0.208371997113104
6.34600000921637 0.222290957830704
6.41699999687262 0.236106815135581
6.49299999349751 0.250025775853181
6.56999999773689 0.26394473657078
6.62499999452848 0.277760593875657
6.70099999115337 0.291679554593257
6.78200001129881 0.305598515310857
6.85599999269471 0.319414372615734
6.92099999287166 0.333333333333333
6.99499998881947 0.347252294050933
7.06200000422541 0.361171254768533
7.13300000643358 0.37498711207341
7.20100000035018 0.388906072791009
7.26499999291264 0.402825033508609
7.3219999903813 0.416640890813486
7.3850000044331 0.430559851531086
7.45999999344349 0.444478812248685
7.53299999632873 0.458294669553562
7.60599999921396 0.472213630271162
7.67099999939092 0.486132590988762
7.74400000227615 0.499948448293638
7.80999999551568 0.513867409011238
7.87899999704678 0.527786369728838
7.94999999925494 0.541705330446438
8.01999999384861 0.555521187751315
8.08899999537971 0.569440148468914
8.16299999132752 0.583359109186514
8.23099999979604 0.597174966491391
8.29999998677522 0.611093927208991
8.36100000014994 0.62501288792659
8.43099999474362 0.638828745231467
8.51100000727456 0.652747705949067
8.57999999425374 0.666666666666667
8.64499999443069 0.680585627384266
8.72699999308679 0.694401484689143
8.79199999326374 0.708320445406743
8.86300001002382 0.722239406124343
8.94100000732578 0.736055263429219
9.00499999988824 0.749974224146819
9.07400000141934 0.763893184864419
9.14100000227336 0.777709042169296
9.20899999618996 0.791628002886896
9.28300000668969 0.805546963604495
9.34099999722093 0.819362820909372
9.40399999672081 0.833281781626972
9.46800000383519 0.847200742344572
9.55500001145992 0.861119703062171
9.62100000469945 0.874935560367048
9.70499998948071 0.888854521084648
9.77000000420958 0.902773481802248
9.83799999812618 0.916589339107124
9.91200000862591 0.930508299824724
9.98299999628216 0.944427260542324
10.0530000054277 0.958243117847201
10.1230000000214 0.972162078564801
10.21700000274 0.9860810392824
199.089999994612 0.999896896587277
209.375999998883 1
};
\addlegendentry{TW=2 N=2}
\addplot [thick, darkorange2471270, opacity=0.8, dotted]
table {%
5.51600000471808 0
5.51600000471808 0.000103082156478714
5.76299999374896 0.0139160911246263
5.81299999612384 0.0278321822492527
5.85400000272784 0.041748273373879
5.88799999968614 0.0555612823420266
5.92100000358187 0.0694773734666529
5.95599999360275 0.0833934645912792
5.99299999885261 0.0973095557159056
6.03699999919627 0.111122564684053
6.07200000376906 0.12503865580868
6.10800000140443 0.138954746933306
6.13799999700859 0.152767755901453
6.17300000158139 0.16668384702608
6.20499999786261 0.180599938150706
6.25599999330007 0.194516029275332
6.30799999635201 0.20832903824348
6.37199998891447 0.222245129368106
6.44099999044556 0.236161220492733
6.51700000162236 0.250077311617359
6.5900000045076 0.263890320585507
6.65100000333041 0.277806411710133
6.72400000621565 0.291722502834759
6.79600000148639 0.305535511802907
6.86900000437163 0.319451602927533
6.94099999964237 0.33336769405216
7.01700001081917 0.347283785176786
7.09600000118371 0.361096794144934
7.16299998748582 0.37501288526956
7.23299999663141 0.388928976394186
7.29599999613129 0.402741985362334
7.35800000256859 0.41665807648696
7.41500000003725 0.430574167611586
7.48900001053698 0.444490258736213
7.5630000064848 0.45830326770436
7.62799999210984 0.472219358828987
7.68999999854714 0.486135449953613
7.77699999161996 0.500051541078239
7.83999999111984 0.513864550046387
7.91000000026543 0.527780641171013
7.98599999689031 0.54169673229564
8.05500001297332 0.555509741263787
8.12199999927543 0.569425832388414
8.19799999590032 0.58334192351304
8.26200000301469 0.597258014637666
8.32799999625422 0.611071023605814
8.38600000133738 0.62498711473044
8.45100000151433 0.638903205855066
8.52500001201406 0.652716214823214
8.59800000034738 0.66663230594784
8.6719999962952 0.680548397072467
8.74499999918044 0.694464488197093
8.81800000206567 0.708277497165241
8.88900000427384 0.722193588289867
8.96300000022165 0.736109679414493
9.02700000733603 0.749922688382641
9.10199999634642 0.763838779507267
9.17699999990873 0.777754870631894
9.23999999940861 0.79167096175652
9.31199999467935 0.805483970724668
9.36999999976251 0.819400061849294
9.43099999858532 0.83331615297392
9.50899999588728 0.847232244098547
9.58699999318924 0.861045253066694
9.64899999962654 0.87496134419132
9.72600000386592 0.888877435315947
9.80200000049081 0.902690444284094
9.87399999576155 0.916606535408721
9.94499999796972 0.930522626533347
10.020000001532 0.944438717657973
10.0900000106776 0.958251726626121
10.1640000066254 0.972167817750747
10.2409999963129 0.986083908875374
42.4799999891547 0.999896917843521
52.8199999971548 1
};
\addlegendentry{TW=2 N=3}
\addplot [thick, darkcyan6167125, opacity=0.8]
table {%
5.45200001215562 0
5.45200001215562 0.000103082156478714
5.94699999783188 0.0139160911246263
6.04499998735264 0.0278321822492527
6.13500000326894 0.041748273373879
6.22399999701884 0.0555612823420266
6.30100000125822 0.0694773734666529
6.36900000972673 0.0833934645912792
6.44300000567455 0.0973095557159056
6.52600001194514 0.111122564684053
6.60100000095554 0.12503865580868
6.66800000180956 0.138954746933306
6.73599999572616 0.152767755901453
6.80999999167398 0.16668384702608
6.87800000014249 0.180599938150706
6.94399999338202 0.194516029275332
7.00800000049639 0.20832903824348
7.07700000202749 0.222245129368106
7.15399999171495 0.236161220492733
7.21199999679811 0.250077311617359
7.27499999629799 0.263890320585507
7.35700000950601 0.277806411710133
7.4319999985164 0.291722502834759
7.51800001307856 0.305535511802907
7.58399999176618 0.319451602927533
7.66099999600556 0.33336769405216
7.727000003797 0.347283785176786
7.80000000668224 0.361096794144934
7.86100000550505 0.37501288526956
7.92400000500493 0.388928976394186
7.98700000450481 0.402741985362334
8.05800000671297 0.41665807648696
8.13200000266079 0.430574167611586
8.19800001045223 0.444490258736213
8.2609999954002 0.45830326770436
8.33799999963958 0.472219358828987
8.40799999423325 0.486135449953613
8.48000000405591 0.500051541078239
8.54799999797251 0.513864550046387
8.62200000847224 0.527780641171013
8.70299999951385 0.54169673229564
8.77000000036787 0.555509741263787
8.83299999986775 0.569425832388414
8.88999999733642 0.58334192351304
8.96400000783615 0.597258014637666
9.03600000310689 0.611071023605814
9.10200001089834 0.62498711473044
9.1649999958463 0.638903205855066
9.22500000160653 0.652716214823214
9.30399999197107 0.66663230594784
9.37400000111666 0.680548397072467
9.44399999571033 0.694464488197093
9.51199998962693 0.708277497165241
9.58199999877252 0.722193588289867
9.65999999607448 0.736109679414493
9.72899999760557 0.749922688382641
9.79300000471994 0.763838779507267
9.86399999237619 0.777754870631894
9.93000000016764 0.79167096175652
9.99400000728201 0.805483970724668
10.0640000018757 0.819400061849294
10.126000008313 0.83331615297392
10.1959999883547 0.847232244098547
10.2619999961462 0.861045253066694
10.3410000010626 0.87496134419132
10.4049999936251 0.888877435315947
10.4849999916041 0.902690444284094
10.5610000027809 0.916606535408721
10.6320000049891 0.930522626533347
10.7049999933224 0.944438717657973
10.7759999955306 0.958251726626121
10.8619999955408 0.972167817750747
10.999999998603 0.986083908875374
17.0099999959348 0.999896917843521
20.753999997396 1
};
\addlegendentry{TW=3 N=1}
\addplot [thick, darkcyan6167125, opacity=0.8, dashed]
table {%
5.5069999943953 0
5.5069999943953 0.000103082156478714
5.90199998987373 0.0139160911246263
5.95500000054017 0.0278321822492527
5.99399999191519 0.041748273373879
6.03300001239404 0.0555612823420266
6.07099999615457 0.0694773734666529
6.10600000072736 0.0833934645912792
6.14300000597723 0.0973095557159056
6.17700000293553 0.111122564684053
6.20899999921676 0.12503865580868
6.25500000023749 0.138954746933306
6.29099999787286 0.152767755901453
6.32900001073722 0.16668384702608
6.3699999882374 0.180599938150706
6.41299999551848 0.194516029275332
6.47200000821613 0.20832903824348
6.53400000010151 0.222245129368106
6.60199999401812 0.236161220492733
6.66600000113249 0.250077311617359
6.74700000672601 0.263890320585507
6.81899998744484 0.277806411710133
6.89399999100715 0.291722502834759
6.96600000082981 0.305535511802907
7.03399999474641 0.319451602927533
7.11099999898579 0.33336769405216
7.18099999357946 0.347283785176786
7.25700000475626 0.361096794144934
7.32300001254771 0.37501288526956
7.39000001340173 0.388928976394186
7.46300000173505 0.402741985362334
7.52499999362044 0.41665807648696
7.59000000834931 0.430574167611586
7.65200000023469 0.444490258736213
7.73899999330752 0.45830326770436
7.8090000024531 0.472219358828987
7.87499999569263 0.486135449953613
7.95199999993201 0.500051541078239
8.0279999965569 0.513864550046387
8.09400000434835 0.527780641171013
8.16599999961909 0.54169673229564
8.23499998659827 0.555509741263787
8.30700001097284 0.569425832388414
8.37199999659788 0.58334192351304
8.43599998916034 0.597258014637666
8.50399999762885 0.611071023605814
8.57099999848288 0.62498711473044
8.63300000492018 0.638903205855066
8.70200000645127 0.652716214823214
8.78499999816995 0.66663230594784
8.85899999411777 0.680548397072467
8.92100000055507 0.694464488197093
9.00000000547152 0.708277497165241
9.06599999871105 0.722193588289867
9.1450000036275 0.736109679414493
9.21399999060668 0.749922688382641
9.27699999010656 0.763838779507267
9.35000000754371 0.777754870631894
9.41100000636652 0.79167096175652
9.48000000789762 0.805483970724668
9.54699999419972 0.819400061849294
9.62100000469945 0.83331615297392
9.68400000419933 0.847232244098547
9.76800000353251 0.861045253066694
9.83399999677204 0.87496134419132
9.9049999989802 0.888877435315947
9.97600000118837 0.902690444284094
10.0450000027195 0.916606535408721
10.1189999986673 0.930522626533347
10.1889999932609 0.944438717657973
10.2699999988545 0.958251726626121
10.3390000003856 0.972167817750747
10.4420000134269 0.986083908875374
61.4849999983562 0.999896917843521
69.5839999971213 1
};
\addlegendentry{TW=3 N=2}
\addplot [thick, darkcyan6167125, opacity=0.8, dotted]
table {%
5.42700001096819 0
5.42700001096819 0.000103082156478714
5.76000000000931 0.0139160911246263
5.8359999966342 0.0278321822492527
5.88099999004044 0.041748273373879
5.91700000222772 0.0555612823420266
5.95899998734239 0.0694773734666529
5.99800000782125 0.0833934645912792
6.03300001239404 0.0973095557159056
6.06700000935234 0.111122564684053
6.10400000005029 0.12503865580868
6.13999999768566 0.138954746933306
6.17599999532104 0.152767755901453
6.2189999880502 0.16668384702608
6.26199999533128 0.180599938150706
6.30999999702908 0.194516029275332
6.35399999737274 0.20832903824348
6.41600000381004 0.222245129368106
6.48599999840371 0.236161220492733
6.54099999519531 0.250077311617359
6.61399999808054 0.263890320585507
6.68099999893457 0.277806411710133
6.75600000249688 0.291722502834759
6.83600000047591 0.305535511802907
6.90899998880923 0.319451602927533
6.97400000353809 0.33336769405216
7.04600001336075 0.347283785176786
7.12799999746494 0.361096794144934
7.19000000390224 0.37501288526956
7.25999999849591 0.388928976394186
7.32599999173544 0.402741985362334
7.38499998988118 0.41665807648696
7.45599999208935 0.430574167611586
7.53300001088064 0.444490258736213
7.6029999909224 0.45830326770436
7.67099999939092 0.472219358828987
7.74099999398459 0.486135449953613
7.80700000177603 0.500051541078239
7.87799998943228 0.513864550046387
7.96300001093186 0.527780641171013
8.03399999858811 0.54169673229564
8.10500000079628 0.555509741263787
8.18099999742117 0.569425832388414
8.24700000521261 0.58334192351304
8.30999999016058 0.597258014637666
8.3769999910146 0.611071023605814
8.44599999254569 0.62498711473044
8.51799998781644 0.638903205855066
8.58000000880565 0.652716214823214
8.65699999849312 0.66663230594784
8.72100000560749 0.680548397072467
8.80899999174289 0.694464488197093
8.87900000088848 0.708277497165241
8.94200000038836 0.722193588289867
9.01400001021102 0.736109679414493
9.08300001174212 0.749922688382641
9.16299999516923 0.763838779507267
9.22999999602325 0.777754870631894
9.29500001075212 0.79167096175652
9.36599999840837 0.805483970724668
9.43099999858532 0.819400061849294
9.50699999521021 0.83331615297392
9.57400001061615 0.847232244098547
9.64600000588689 0.861045253066694
9.70500000403263 0.87496134419132
9.78799999575131 0.888877435315947
9.85899999795947 0.902690444284094
9.93800000287592 0.916606535408721
10.0099999981467 0.930522626533347
10.0899999961257 0.944438717657973
10.1680000079796 0.958251726626121
10.2509999996983 0.972167817750747
10.3720000042813 0.986083908875374
20.183000000543 0.999896917843521
23.0460000020685 1
};
\addlegendentry{TW=3 N=3}
\end{axis}

\end{tikzpicture}

%% file: figures/freqsel_runtime_preview.tex
\begin{tikzpicture}
\definecolor{darkcyan6167125}{RGB}{6,167,125}
\definecolor{darkgrey176}{RGB}{176,176,176}
\definecolor{darkorange2471270}{RGB}{247,127,0}
\definecolor{lightgrey204}{RGB}{204,204,204}
\definecolor{seagreen8416275}{RGB}{84,162,75}
\definecolor{steelblue46134171}{RGB}{46,134,171}
\begin{axis}[
height=0.54\linewidth,
legend cell align={left},
legend columns=2,
legend style={
  fill opacity=0.8,
  draw opacity=1,
  text opacity=1,
  at={(0.5,1.02)},
  anchor=south,
  draw=lightgrey204
},
log basis y={10},
minor xtick={},
minor ytick={},
tick align=inside,
tick pos=left,
width=\linewidth,
x grid style={darkgrey176},
xlabel={Candidate UEs (\(\displaystyle K=8\) scheduled)},
xmin=-0.65, xmax=5.4,
xtick style={color=black},
xtick={0,1,2,3,4,5},
xticklabels={8,12,16,32,64,128},
y grid style={darkgrey176},
ylabel={Allocator runtime (\(\displaystyle \mu\)s, log scale)},
ymajorgrids,
ymin=6, ymax=8000000,
yminorgrids,
ymode=log,
ytick style={color=black},
ytick={0.1,1,10,100,1000,10000,100000,1000000,10000000,100000000},
yticklabels={
  \(\displaystyle {10^{-1}}\),
  \(\displaystyle {10^{0}}\),
  \(\displaystyle {10^{1}}\),
  \(\displaystyle {10^{2}}\),
  \(\displaystyle {10^{3}}\),
  \(\displaystyle {10^{4}}\),
  \(\displaystyle {10^{5}}\),
  \(\displaystyle {10^{6}}\),
  \(\displaystyle {10^{7}}\),
  \(\displaystyle {10^{8}}\)
}
]
\addlegendimage{area legend,draw=black,fill=steelblue46134171,opacity=0.7,postaction={pattern=horizontal lines,pattern color=black}}
\addlegendentry{\freqselfixed}
\addlegendimage{area legend,draw=black,fill=darkorange2471270,opacity=0.7,postaction={pattern=north east lines,pattern color=black}}
\addlegendentry{\freqselrefined}
\addlegendimage{area legend,draw=black,fill=darkcyan6167125,opacity=0.7,postaction={pattern=north west lines,pattern color=black}}
\addlegendentry{exact optimum}
\addlegendimage{black,dashed}
\addlegendentry{slot duration ($500\,\mu$s)}
\draw[draw=black,fill=steelblue46134171,opacity=0.7,line width=0.22pt,postaction={pattern=horizontal lines, fill opacity=0.7}] (axis cs:-0.35,6) rectangle (axis cs:-0.15,10);
\draw[draw=black,fill=steelblue46134171,opacity=0.7,line width=0.22pt,postaction={pattern=horizontal lines, fill opacity=0.7}] (axis cs:0.65,6) rectangle (axis cs:0.85,13);
\draw[draw=black,fill=steelblue46134171,opacity=0.7,line width=0.22pt,postaction={pattern=horizontal lines, fill opacity=0.7}] (axis cs:1.65,6) rectangle (axis cs:1.85,17);
\draw[draw=black,fill=steelblue46134171,opacity=0.7,line width=0.22pt,postaction={pattern=horizontal lines, fill opacity=0.7}] (axis cs:2.65,6) rectangle (axis cs:2.85,35);
\draw[draw=black,fill=steelblue46134171,opacity=0.7,line width=0.22pt,postaction={pattern=horizontal lines, fill opacity=0.7}] (axis cs:3.65,6) rectangle (axis cs:3.85,70);
\draw[draw=black,fill=steelblue46134171,opacity=0.7,line width=0.22pt,postaction={pattern=horizontal lines, fill opacity=0.7}] (axis cs:4.65,6) rectangle (axis cs:4.85,145);
\draw[draw=black,fill=darkorange2471270,opacity=0.7,line width=0.22pt,postaction={pattern=north east lines, fill opacity=0.7}] (axis cs:-0.1,6) rectangle (axis cs:0.1,245);
\draw[draw=black,fill=darkorange2471270,opacity=0.7,line width=0.22pt,postaction={pattern=north east lines, fill opacity=0.7}] (axis cs:0.9,6) rectangle (axis cs:1.1,251);
\draw[draw=black,fill=darkorange2471270,opacity=0.7,line width=0.22pt,postaction={pattern=north east lines, fill opacity=0.7}] (axis cs:1.9,6) rectangle (axis cs:2.1,252);
\draw[draw=black,fill=darkorange2471270,opacity=0.7,line width=0.22pt,postaction={pattern=north east lines, fill opacity=0.7}] (axis cs:2.9,6) rectangle (axis cs:3.1,267);
\draw[draw=black,fill=darkorange2471270,opacity=0.7,line width=0.22pt,postaction={pattern=north east lines, fill opacity=0.7}] (axis cs:3.9,6) rectangle (axis cs:4.1,298);
\draw[draw=black,fill=darkorange2471270,opacity=0.7,line width=0.22pt,postaction={pattern=north east lines, fill opacity=0.7}] (axis cs:4.9,6) rectangle (axis cs:5.1,360);
\draw[draw=black,fill=darkcyan6167125,opacity=0.7,line width=0.22pt,postaction={pattern=north west lines, fill opacity=0.7}] (axis cs:0.15,6) rectangle (axis cs:0.35,3282);
\draw[draw=black,fill=darkcyan6167125,opacity=0.7,line width=0.22pt,postaction={pattern=north west lines, fill opacity=0.7}] (axis cs:1.15,6) rectangle (axis cs:1.35,246523);
\draw[draw=black,fill=darkcyan6167125,opacity=0.7,line width=0.22pt,postaction={pattern=north west lines, fill opacity=0.7}] (axis cs:2.15,6) rectangle (axis cs:2.35,3140000);
\path [draw=black, line width=0.32pt]
(axis cs:-0.25,10)
--(axis cs:-0.25,11);
\path [draw=black, line width=0.32pt]
(axis cs:0.75,13)
--(axis cs:0.75,14);
\path [draw=black, line width=0.32pt]
(axis cs:1.75,17)
--(axis cs:1.75,18);
\path [draw=black, line width=0.32pt]
(axis cs:2.75,35)
--(axis cs:2.75,38);
\path [draw=black, line width=0.32pt]
(axis cs:3.75,70)
--(axis cs:3.75,89);
\path [draw=black, line width=0.32pt]
(axis cs:4.75,145)
--(axis cs:4.75,168);
\addplot [semithick, black, mark=-, mark size=2, mark options={solid}, only marks]
table {%
-0.25 10
0.75 13
1.75 17
2.75 35
3.75 70
4.75 145
};
\addplot [semithick, black, mark=-, mark size=2, mark options={solid}, only marks]
table {%
-0.25 11
0.75 14
1.75 18
2.75 38
3.75 89
4.75 168
};
\path [draw=black, line width=0.32pt]
(axis cs:0,245)
--(axis cs:0,281);
\path [draw=black, line width=0.32pt]
(axis cs:1,251)
--(axis cs:1,276);
\path [draw=black, line width=0.32pt]
(axis cs:2,252)
--(axis cs:2,292);
\path [draw=black, line width=0.32pt]
(axis cs:3,267)
--(axis cs:3,293);
\path [draw=black, line width=0.32pt]
(axis cs:4,298)
--(axis cs:4,332);
\path [draw=black, line width=0.32pt]
(axis cs:5,360)
--(axis cs:5,394);
\addplot [semithick, black, mark=-, mark size=2, mark options={solid}, only marks]
table {%
0 245
1 251
2 252
3 267
4 298
5 360
};
\addplot [semithick, black, mark=-, mark size=2, mark options={solid}, only marks]
table {%
0 281
1 276
2 292
3 293
4 332
5 394
};
\path [draw=black, line width=0.32pt]
(axis cs:0.25,3282)
--(axis cs:0.25,3316);
\path [draw=black, line width=0.32pt]
(axis cs:1.25,246523)
--(axis cs:1.25,246523);
\path [draw=black, line width=0.32pt]
(axis cs:2.25,3140000)
--(axis cs:2.25,3140000);
\addplot [semithick, black, mark=-, mark size=2, mark options={solid}, only marks]
table {%
0.25 3282
1.25 246523
2.25 3140000
};
\addplot [semithick, black, mark=-, mark size=2, mark options={solid}, only marks]
table {%
0.25 3316
1.25 246523
2.25 3140000
};
\addplot [black, dashed]
table {%
-0.65 500
5.4 500
};
\draw (axis cs:3.25,6928.20323027551) node[
  text=seagreen8416275,
  rotate=90.0
]{DNF};
\draw (axis cs:4.25,6928.20323027551) node[
  text=seagreen8416275,
  rotate=90.0
]{DNF};
\draw (axis cs:5.25,6928.20323027551) node[
  text=seagreen8416275,
  rotate=90.0
]{DNF};
\end{axis}
\end{tikzpicture}

%% file: figures/f10_bound_vs_pl.tex
\begin{tikzpicture}
\definecolor{darkgrey176}{RGB}{176,176,176}
\definecolor{darkorange2471270}{RGB}{247,127,0}
\definecolor{lightgrey204}{RGB}{204,204,204}
\definecolor{steelblue46134171}{RGB}{46,134,171}
\begin{axis}[
height=0.54\linewidth,
legend cell align={left},
legend style={fill opacity=0.8, draw opacity=1, text opacity=1, draw=lightgrey204},
minor xtick={},
minor ytick={},
tick align=inside,
tick pos=left,
width=\linewidth,
x grid style={darkgrey176},
xlabel={Path loss (dB)},
xmajorgrids,
xmin=94.95, xmax=118.05,
xtick style={color=black},
xtick={90,95,100,105,110,115,120},
y grid style={darkgrey176},
ylabel={Expected cell spectral efficiency (bit/s/Hz)},
ymajorgrids,
ymin=2.3226688, ymax=11.3941092,
ytick style={color=black},
ytick={2,4,6,8,10,12}
]
\addlegendimage{line width=1pt,steelblue46134171,opacity=0.9,mark=*,mark size=2.2,mark options={solid}}
\addlegendentry{\freqselfixed}
\addlegendimage{line width=1pt,darkorange2471270,opacity=0.9,dashed,mark=square*,mark size=2.2,mark options={solid}}
\addlegendentry{\freqselrefined}
\addplot [line width=1pt, steelblue46134171, opacity=0.9, mark=*, mark size=2.2, mark options={solid}]
table {%
96 10.267492
100 9.529944
103 8.418986
106 7.272491
110 5.459256
113 4.033141
117 2.735007
};
\addplot [line width=1pt, darkorange2471270, opacity=0.9, dashed, mark=square*, mark size=2.2, mark options={solid}]
table {%
96 10.981771
100 10.114053
103 9.376344
106 8.264938
110 6.288111
113 4.831376
117 2.907888
};
\end{axis}
\end{tikzpicture}

%% file: figures/f13_pl_sweep.tex
\begin{tikzpicture}
\definecolor{darkgrey176}{RGB}{176,176,176}
\definecolor{darkorange2471270}{RGB}{247,127,0}
\definecolor{dimgrey102}{RGB}{102,102,102}
\definecolor{steelblue46134171}{RGB}{46,134,171}
\begin{groupplot}[group style={group size=2 by 1, horizontal sep=0.80cm}]
\nextgroupplot[
height=0.47\linewidth,
minor xtick={},
minor ytick={},
tick align=inside,
tick pos=left,
title={96 dB path loss},
title style={yshift=-0.12cm},
width=0.44\linewidth,
x grid style={darkgrey176},
xmin=-0.621, xmax=5.121,
xtick style={color=black},
xtick={1,4},
xticklabels={Default,{MCS\\Oracle}},
xticklabel style={align=center},
y grid style={darkgrey176},
ylabel={UL throughput (Mbit/s)},
ymajorgrids,
ymin=31.5, ymax=36.7,
ytick style={color=black},
ytick={32,33,34,35,36},
legend columns=3,
legend to name=freqselpolicylegend,
legend style={draw=lightgray, fill=white, font=\scriptsize,
  /tikz/every even column/.append style={column sep=0.20cm}},
]
\addlegendimage{area legend,draw=black,fill=dimgrey102,opacity=0.7}
\addlegendentry{PF}
\addlegendimage{area legend,draw=black,fill=steelblue46134171,opacity=0.7,postaction={pattern=horizontal lines,pattern color=black}}
\addlegendentry{\freqselfixed}
\addlegendimage{area legend,draw=black,fill=darkorange2471270,opacity=0.7,postaction={pattern=north east lines,pattern color=black}}
\addlegendentry{\freqselrefined}
\draw[draw=black,fill=dimgrey102,opacity=0.7,line width=0.22pt] (axis cs:-0.288,0) rectangle (axis cs:0.288,31.7366666666667);
\draw[draw=black,fill=steelblue46134171,opacity=0.7,line width=0.22pt,postaction={pattern=horizontal lines, fill opacity=0.7}] (axis cs:0.712,0) rectangle (axis cs:1.288,33.3166666666667);
\draw[draw=black,fill=darkorange2471270,opacity=0.7,line width=0.22pt,postaction={pattern=north east lines, fill opacity=0.7}] (axis cs:1.712,0) rectangle (axis cs:2.288,32.9733333333333);
\draw[draw=black,fill=steelblue46134171,opacity=0.7,line width=0.22pt,postaction={pattern=horizontal lines, fill opacity=0.7}] (axis cs:3.212,0) rectangle (axis cs:3.788,36.3633333333333);
\draw[draw=black,fill=darkorange2471270,opacity=0.7,line width=0.22pt,postaction={pattern=north east lines, fill opacity=0.7}] (axis cs:4.212,0) rectangle (axis cs:4.788,35.5933333333333);
\path [draw=black, semithick]
(axis cs:0,31.7077991532072)
--(axis cs:0,31.7655341801261);
\addplot [semithick, black, mark=-, mark size=3, mark options={solid}, only marks]
table {%
0 31.7077991532072
};
\addplot [semithick, black, mark=-, mark size=3, mark options={solid}, only marks]
table {%
0 31.7655341801261
};
\path [draw=black, semithick]
(axis cs:1,33.3013914143501)
--(axis cs:1,33.3319419189832);
\addplot [semithick, black, mark=-, mark size=3, mark options={solid}, only marks]
table {%
1 33.3013914143501
};
\addplot [semithick, black, mark=-, mark size=3, mark options={solid}, only marks]
table {%
1 33.3319419189832
};
\path [draw=black, semithick]
(axis cs:2,32.9329188144901)
--(axis cs:2,33.0137478521766);
\addplot [semithick, black, mark=-, mark size=3, mark options={solid}, only marks]
table {%
2 32.9329188144901
};
\addplot [semithick, black, mark=-, mark size=3, mark options={solid}, only marks]
table {%
2 33.0137478521766
};
\path [draw=black, semithick]
(axis cs:3.5,36.3381672185491)
--(axis cs:3.5,36.3884994481176);
\addplot [semithick, black, mark=-, mark size=3, mark options={solid}, only marks]
table {%
3.5 36.3381672185491
};
\addplot [semithick, black, mark=-, mark size=3, mark options={solid}, only marks]
table {%
3.5 36.3884994481176
};
\path [draw=black, semithick]
(axis cs:4.5,35.5702393225657)
--(axis cs:4.5,35.6164273441009);
\addplot [semithick, black, mark=-, mark size=3, mark options={solid}, only marks]
table {%
4.5 35.5702393225657
};
\addplot [semithick, black, mark=-, mark size=3, mark options={solid}, only marks]
table {%
4.5 35.6164273441009
};
\addplot [line width=0.32pt, darkgrey176]
table {%
2.75 31.5
2.75 36.7
};
\nextgroupplot[
height=0.47\linewidth,
minor xtick={},
minor ytick={},
tick align=inside,
tick pos=left,
title={110 dB path loss},
title style={yshift=-0.12cm},
width=0.44\linewidth,
x grid style={darkgrey176},
xmin=-0.621, xmax=5.121,
xtick style={color=black},
xtick={1,4},
xticklabels={Default,{MCS\\Oracle}},
xticklabel style={align=center},
y grid style={darkgrey176},
ymajorgrids,
ymin=5, ymax=7.4,
ytick style={color=black},
ytick={5,5.5,6,6.5,7}
]
\draw[draw=black,fill=dimgrey102,opacity=0.7,line width=0.22pt] (axis cs:-0.288,0) rectangle (axis cs:0.288,5.21666666666667);
\draw[draw=black,fill=steelblue46134171,opacity=0.7,line width=0.22pt,postaction={pattern=horizontal lines, fill opacity=0.7}] (axis cs:0.712,0) rectangle (axis cs:1.288,5.58333333333333);
\draw[draw=black,fill=darkorange2471270,opacity=0.7,line width=0.22pt,postaction={pattern=north east lines, fill opacity=0.7}] (axis cs:1.712,0) rectangle (axis cs:2.288,5.38666666666667);
\draw[draw=black,fill=steelblue46134171,opacity=0.7,line width=0.22pt,postaction={pattern=horizontal lines, fill opacity=0.7}] (axis cs:3.212,0) rectangle (axis cs:3.788,7.2);
\draw[draw=black,fill=darkorange2471270,opacity=0.7,line width=0.22pt,postaction={pattern=north east lines, fill opacity=0.7}] (axis cs:4.212,0) rectangle (axis cs:4.788,6.775);
\path [draw=black, semithick]
(axis cs:0,5.21089316397477)
--(axis cs:0,5.22244016935856);
\addplot [semithick, black, mark=-, mark size=3, mark options={solid}, only marks]
table {%
0 5.21089316397477
};
\addplot [semithick, black, mark=-, mark size=3, mark options={solid}, only marks]
table {%
0 5.22244016935856
};
\path [draw=black, semithick]
(axis cs:1,5.57755983064144)
--(axis cs:1,5.58910683602523);
\addplot [semithick, black, mark=-, mark size=3, mark options={solid}, only marks]
table {%
1 5.57755983064144
};
\addplot [semithick, black, mark=-, mark size=3, mark options={solid}, only marks]
table {%
1 5.58910683602523
};
\path [draw=black, semithick]
(axis cs:2,5.37511966128287)
--(axis cs:2,5.39821367205046);
\addplot [semithick, black, mark=-, mark size=3, mark options={solid}, only marks]
table {%
2 5.37511966128287
};
\addplot [semithick, black, mark=-, mark size=3, mark options={solid}, only marks]
table {%
2 5.39821367205046
};
\path [draw=black, semithick]
(axis cs:3.5,7.2)
--(axis cs:3.5,7.2);
\addplot [semithick, black, mark=-, mark size=3, mark options={solid}, only marks]
table {%
3.5 7.2
};
\addplot [semithick, black, mark=-, mark size=3, mark options={solid}, only marks]
table {%
3.5 7.2
};
\path [draw=black, semithick]
(axis cs:4.5,6.76792893218813)
--(axis cs:4.5,6.78207106781187);
\addplot [semithick, black, mark=-, mark size=3, mark options={solid}, only marks]
table {%
4.5 6.76792893218813
};
\addplot [semithick, black, mark=-, mark size=3, mark options={solid}, only marks]
table {%
4.5 6.78207106781187
};
\addplot [line width=0.32pt, darkgrey176]
table {%
2.75 5
2.75 7.4
};
\end{groupplot}
\node[anchor=south] at ($(group c1r1.north west)!0.5!(group c2r1.north east)+(0,0.42cm)$) {\ref{freqselpolicylegend}};
\end{tikzpicture}

%% file: figures/f2_gain_vs_speed.tex
\begin{tikzpicture}
\definecolor{darkgrey176}{RGB}{176,176,176}
\definecolor{dimgrey102}{RGB}{102,102,102}
\definecolor{lightgrey204}{RGB}{204,204,204}
\definecolor{steelblue46134171}{RGB}{46,134,171}
\begin{axis}[
height=0.56\linewidth,
legend cell align={left},
legend style={fill opacity=0.8, draw opacity=1, text opacity=1, draw=lightgrey204},
minor xtick={},
minor ytick={},
tick align=inside,
tick pos=left,
width=\linewidth,
x grid style={darkgrey176},
xlabel={UE mobility},
xmin=-0.518, xmax=2.518,
xtick style={color=black},
xtick={0,1,2},
xticklabels={static
(0.5 Hz),1 km/h,5 km/h},
y grid style={darkgrey176},
ylabel={UL throughput (Mbit/s)},
ymajorgrids,
ymin=30.5, ymax=35,
ytick style={color=black},
ytick={31,32,33,34,35}
]
\addlegendimage{area legend,draw=black,fill=dimgrey102,opacity=0.7}
\addlegendentry{PF}
\addlegendimage{area legend,draw=black,fill=steelblue46134171,opacity=0.7,postaction={pattern=horizontal lines,pattern color=black}}
\addlegendentry{\freqselfixed}
\draw[draw=black,fill=dimgrey102,opacity=0.7,line width=0.22pt] (axis cs:-0.342,0) rectangle (axis cs:-0.038,31.8833333333333);
\draw[draw=black,fill=dimgrey102,opacity=0.7,line width=0.22pt] (axis cs:0.658,0) rectangle (axis cs:0.962,31.55);
\draw[draw=black,fill=dimgrey102,opacity=0.7,line width=0.22pt] (axis cs:1.658,0) rectangle (axis cs:1.962,31.56);
\draw[draw=black,fill=steelblue46134171,opacity=0.7,line width=0.22pt,postaction={pattern=horizontal lines, fill opacity=0.7}] (axis cs:0.038,0) rectangle (axis cs:0.342,34.6425);
\draw[draw=black,fill=steelblue46134171,opacity=0.7,line width=0.22pt,postaction={pattern=horizontal lines, fill opacity=0.7}] (axis cs:1.038,0) rectangle (axis cs:1.342,32.6825);
\draw[draw=black,fill=steelblue46134171,opacity=0.7,line width=0.22pt,postaction={pattern=horizontal lines, fill opacity=0.7}] (axis cs:2.038,0) rectangle (axis cs:2.342,31.875);
\path [draw=black, semithick]
(axis cs:-0.19,31.8511878307967)
--(axis cs:-0.19,31.91547883587);
\path [draw=black, semithick]
(axis cs:0.81,31.53)
--(axis cs:0.81,31.57);
\path [draw=black, semithick]
(axis cs:1.81,31.5225834261323)
--(axis cs:1.81,31.5974165738677);
\addplot [semithick, black, mark=-, mark size=3, mark options={solid}, only marks]
table {%
-0.19 31.8511878307967
0.81 31.53
1.81 31.5225834261323
};
\addplot [semithick, black, mark=-, mark size=3, mark options={solid}, only marks]
table {%
-0.19 31.91547883587
0.81 31.57
1.81 31.5974165738677
};
\path [draw=black, semithick]
(axis cs:0.19,34.6299169426079)
--(axis cs:0.19,34.6550830573921);
\path [draw=black, semithick]
(axis cs:1.19,32.6413701244025)
--(axis cs:1.19,32.7236298755975);
\path [draw=black, semithick]
(axis cs:2.19,31.8679289321881)
--(axis cs:2.19,31.8820710678119);
\addplot [semithick, black, mark=-, mark size=3, mark options={solid}, only marks]
table {%
0.19 34.6299169426079
1.19 32.6413701244025
2.19 31.8679289321881
};
\addplot [semithick, black, mark=-, mark size=3, mark options={solid}, only marks]
table {%
0.19 34.6550830573921
1.19 32.7236298755975
2.19 31.8820710678119
};
\end{axis}
\end{tikzpicture}

%% file: figures/f5_proportion_law.tex
\begin{tikzpicture}
\definecolor{darkgrey176}{RGB}{176,176,176}
\definecolor{dimgrey102}{RGB}{102,102,102}
\definecolor{lightgrey204}{RGB}{204,204,204}
\definecolor{steelblue46134171}{RGB}{46,134,171}
\begin{axis}[
height=0.55\linewidth,
legend cell align={left},
legend style={fill opacity=0.8, draw opacity=1, text opacity=1, draw=lightgrey204},
minor xtick={},
minor ytick={},
tick align=inside,
tick pos=left,
width=\linewidth,
x grid style={darkgrey176},
xlabel={mobile UEs (of 32)},
xmajorgrids,
xmin=-1.6, xmax=33.6,
xtick style={color=black},
xtick={0,8,16,24,32},
y grid style={darkgrey176},
ylabel={UL throughput (Mbit/s)},
ymajorgrids,
ymin=28.0698809043386, ymax=31.2158649062998,
ytick style={color=black},
ytick={28,28.5,29,29.5,30,30.5,31,31.5}
]
\addlegendimage{line width=1pt,dimgrey102,opacity=0.9,mark=square*,mark size=2.2,mark options={solid}}
\addlegendentry{PF}
\addlegendimage{line width=1pt,steelblue46134171,opacity=0.9,mark=*,mark size=2.2,mark options={solid}}
\addlegendentry{\freqselfixed}
\path [draw=dimgrey102, semithick]
(axis cs:0,28.432443871707)
--(axis cs:0,29.639556128293);
\path [draw=dimgrey102, semithick]
(axis cs:8,28.2179166717424)
--(axis cs:8,29.5860833282576);
\path [draw=dimgrey102, semithick]
(axis cs:16,28.4050779152572)
--(axis cs:16,29.4189220847428);
\path [draw=dimgrey102, semithick]
(axis cs:24,28.5987893035715)
--(axis cs:24,29.8852106964285);
\path [draw=dimgrey102, semithick]
(axis cs:32,28.212880177155)
--(axis cs:32,29.423119822845);
\addplot [semithick, dimgrey102, mark=-, mark size=3, mark options={solid}, only marks]
table {%
0 28.432443871707
8 28.2179166717424
16 28.4050779152572
24 28.5987893035715
32 28.212880177155
};
\addplot [semithick, dimgrey102, mark=-, mark size=3, mark options={solid}, only marks]
table {%
0 29.639556128293
8 29.5860833282576
16 29.4189220847428
24 29.8852106964285
32 29.423119822845
};
\path [draw=steelblue46134171, semithick]
(axis cs:0,30.9911343665166)
--(axis cs:0,31.0728656334834);
\path [draw=steelblue46134171, semithick]
(axis cs:8,30.4199271110042)
--(axis cs:8,30.6120728889958);
\path [draw=steelblue46134171, semithick]
(axis cs:16,29.6763469253667)
--(axis cs:16,29.8316530746333);
\path [draw=steelblue46134171, semithick]
(axis cs:24,28.9366501287923)
--(axis cs:24,29.1113498712077);
\path [draw=steelblue46134171, semithick]
(axis cs:32,28.3110797562396)
--(axis cs:32,28.4489202437605);
\addplot [semithick, steelblue46134171, mark=-, mark size=3, mark options={solid}, only marks]
table {%
0 30.9911343665166
8 30.4199271110042
16 29.6763469253667
24 28.9366501287923
32 28.3110797562396
};
\addplot [semithick, steelblue46134171, mark=-, mark size=3, mark options={solid}, only marks]
table {%
0 31.0728656334834
8 30.6120728889958
16 29.8316530746333
24 29.1113498712077
32 28.4489202437605
};
\addplot [line width=1pt, dimgrey102, opacity=0.9, mark=square*, mark size=2.2, mark options={solid}]
table {%
0 29.036
8 28.902
16 28.912
24 29.242
32 28.818
};
\addplot [line width=1pt, steelblue46134171, opacity=0.9, mark=*, mark size=2.2, mark options={solid}]
table {%
0 31.032
8 30.516
16 29.754
24 29.024
32 28.38
};
\end{axis}
\end{tikzpicture}

%% file: figures/f8_class_cdf.tex
\begin{tikzpicture}
\definecolor{darkgrey176}{RGB}{176,176,176}
\definecolor{dimgrey102}{RGB}{102,102,102}
\definecolor{lightgrey204}{RGB}{204,204,204}
\definecolor{steelblue46134171}{RGB}{46,134,171}
\begin{axis}[
height=0.55\linewidth,
legend cell align={left},
legend columns=2,
legend style={
  font=\fontsize{6.2}{7.1}\selectfont,
  fill opacity=0.8,
  draw opacity=1,
  text opacity=1,
  at={(0.5,1.02)},
  anchor=south,
  draw=lightgrey204
},
minor xtick={},
minor ytick={},
tick align=inside,
tick pos=left,
width=\linewidth,
x grid style={darkgrey176},
xlabel={per-UE UL throughput (Mbit/s)},
xmajorgrids,
xmin=0.82395, xmax=1.04505,
xtick style={color=black},
xtick={0.8,0.85,0.9,0.95,1,1.05},
y grid style={darkgrey176},
ylabel={CDF},
ymajorgrids,
ymin=0, ymax=1,
ytick style={color=black},
ytick={0,0.2,0.4,0.6,0.8,1}
]
\addlegendimage{line width=1pt,dimgrey102,opacity=0.9}
\addlegendentry{PF, static (0.5 Hz)}
\addlegendimage{line width=1pt,steelblue46134171,opacity=0.9}
\addlegendentry{\freqselfixed, static (0.5 Hz)}
\addlegendimage{line width=1pt,dimgrey102,opacity=0.9,dashed}
\addlegendentry{PF, mobile}
\addlegendimage{line width=1pt,steelblue46134171,opacity=0.9,dashed}
\addlegendentry{\freqselfixed, mobile}
\addplot [line width=1pt, dimgrey102, opacity=0.9]
table {%
0.839 0
0.839 0.0125
0.84 0.025
0.841 0.0375
0.863 0.05
0.864 0.0625
0.864 0.075
0.865 0.0875
0.865 0.1
0.866 0.1125
0.867 0.125
0.87 0.1375
0.87 0.15
0.872 0.1625
0.877 0.175
0.877 0.1875
0.879 0.2
0.88 0.2125
0.881 0.225
0.881 0.2375
0.882 0.25
0.882 0.2625
0.884 0.275
0.884 0.2875
0.889 0.3
0.891 0.3125
0.893 0.325
0.893 0.3375
0.896 0.35
0.897 0.3625
0.899 0.375
0.9 0.3875
0.9 0.4
0.903 0.4125
0.904 0.425
0.905 0.4375
0.906 0.45
0.907 0.4625
0.908 0.475
0.909 0.4875
0.91 0.5
0.912 0.5125
0.913 0.525
0.914 0.5375
0.915 0.55
0.915 0.5625
0.916 0.575
0.921 0.5875
0.922 0.6
0.923 0.6125
0.924 0.625
0.925 0.6375
0.928 0.65
0.932 0.6625
0.935 0.675
0.937 0.6875
0.937 0.7
0.938 0.7125
0.939 0.725
0.941 0.7375
0.942 0.75
0.943 0.7625
0.947 0.775
0.949 0.7875
0.949 0.8
0.95 0.8125
0.953 0.825
0.954 0.8375
0.955 0.85
0.956 0.8625
0.956 0.875
0.957 0.8875
0.96 0.9
0.964 0.9125
0.967 0.925
0.967 0.9375
0.968 0.95
0.974 0.9625
0.975 0.975
0.998 0.9875
1.012 1
};
\addplot [line width=1pt, dimgrey102, opacity=0.9, dashed]
table {%
0.834 0
0.834 0.0125
0.839 0.025
0.844 0.0375
0.847 0.05
0.848 0.0625
0.855 0.075
0.856 0.0875
0.859 0.1
0.86 0.1125
0.86 0.125
0.861 0.1375
0.862 0.15
0.863 0.1625
0.863 0.175
0.868 0.1875
0.869 0.2
0.869 0.2125
0.869 0.225
0.871 0.2375
0.872 0.25
0.873 0.2625
0.874 0.275
0.874 0.2875
0.875 0.3
0.876 0.3125
0.876 0.325
0.878 0.3375
0.878 0.35
0.881 0.3625
0.883 0.375
0.883 0.3875
0.886 0.4
0.887 0.4125
0.887 0.425
0.888 0.4375
0.888 0.45
0.889 0.4625
0.889 0.475
0.889 0.4875
0.892 0.5
0.893 0.5125
0.894 0.525
0.897 0.5375
0.897 0.55
0.897 0.5625
0.898 0.575
0.898 0.5875
0.898 0.6
0.898 0.6125
0.898 0.625
0.899 0.6375
0.9 0.65
0.902 0.6625
0.902 0.675
0.903 0.6875
0.904 0.7
0.905 0.7125
0.905 0.725
0.913 0.7375
0.915 0.75
0.916 0.7625
0.916 0.775
0.917 0.7875
0.917 0.8
0.918 0.8125
0.921 0.825
0.923 0.8375
0.923 0.85
0.926 0.8625
0.926 0.875
0.926 0.8875
0.928 0.9
0.929 0.9125
0.936 0.925
0.938 0.9375
0.939 0.95
0.945 0.9625
0.947 0.975
0.97 0.9875
0.981 1
};
\addplot [line width=1pt, steelblue46134171, opacity=0.9]
table {%
0.86 0
0.86 0.0125
0.869 0.025
0.871 0.0375
0.873 0.05
0.873 0.0625
0.877 0.075
0.888 0.0875
0.895 0.1
0.896 0.1125
0.899 0.125
0.912 0.1375
0.921 0.15
0.923 0.1625
0.923 0.175
0.93 0.1875
0.932 0.2
0.932 0.2125
0.933 0.225
0.936 0.2375
0.936 0.25
0.94 0.2625
0.94 0.275
0.942 0.2875
0.942 0.3
0.943 0.3125
0.943 0.325
0.951 0.3375
0.951 0.35
0.955 0.3625
0.96 0.375
0.963 0.3875
0.963 0.4
0.963 0.4125
0.964 0.425
0.964 0.4375
0.964 0.45
0.965 0.4625
0.967 0.475
0.967 0.4875
0.967 0.5
0.972 0.5125
0.972 0.525
0.973 0.5375
0.973 0.55
0.973 0.5625
0.974 0.575
0.975 0.5875
0.976 0.6
0.976 0.6125
0.977 0.625
0.978 0.6375
0.98 0.65
0.981 0.6625
0.983 0.675
0.983 0.6875
0.985 0.7
0.985 0.7125
0.986 0.725
0.988 0.7375
0.991 0.75
1.007 0.7625
1.009 0.775
1.01 0.7875
1.01 0.8
1.011 0.8125
1.013 0.825
1.014 0.8375
1.015 0.85
1.017 0.8625
1.018 0.875
1.021 0.8875
1.021 0.9
1.022 0.9125
1.024 0.925
1.024 0.9375
1.028 0.95
1.028 0.9625
1.029 0.975
1.031 0.9875
1.035 1
};
\addplot [line width=1pt, steelblue46134171, opacity=0.9, dashed]
table {%
0.839 0
0.839 0.0125
0.849 0.025
0.849 0.0375
0.85 0.05
0.851 0.0625
0.855 0.075
0.855 0.0875
0.857 0.1
0.858 0.1125
0.859 0.125
0.859 0.1375
0.861 0.15
0.862 0.1625
0.862 0.175
0.862 0.1875
0.863 0.2
0.865 0.2125
0.865 0.225
0.865 0.2375
0.869 0.25
0.872 0.2625
0.872 0.275
0.872 0.2875
0.874 0.3
0.874 0.3125
0.875 0.325
0.875 0.3375
0.875 0.35
0.876 0.3625
0.876 0.375
0.876 0.3875
0.877 0.4
0.878 0.4125
0.878 0.425
0.879 0.4375
0.88 0.45
0.88 0.4625
0.883 0.475
0.884 0.4875
0.889 0.5
0.896 0.5125
0.898 0.525
0.899 0.5375
0.9 0.55
0.902 0.5625
0.902 0.575
0.903 0.5875
0.904 0.6
0.905 0.6125
0.905 0.625
0.908 0.6375
0.91 0.65
0.91 0.6625
0.912 0.675
0.916 0.6875
0.916 0.7
0.917 0.7125
0.919 0.725
0.919 0.7375
0.92 0.75
0.923 0.7625
0.923 0.775
0.925 0.7875
0.926 0.8
0.927 0.8125
0.927 0.825
0.927 0.8375
0.928 0.85
0.93 0.8625
0.932 0.875
0.933 0.8875
0.933 0.9
0.933 0.9125
0.948 0.925
0.952 0.9375
0.953 0.95
0.957 0.9625
0.957 0.975
0.958 0.9875
0.96 1
};
\end{axis}
\end{tikzpicture}

%% file: figures/f9_levy_state_cdf.tex
\begin{tikzpicture}
\definecolor{darkgrey176}{RGB}{176,176,176}
\definecolor{dimgrey102}{RGB}{102,102,102}
\definecolor{lightgrey204}{RGB}{204,204,204}
\definecolor{steelblue46134171}{RGB}{46,134,171}
\begin{axis}[
height=0.55\linewidth,
legend cell align={left},
legend columns=2,
legend style={
  font=\fontsize{6.2}{7.1}\selectfont,
  fill opacity=0.8,
  draw opacity=1,
  text opacity=1,
  at={(0.5,1.02)},
  anchor=south,
  draw=lightgrey204
},
minor xtick={},
minor ytick={},
tick align=inside,
tick pos=left,
width=\linewidth,
x grid style={darkgrey176},
xlabel={per-grant MCS (initial transmissions)},
xmajorgrids,
xmin=19, xmax=29,
xtick style={color=black},
xtick={18,20,22,24,26,28,30},
y grid style={darkgrey176},
ylabel={CDF},
ymajorgrids,
ymin=0, ymax=1,
ytick style={color=black},
ytick={0,0.2,0.4,0.6,0.8,1}
]
\addlegendimage{line width=1pt,dimgrey102,opacity=0.9}
\addlegendentry{PF, paused (0.5 Hz)}
\addlegendimage{line width=1pt,steelblue46134171,opacity=0.9}
\addlegendentry{\freqselfixed, paused (0.5 Hz)}
\addlegendimage{line width=1pt,dimgrey102,opacity=0.9,dashed}
\addlegendentry{PF, moving (5 km/h)}
\addlegendimage{line width=1pt,steelblue46134171,opacity=0.9,dashed}
\addlegendentry{\freqselfixed, moving (5 km/h)}
\addplot [line width=1pt, dimgrey102, opacity=0.9]
table {%
1 0
1 1.87534224996062e-06
1 0.000390071187991809
19 0.000391946530241769
19 0.000405073925991493
20 0.000406949268241454
20 0.000645117733986453
21 0.000646993076236413
21 0.00603672670262323
22 0.00603860204487319
22 0.0489651861464717
23 0.0489670614887217
23 0.204472316197706
24 0.204474191539956
24 0.566604655349601
25 0.566606530691851
25 0.886721826733379
26 0.886723702075629
26 0.983108792354605
27 0.983110667696855
27 0.995480425177595
28 0.995482300519845
28 1
};
\addplot [line width=1pt, dimgrey102, opacity=0.9, dashed]
table {%
1 0
1 3.67443073881779e-06
1 0.000121256214380987
21 0.000124930645119805
21 0.000334373197232419
22 0.000338047627971237
22 0.0195994135608541
23 0.0196030879915929
23 0.198095909991145
24 0.198099584421883
24 0.742510591546605
25 0.742514265977343
25 0.988091169975492
26 0.98809484440623
26 0.999261439421498
27 0.999265113852236
27 0.999992651138522
28 0.999996325569261
28 1
};
\addplot [line width=1pt, steelblue46134171, opacity=0.9]
table {%
1 0
1 1.88000315840531e-06
1 0.000511360859086243
20 0.000513240862244649
20 0.000530160890670296
21 0.000532040893828702
21 0.00125772211297315
22 0.00125960211613156
22 0.00961433615208473
23 0.00961621615524314
23 0.0466240783284516
24 0.04662595833161
24 0.195595528600488
25 0.195597408603646
25 0.516654947980313
26 0.516656827983471
26 0.807247036175021
27 0.807248916178179
27 0.972300033464056
28 0.972301913467215
28 1
};
\addplot [line width=1pt, steelblue46134171, opacity=0.9, dashed]
table {%
1 0
1 3.65573237114457e-06
1 0.000230311139382108
21 0.000233966871753253
21 0.00047158947587765
22 0.000475245208248795
22 0.0157050262664371
23 0.0157086819988082
23 0.172397027158436
24 0.172400682890807
24 0.713715942283297
25 0.713719598015668
25 0.981940682086546
26 0.981944337818917
26 0.998515772657315
27 0.998519428389686
27 0.999963442676289
28 0.99996709840866
28 1
};
\end{axis}
\end{tikzpicture}